\documentclass[useAMS]{mn2e}
\usepackage{mnras_cite}
\usepackage{epsfig}

\def\gsim{\ga}
\def\lsim{\la}
\def\d{{\rm d}}

\begin{document}

\title[Luminosity and stellar mass functions of disks and spheroids in the
SDSS and the SMBH mass function]{Luminosity and stellar mass functions
of disks and spheroids in the SDSS and the supermassive black hole mass function}  
\author[A.~J.~Benson, Dajana D\v{z}anovi\'c, C. S. Frenk, Ray
Sharples]{A.~J.~Benson$^1$, Dajana D\v{z}anovi\'c$^2$, 
C. S. Frenk$^2$, Ray Sharples$^2$\\ 
1. Theoretical Astrophysics, Caltech, MC130-33, 1200 E. California
Blvd., Pasadena, CA 91125, U.S.A. (e-mail: abenson@its.caltech.edu)\\ 
2. Institute for Computational Cosmology, University of Durham,
Science Laboratories, South Road, Durham, DH1 3LE, U.K.\\ 
} 

\maketitle

\begin{abstract}
Using the {\sc Galactica} code of Benson et al., we obtain
quantitative measurements of spheroid-to-disk ratios for a sample of
8839 galaxies observed in the Sloan Digital Sky Survey. We carry out
extensive tests of this code and of {\sc Gim2D}, finding that they
perform similarly in all respects. From the spheroid and disk
luminosities, we construct luminosity and stellar mass functions for
each component and estimate the relative luminosity and stellar mass
densities of disks and spheroids in the local Universe. Assuming a
simple one-to-one mapping between between spheroid mass and the mass of
a central supermassive black hole, we provide the most accurate
determination so far of the black hole mass function in the local
universe. From this, we infer a cosmological mass density of black
holes of $\rho_{\bullet}=(3.77 \pm 0.97) \times 10^5 h  M_{\odot} {\rm
Mpc}^{-3}$. We compare our results to
predictions from current hierarchical models of galaxy formation and
these are found to fare well in predicting the qualitative trends
observed. We find that stars in disks contribute 35--51\% of the local stellar mass density.
\end{abstract}

\begin{keywords}
galaxies: structure; galaxies: abundances; galaxies: bulges; galaxies: luminosity function, mass function; galaxies: statistics
\end{keywords}

\section{Introduction}

The distinction between disks and spheroids is one of the defining
properties of galaxies. Determining the relative importance of these
two basic types of galactic component is fundamental to a broad
characterization of the galaxy population. Yet, this is a complicated
task which requires not only high quality imaging for large samples,
but also software capable of decomposing the light from each object
into a disk and a spheroid\footnote{We will use the term ``spheroid''
throughout to refer to both elliptical galaxies and the bulges of
spiral galaxies.}. The determination of spheroid luminosities has
recently received even more prominence since the discovery that
perhaps all galaxies harbour a supermassive black hole at
their centre whose mass is proportional to the luminosity of the
spheroid or bulge \cite{kr95,mag98,mf01,mh03,harrix}.

From a theoretical point of view, explaining why most of the stars in
the Universe end up either in disks or in spheroids and understanding
the physical processes that result in the formation of one or the
other of these morphological structures is a major challenge. The
current theoretical framework used to investigate galaxy formation 
is the cold dark matter model (Peebles 1982, Blumenthal et
al. 1984, Davis et al. 1985), in
which galaxies build up hierarchically. Within this model, the basic
processes thought to be responsible for the distinction between disks
and spheroids were identified over twenty years ago (Fall 1979, Frenk et al. 1985): disks result from the collapse of
rotating gas cooling within dark matter halos whereas spheroids result
from major mergers or disk instabilities \cite{fallef,bar92,mmw}. The traditional categories of galaxy morphology, spirals,
irregulars, etc, are too detailed for current theoretical models to
explain, but the relative luminosities and stellar masses of spheroids
and disks can readily be predicted
\cite{kwg93,baugh96a,baugh96b,kcw96,kc98,sp99,hatton03}. Thus, accurate
measurements of these quantities, for example, as a function of
absolute magnitude and in different environments provides a powerful
test of models of galaxy formation and evolution.

An early attempt to determine the relative contributions of spheroids
and disks to the luminosity density of the Universe was made by
\scite{sd87}. They studied a magnitude limited sample of $\sim 200$
galaxies brighter than $V=16.5$, drawn from the catalogue of
\scite{d80cat}, and determined spheroid-to-disk ratios
by visual inspection. From this, they derived the distribution of
spheroid-to-disk ratios, as a function of absolute magnitude, and
found the overall spheroid-to-disk ratio to be higher in high density
environments (galaxy clusters) than in low density environments (the
``field''). \scite{sd87} found that disks appear to contribute roughly
twice as much as spheroids to the mean luminosity density of the
Universe. Since a large fraction of the disk light comes from a
relatively small number of young stars, \scite{sd87} concluded that
the relative contribution of spheroid and disk components to the mean
stellar mass density of the Universe is very nearly equal.

More recently, \scite{benson02} developed a quantitative method to
determine galaxy morphology, specifically to estimate
spheroid-to-total (S/T) light ratios. This method is implemented in
the code {\sc Galactica} (GALaxy Automated ComponenT Image Construction Algorithm). \scite{benson02} analyzed a
magnitude-limited sample of $\sim 100$ field galaxies brighter than
$I=16.0$ and found the luminosity functions of spheroids and disks to
be remarkably similar. They provisionally concluded that spheroids and
disks contribute almost equally to the total stellar mass density in
the Universe but stressed the significant uncertainties in their
result arising from the small sample size.  A larger sample of 1800
galaxies drawn from the Sloan Digital Sky Survey (SDSS;
\pcite{sdssedr}) was analyzed by \scite{tw2} using the publicly
available code {\sc Gim2D}. They found that $54 \pm 2\%$ of the local cosmic
luminosity density in both r and i bands comes from disks, $32 \pm 2\%$ from
``pure bulge'' systems and the remaining $14 \pm 2\%$ from bulges in
galaxies with detectable disks.

In this paper, we perform spheroid/disk decompositions from r-band
images of a much larger sample of galaxies in the SDSS. The structure
of the paper is as follows. In \S\ref{sec:SDSS}, we describe our
dataset and how it is processed. In \S\ref{sec:decomp}, we present
results from the spheroid/disk decomposition. In
\S\ref{sec:lfs}, we derive  the luminosity functions of disks and
spheroids, and present stellar mass functions and also the supermassive
black hole mass function. Finally, in
\S\ref{sec:conc}, we give our conclusions. Appendices describe
extensive tests of the reliability of both {\sc Galactica} and {\sc
Gim2D} (Appendix~\ref{Two}), technicalities of the fitting process
(Appendix~\ref{sec:tech}) and comparisons of our S/T ratio
morphologies with more traditional morphological measures
(Appendix~\ref{sec:bt_morph}).  A cosmological model with
$\Omega_0=0.3$, $\Lambda_0=0.7$ is adopted throughout and the Hubble
constant is defined to be $H_0=100h$kms$^{-1}$ Mpc$^{-1}$.

\section{Data: Sloan Digital Sky Survey}
\label{sec:SDSS}

\subsection{Basic properties}

The Sloan Digital Sky Survey (SDSS) is the largest imaging and
spectroscopic survey to date. The SDSS Early Data Release (EDR), made
publicly available in 2001, consists of a $462$ square degree area
imaged in five passbands ($u, g, r, i$ and $z$) and also covered
spectroscopically. The SDSS EDR galaxy catalogue is spectroscopically
complete down to $r = 17.7$ and contains measurements of various
galaxy parameters \cite{sdssedr}. The imaging data were taken with a
dedicated $2.5$m telescope in the drift-scan (time-delay) integration
mode with an effective exposure time of $54$s. The data used in this
study are the $r$-band imaging frames with corrections for bias, flat
field, cosmic ray and pixel defects \cite{lup01}. Each imaging frame
is a $2048 \times 1489$ pixel array with a pixel size of $0.394''$.

\subsection{SDSS apparent magnitude limit}

\scite{benson02} measured spheroid-to-total (S/T) light ratios for the
field galaxy sample of \scite{gard96} using $I$-band imaging. The data
were originally obtained to determine the $K$-band luminosity function
and \scite{benson02} showed that they could be used reliably to
estimate S/T ratios for galaxies brighter than $I_{\rm Gar}=16.0$ with
an rms accuracy of $\sigma_{\rm rms}\sim0.1$. Unfortunately, the area 
covered by this sample is rather small (4.4 deg$^2$).

The SDSS imaging data were obtained with a larger telescope but using
shorter exposure times than those of \scite{gard96}.  From Monte-Carlo
simulations, \scite{benson02} established the signal-to-noise required
to obtain reliable measurements of S/T using the {\sc Galactica}
decomposition code. Assuming that, when applied to the SDSS data, the
code will be reliable to the same overall signal-to-noise level, we
find that the limiting magnitude required for our SDSS sample is
$I_{\rm SDSS} - I_{\rm Gar} = 0.4$. Using the mean galaxy colours of
\scite{fuku95}, the transformation between the $I_{\rm SDSS}$ and $r$
bands is $r - I_{\rm SDSS} = 0.9$, making the total difference equal
to $r-I_{\rm Gar}=1.3$ magnitudes. We therefore select EDR galaxies
with $r \le 17.3$---this is 1.4 magnitudes fainter than the sample used in a similar study by \scite{tw}.

\subsection{SDSS data selection and galaxy catalogue}

\begin{figure}
 \begin{center}
 \psfig{file=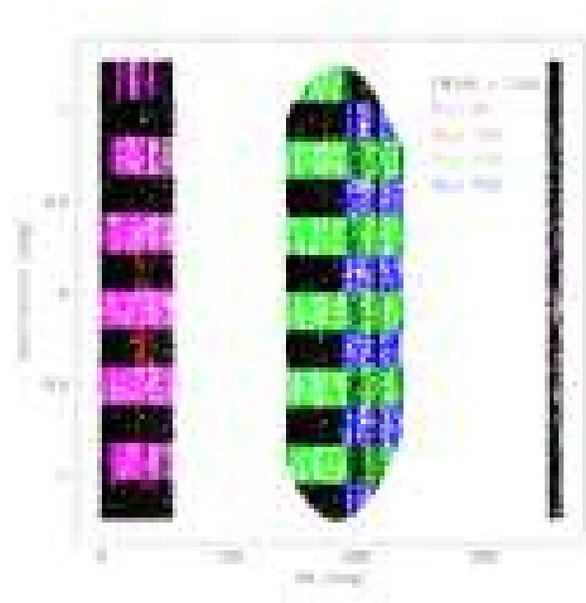,width=8.cm}
 \caption{SDSS galaxies that meet our selection criteria. Black points
 correspond to galaxies imaged when the seeing was greater than
 \protect $1.55''$.}  
\label{fig:sol_ang} 
\end{center}
\end{figure}

Galaxies with $r \le 17.3$ in the SDSS EDR equatorial strip are
plotted in Fig.~\ref{fig:sol_ang}, colour-coded according to the SDSS
run number ($94$, $125$, $752$ and $756$). The black points represent
imaging taken in `poor' seeing conditions (${\rm PSF}_{\rm
FWHM}>1.55''$, where ${\rm PSF}_{\rm FWHM}$ is the full-width at
half-maximum of the point spread function). This cut on of the seeing is used to
impose a second galaxy selection criterion since reliable
spheroid-to-disk decompositions require that the seeing be less than
a typical galaxy half-light radius \cite{bei99}.

The final galaxy selection criterion is redshift. To avoid
contamination of the measured redshift by the local galaxy infall
velocity, a low redshift cut, $z=0.02$, is imposed. Since the total
SDSS sample begins to tail off at large distance, a high redshift cut,
$z=0.3$, was also imposed. 

The selection leads to a total of $8839$ SDSS EDR galaxies.

\subsection{Sample solid angle}

To calculate the solid angle  covered by our sample, the galaxy
coordinates were accumulated in $0.2^\circ$ bins. All the areas which
contain at least one galaxy residing in a particular bin were summed
to give the total solid angle. The bin size was chosen such that the
derived solid angle was insensitive to small changes in the bin
size. As an additional check, this same bin size was used to reproduce
the SA of the entire SDSS EDR. For our chosen bin size, the solid
angle subtended by our sample is $165.5$ square degrees.

\subsection{Object detection and astrometry}

Object detection was performed using {\sc SExtractor} v2.2.2
\cite{ber96}. The {\sc SExtractor} world coordinates of the object
centroid positions ($x,y$) were used to identify the catalogued
galaxies within the SDSS frames. The {\sc Galactica} code
(Appendix~\ref{fitg}) was run on the extracted postage stamps whose
size was set equal to $(2 \times R_{p})\times (2
\times R_{p})$, where $R_{\rm P}$ is the Petrosian radius (Lupton et
al. 2001). This is large enough to contain many background pixels but
sufficiently small to ensure a reasonable convergence time for the
fitting procedure. Prior to decomposition, the {\sc SExtractor} estimate of
the local sky background was subtracted from every postage-stamp to
ensure that the background level was close to zero (see
Appendix~\ref{Two}).

\subsection{SDSS point spread function (PSF)}

\begin{figure}
  \begin{center}
\begin{tabular}{cc}
 \psfig{file=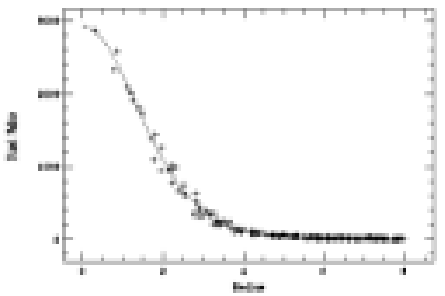,width=4.cm} &
\psfig{file=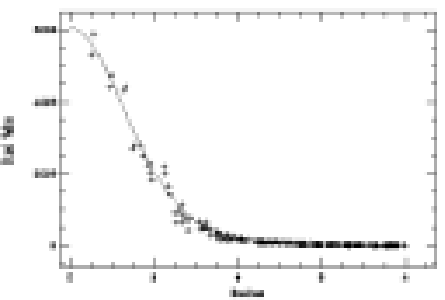,width=4.cm} \\ 
\end{tabular}
	\caption{The radial Moffat profile ($\beta=4.5$) fits (solid
	line) to the SDSS stellar light profiles (points) for stars
	found at various positions within several SDSS frames. The radius
	is in pixels and the Pixel Value are counts. Results are shown for two different frames.} 
\label{fig:psf_moff}
\end{center}
\end{figure}

Before beginning the decomposition procedure, it is necessary to
ensure that the PSF analytic model assumed by the {\sc Galactica}
code (Appendix~\ref{fitg}) is a realistic representation of the SDSS
PSF.  To demonstrate that the SDSS stars are well represented by the
analytic Moffat profile assumed by the {\sc Galactica} code, the {\sc
iraf} task {\tt IMEXAMINE} was used to fit radial Moffat profiles to a
sample of stars imaged on different SDSS frames and at different
positions within every frame. Fig.~\ref{fig:psf_moff} shows radial
fits to stellar light profiles obtained using $\beta=4.5$ and
demonstrates that a Moffat star with this value of $\beta$ is a good
analytic representation of the SDSS PSF.

\subsubsection{PSF variation}

\begin{figure}
  \begin{center}
  \psfig{file=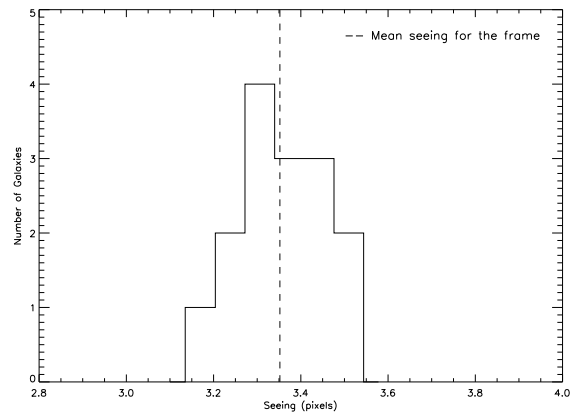,width=8.cm}
  \caption{The variation of the PSF across an SDSS frame. The dashed
  line represents the mean value of the seeing for the frame. The
  seeing appears not to vary by more than $\pm5\%$ from the mean
  value. A similar inspection of other frames shows this to be true in
  general.}  \label{fig:psf_hist} \end{center}
\end{figure}

The {\sc Galactica} code assumes a starting value for the PSF equal
to the measured value of the seeing in the SDSS and allows the value
to fluctuate by $\pm5\%$ (Appendix~\ref{fitg}). The $\pm5\%$ variation
is set from the observed variation of the seeing across a typical SDSS
frame, as shown in Fig.~\ref{fig:psf_hist}, which demonstrates that
for stars imaged at various positions in a given SDSS frame, the FWHM
does not change by more than $\pm5\%$. {\sc Galactica} assumes a circularly symmetric PSF, although this may not be precisely true for drift scan observations such as those of the SDSS \cite{bern02}.

\begin{figure*}
  \begin{center}
\begin{tabular}{cc}
\psfig{file=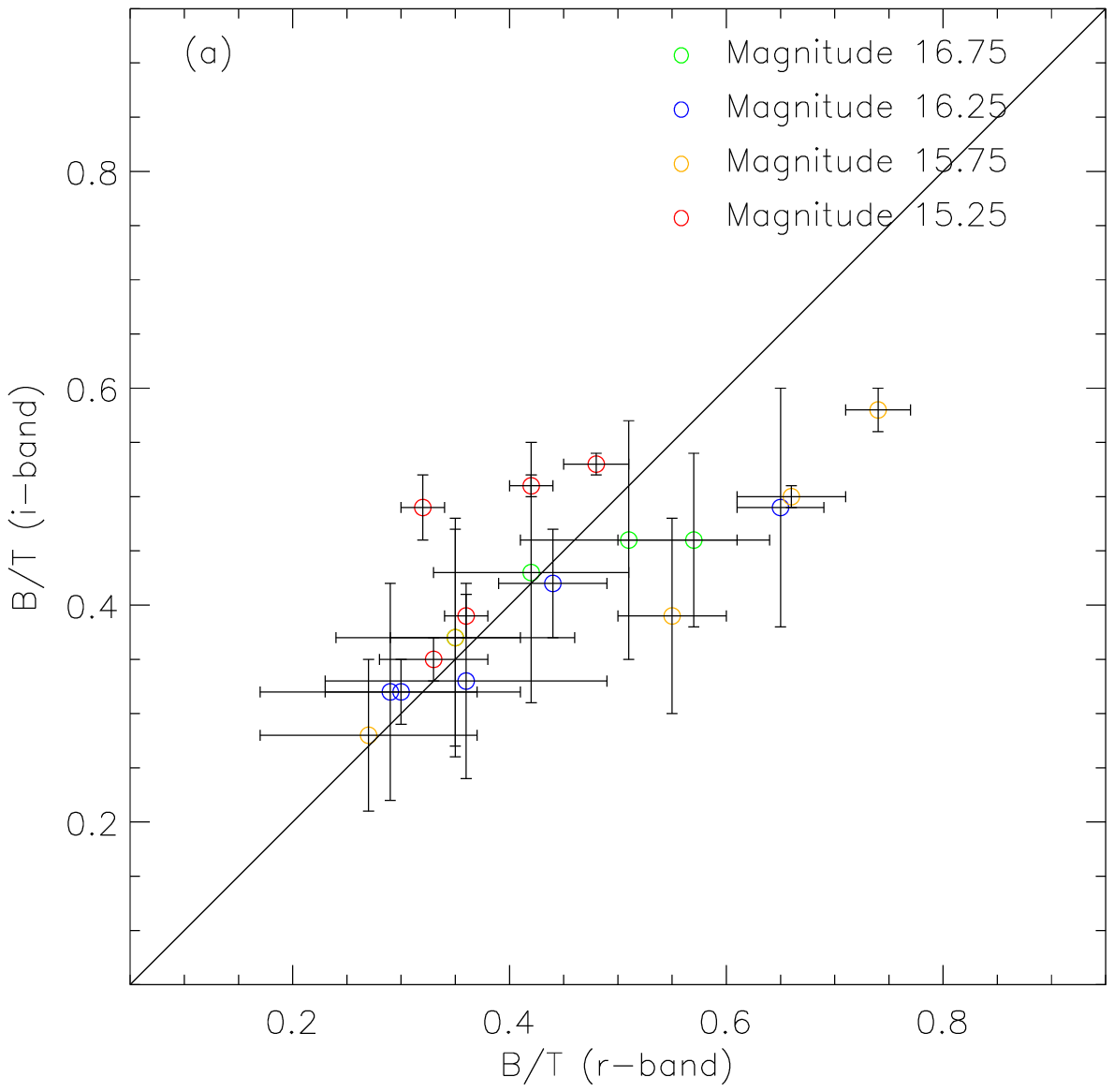,width=11.25cm} &
\hspace{-3cm}\psfig{file=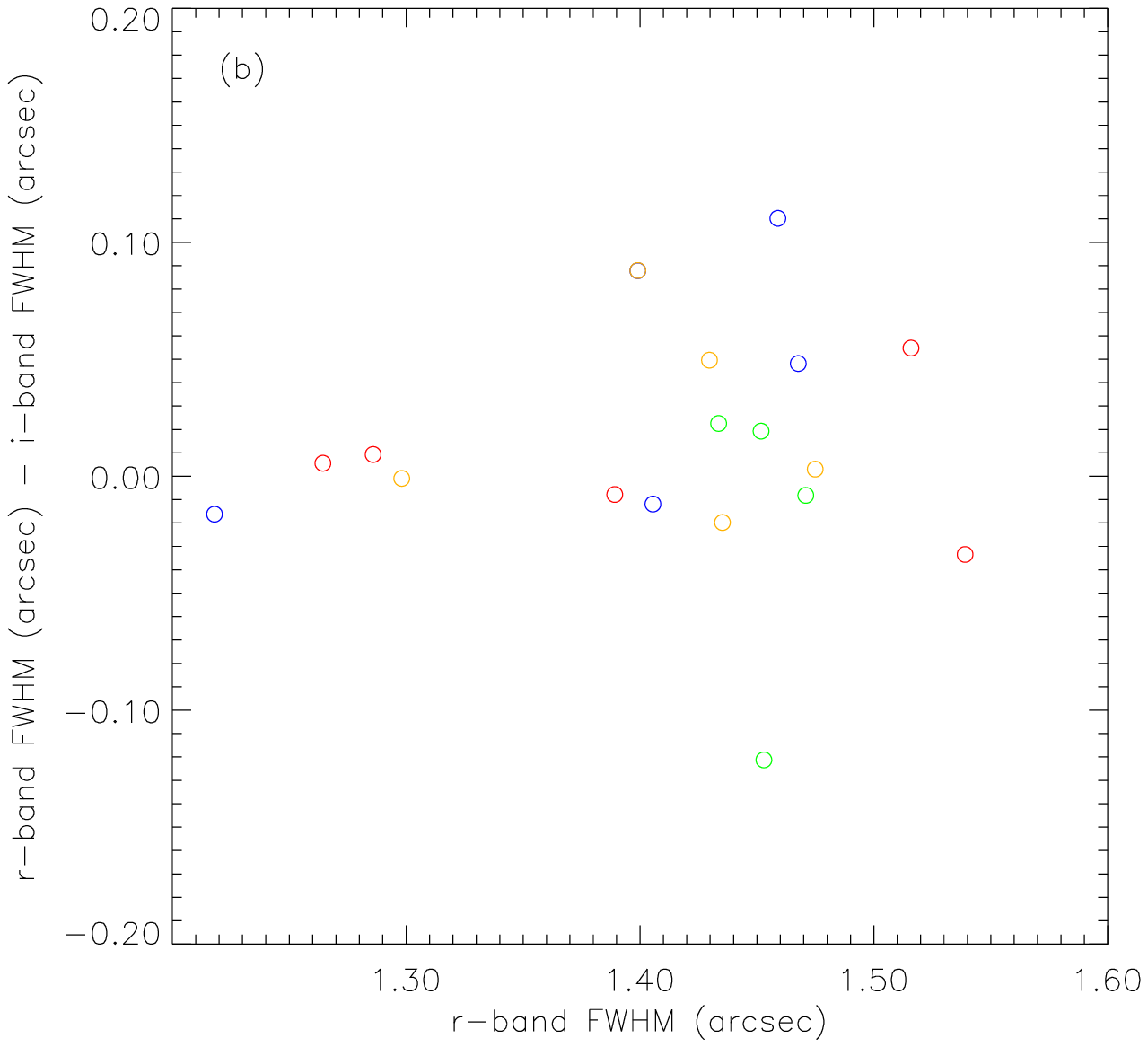,width=11.25cm} 
\end{tabular}
	\caption{(a) Correlation between S/T ratios obtained for the
	same galaxies in the $r$ and $i$ bands. The good correlation
	demonstrates that the S/T ratios are accurately determined for
	different representative PSFs and across the apparent
	magnitude range. The recovered S/T ratios show no obvious
	dependence on galaxy apparent magnitude, indicating that the
	decompositions are not affected by the variation in the
	signal-to-noise ratio. (b) The difference in the output {\sc
	Galactica} seeing for the same set of galaxies observed in
	the $r$ and $i$ bands. The lack of a trend demonstrates that
	the {\sc Galactica} code recovers the representative PSF for
	each galaxy well, without biasing the recovered S/T ratios.}  
\label{fig:psf_ri} 
\end{center}
\end{figure*} 

The small allowed change in the seeing ensures that the {\sc
Galactica} code can find the representative value of the seeing at
each galaxy position.  However, it is important to test how
consistently the {\sc Galactica} code recovers the `correct'
representative PSF for a given galaxy and quantify the effect this has
on the recovered S/T ratios. The observed galaxy properties are
expected to vary little between the $r$ and $i$ bands but the PSF
signatures for these observations will be somewhat
different. Fig.~\ref{fig:psf_ri} shows a good correlation between the
S/T ratios obtained for the same set of galaxies imaged in the two
bands. The code finds consistent S/T ratios across a range of apparent
magnitudes independently of the seeing. Error bars are obtained from
30 Monte Carlo realizations of each of the model fits, assuming the
noise appropriate to each image.

\section{{\sc Galactica} decompositions and galaxy morphologies}
\label{sec:decomp}

\subsection{{\sc Galactica} decomposition outputs}

\begin{figure}
 \begin{center}
\psfig{file=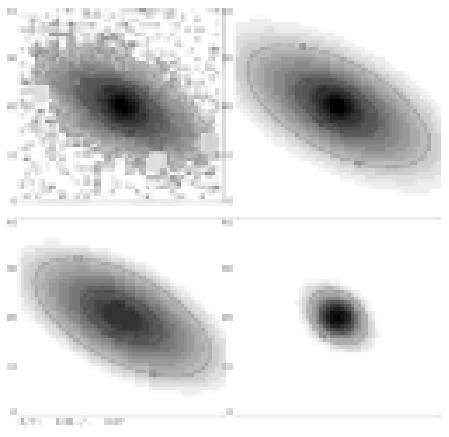,width=8.cm,bbllx=20mm,bblly=140mm,bburx=220mm,bbury=330mm,clip=}
S/T$=0.30 \pm 0.07$ 
  \caption{Top: real (left) and model (right) images. Bottom: disk (left) and
spheroid (right) component fits. The cross-hatched regions represent
potential contamination from overlapping objects (or regions where
data were unavailable after the image was recentred by {\sc
Galactica}) and are excluded from the fitting. The contours
indicate the pixel values in ADU/s.} 
  \label{fig:fit}
 \end{center}
\end{figure}

\begin{figure}
 \begin{center}
\psfig{file=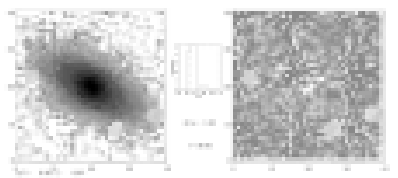,width=8.cm,bbllx=30mm,bblly=132mm,bburx=300mm,bbury=260mm,clip=}
S/T$=0.30\pm 0.07$
  \caption{Real (left) and residual (right) images. The inset shows the
  distribution of S/T ratios from 30 Monte Carlo realizations with the
vertical dashed line indicating the best-fit S/T. The value 
  of $\chi^{2}_{\nu}$ is acceptably small and the residual image also
shows a good fit to the data.} 
  \label{fig:res}
 \end{center}
\end{figure}

Fig.~\ref{fig:fit} demonstrates a typical fit to a galaxy light
profile. The figure shows the postage-stamp of a real galaxy, a
noise-free model generated from the best-fit parameters, along with the
individual model disk and spheroid components. In this study, a galaxy
is deemed to be sufficiently well represented by the model if the
$\chi^2$ per degree of freedom satisfies $\chi^{2}_{\nu} < 2.0$, and if
there are no obvious structures left in the residual image. Obviously for well-fit data and correctly estimate noise $\chi^{2}_{\nu}$ should be very close to unity. Allowing for larger values of $\chi^{2}_{\nu}$ allows galaxies with small (yet significant) departures from our photometric model to be included in our final sample.  An example
of a well fit galaxy is shown in Fig.~\ref{fig:res}. The
cross-hatched areas represent potential contamination from
overlapping objects as determined by the {\sc Galactica} masking
procedure (see Appendix~\ref{fitg}) and are excluded from the
fitting. The inset in Fig.~\ref{fig:res} shows a histogram of ${\rm
d}P/{\rm d}(S/T)$ ---the distribution of the spheroid-to-total ratio
from 30 Monte Carlo realizations, with the vertical dashed line
indicating the best-fit S/T value for this galaxy.

\subsection{Correlations of S/T with other fit parameters}\label{sec:cosi_bias}

Understanding the properties of this large statistical sample of
galaxies is important since it may reveal features which otherwise
would not be discovered in smaller samples such as those discussed in
Appendix~\ref{sec:bt_morph}. Equally, any unexpected correlations between
parameters could help discover and reduce possible biases introduced
by the fitting routine.

\begin{figure*}
 \begin{center}
 $\chi^2<2$; 7493/8839 galaxies \\
 \begin{tabular}{cc}
\psfig{file=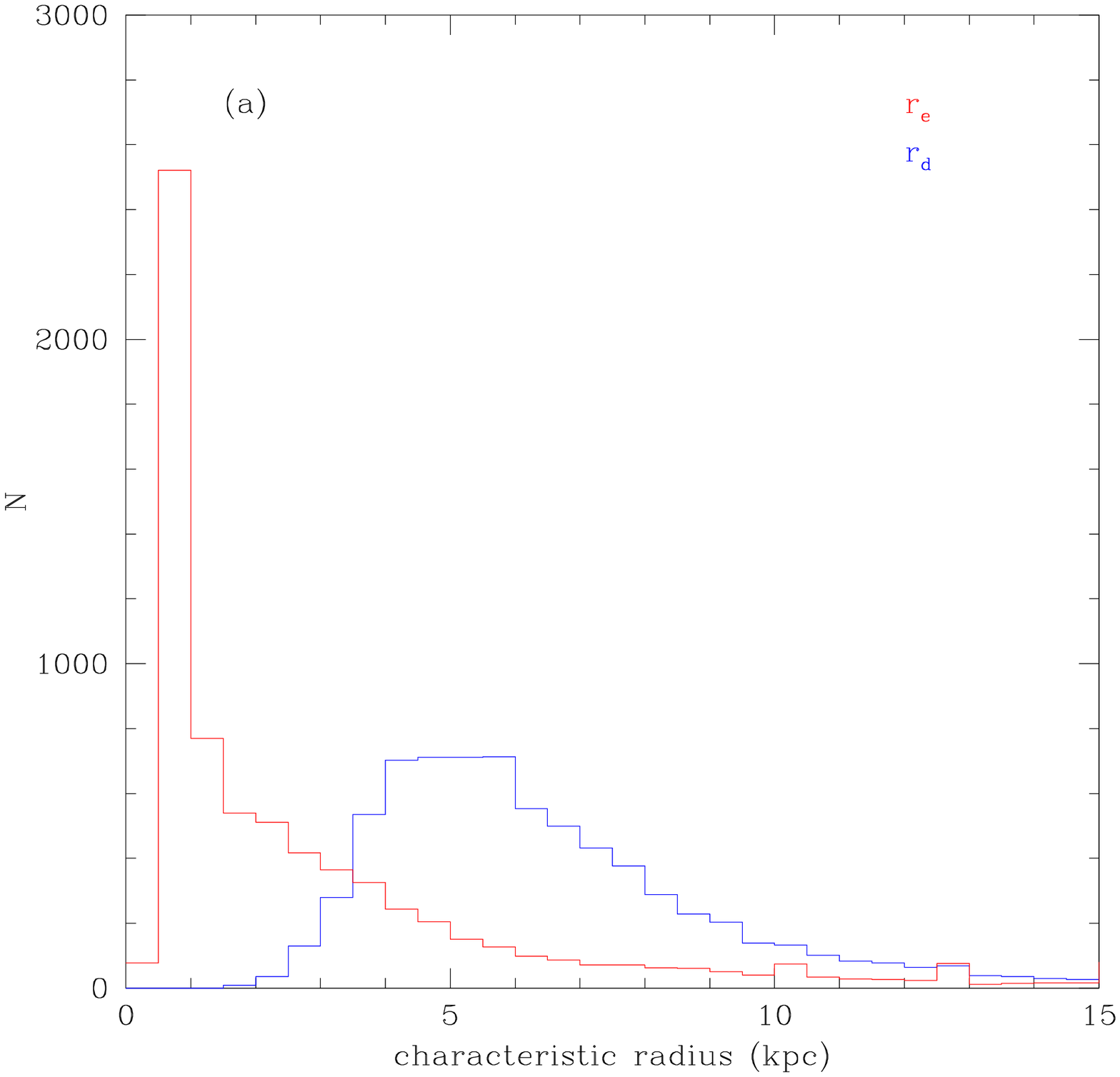,width=7.88cm} &
\psfig{file=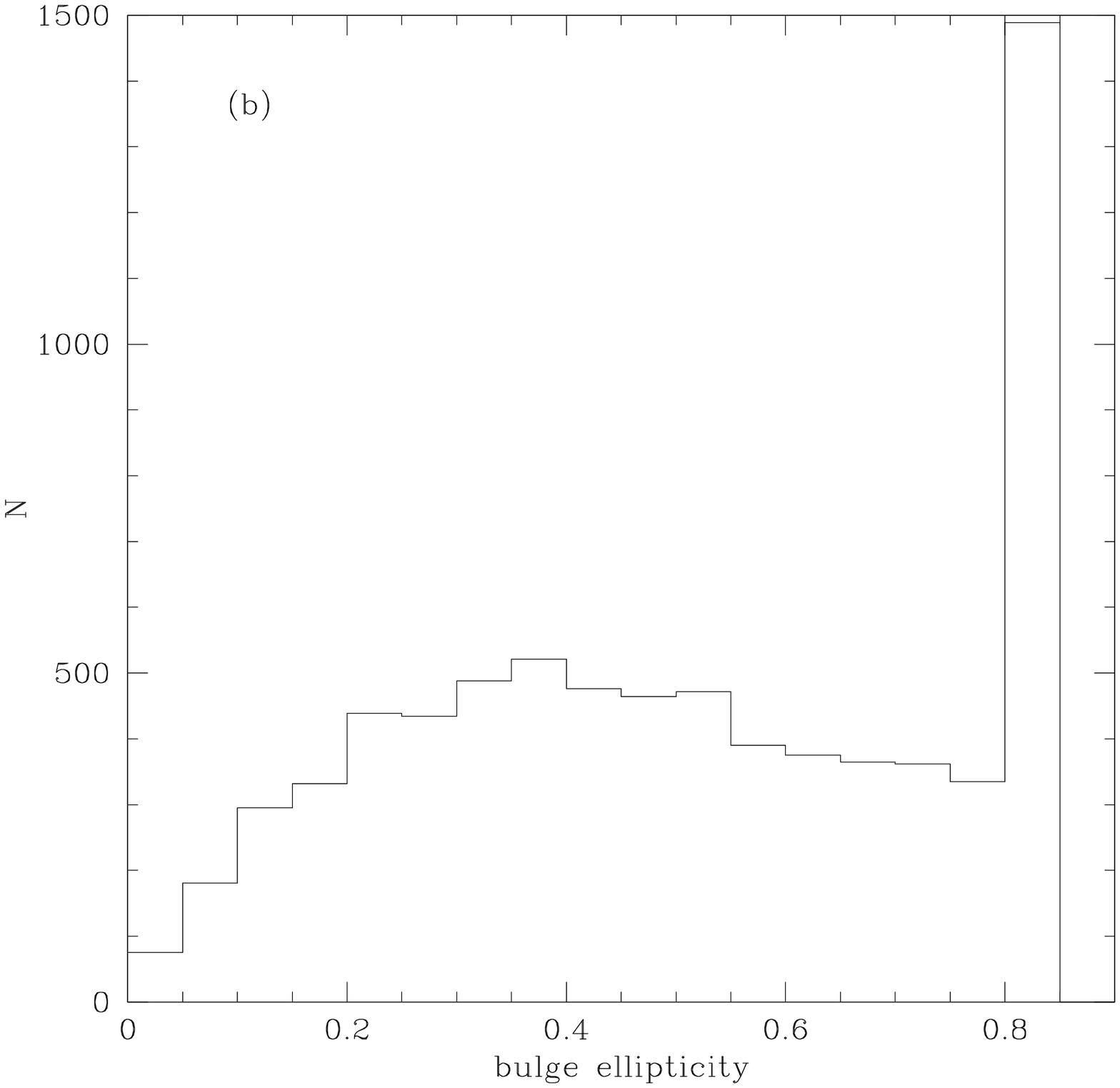,width=7.88cm} \\
\psfig{file=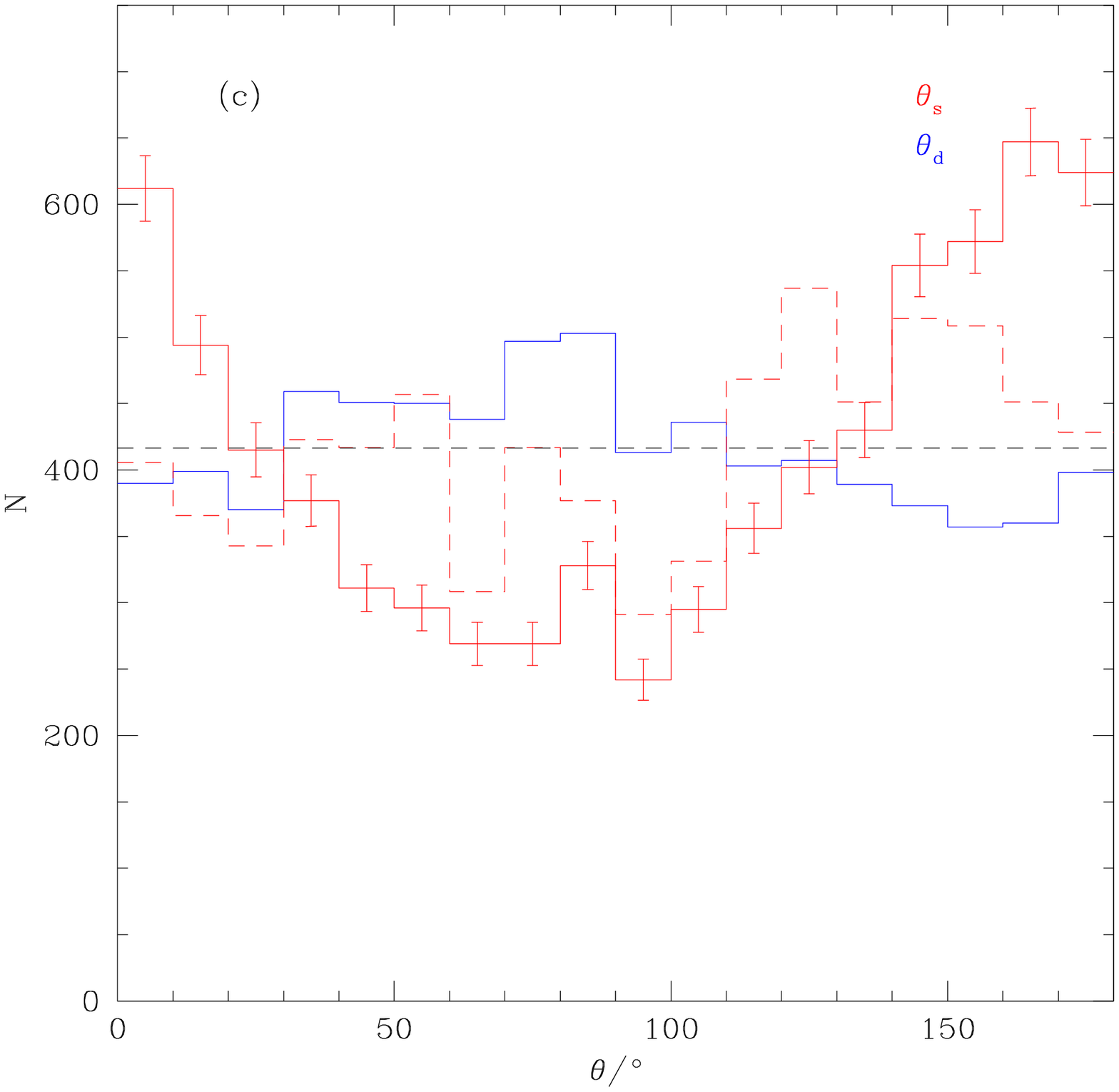,width=7.88cm} &
\psfig{file=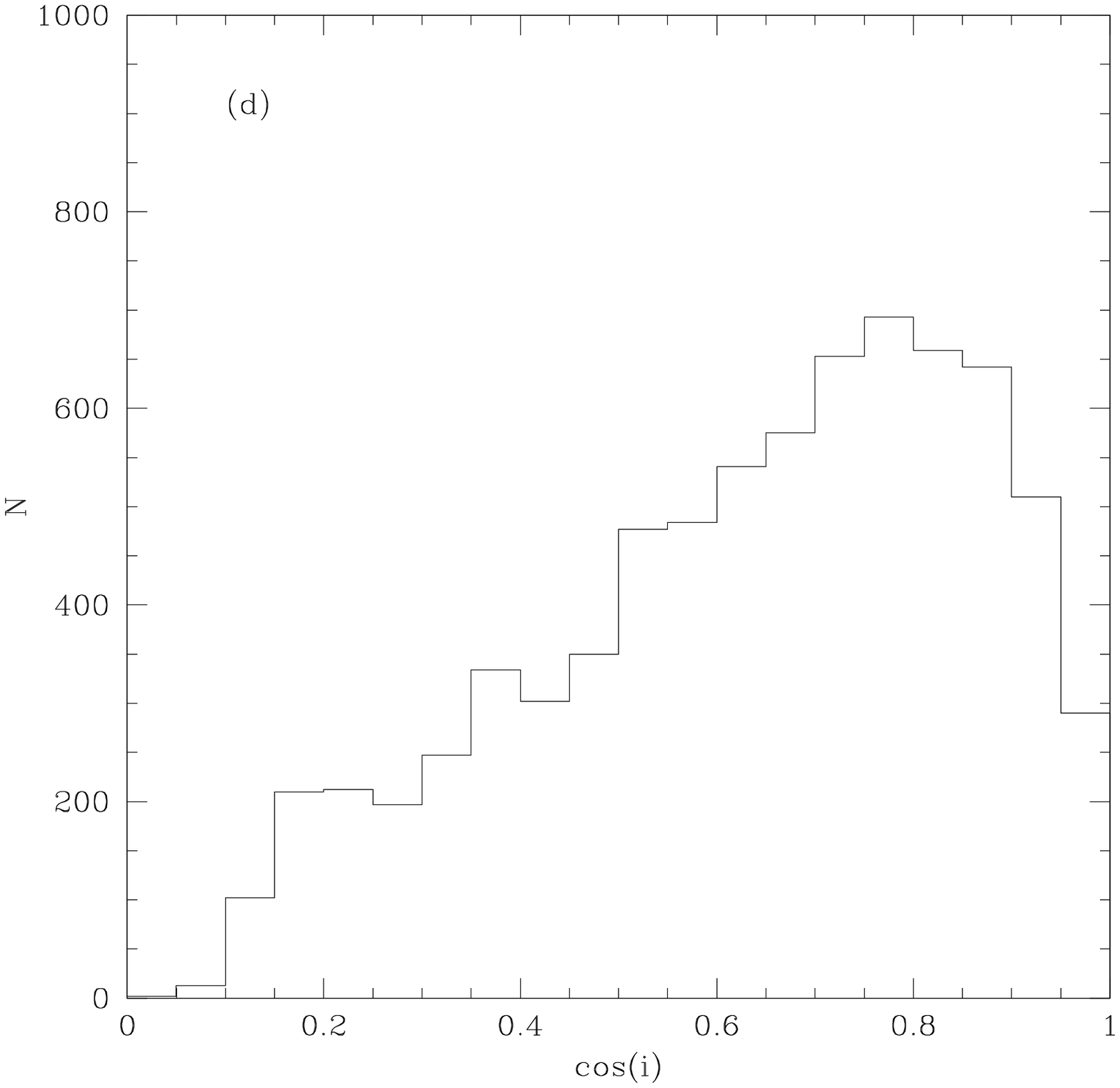,width=7.88cm}
  \end{tabular}
  \caption{Histograms of parameters recovered by {\sc Galactica} from
our sample of SDSS galaxies. a) Characteristic sizes of the spheroidal and
disk components. b) Ellipticity of the spheroidal components. c)
Position angles of disks and spheroids. The dashed histogram shows the position angles of well resolved spheroids (those with effective radii greater than three times the PSF FWHM). d) Cosine of the
inclination angle of the disks. Where appropriate, spheroids are
represented in red and disks in blue. } 
  \label{fig:fitg_distns}
 \end{center}
\end{figure*}

Histograms of various properties of our SDSS galaxies inferred from
the {\sc Galactica} decompositions are shown in
Fig.~\ref{fig:fitg_distns}. These plots reveal the following:
\begin{enumerate}
 \item a large number of highly elliptical spheroids;
 \item an excess in the number of galaxies with spheroid position
angle, $\theta_{\rm s}$, equal to $0^\circ$ and $180^\circ$; and,
 \item a non-uniform distribution of the cosine of the disks' 
inclination, $\cos(i)$. 
\end{enumerate}

In the remainder of this section, we explore the possible origins of
these unexpected distributions and their influence on the recovered
values of the S/T ratio. 

\subsubsection{S/T vs. Ellipticity}\label{sec:ellip}

\begin{figure}
 \begin{center}
\psfig{file=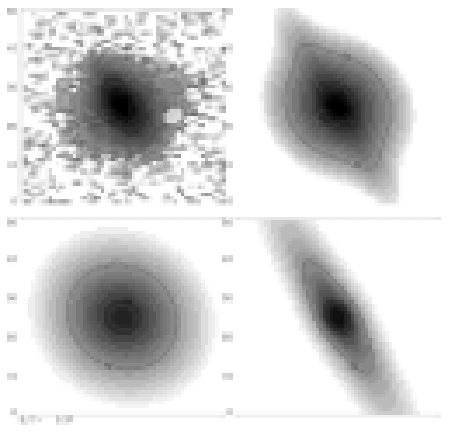,width=8.cm,bbllx=20mm,bblly=142mm,bburx=220mm,bbury=330mm,clip=}
\psfig{file=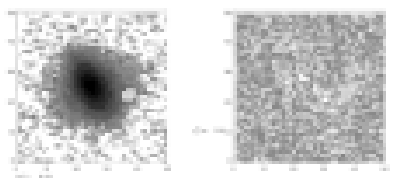,width=8.cm,bbllx=30mm,bblly=130mm,bburx=300mm,bbury=260mm,clip=}
S/T$=0.37$ 
  \caption{An example of a galaxy with a highly elliptical
spheroid. This galaxy demonstrates how the central bar-like 
structure in the galaxy results in the detection of a highly
elliptical spheroid along the same direction. The top row shows the
real (left) and model (right) images. The middle row shows the disk
(left) and spheroidal (right) components. The bottom row shows the
real (left) and residual (right) images.} 
  \label{fig:fitg_ell}
 \end{center}
\end{figure}

Around $15\%$ of galaxies appear to have a highly elliptical spheroid
component whose ellipticity has reached the imposed upper
limit\footnote{The upper limit for the ellipticity corresponds
approximately to that of the most elliptical observed galaxies
\protect\cite{lml92}.} of 
$e=0.83$. A large number of
frames have been inspected by eye and show that these galaxies
generally exhibit bar-like structures in the direction of the detected
highly elongated spheroidal component. In these cases, the existence
of this extra component, which is not part of the fitted model, drives
the code to fit small and highly elliptical spheroids
(Fig.~\ref{fig:fitg_ell}). While in principle it is possible to include additional components, such as bars, in the photometric model (see, for example, \pcite{gadotti07}) we have not done so here due both to the fact that fitting them would result in prohibitively long times to fit each image and that, for poorly resolved galaxies, additional components can cause further systematic errors such as the one described above. These galaxies are generally disk-dominated (with mean S/T ratio of $0.14$)
and thus we expect this shortcoming of the model to introduce only a
small bias on the overall S/T ratio. (It should be noted, however, that this problem may be occurring even in cases where the fitted ellipticity is less than $0.83$ if seeing has made the bar component appear rounder.) Any bias that is introduced would
increase this ratio, resulting in a slight overestimation of the
spheroid luminosity density in \S\ref{sec:lfs}.

Similar problems of this type (i.e. fitting of additional photometric components of galaxies such as bars or isophotal twists by a component of the photometric model) have been noted and discussed by \scite{tw} and \scite{sim02}. In such cases, the S/T ratio will be incorrectly estimated. We return to this problem in \S\ref{sec:lfs}.

\subsubsection{S/T vs. spheroid position angle}

Many galaxies appear to have $\theta_{\rm s}\sim0^\circ$ (or,
equivalently, $\theta_{\rm s}\sim180^\circ$). This could 
be due to either: 
\begin{enumerate} 
\item some feature intrinsic to the code such as the initial estimate
of $\theta_{\rm s}$; or,
 \item a feature intrinsic to the data.
\end{enumerate} 
Such a biased distribution does not arise when fitting mock images
constructed either internally by {\sc Galactica} or externally by {\sc
IRAF} (see Appendix~\ref{model_tests} for details of these tests),
suggesting that (i) is not the correct explanation.

\begin{figure}
 \psfig{file=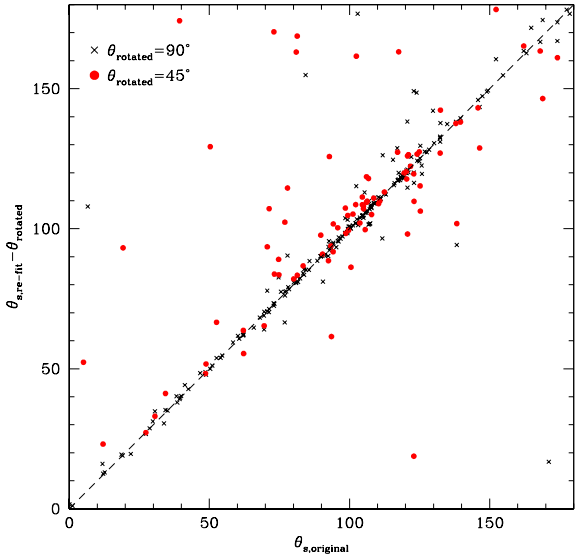,width=8cm}
\caption{A comparison of the bulge position angles recovered for a sample of $\sim 100$ SDSS galaxies fit before (x-axis) and after (y-axis) rotation by $\theta_{\rm rotated}=45^\circ$ and $90^\circ$. When rotating images by $45^\circ$ we crop them to the largest square which fits entirely within the rotated image. As a result there are fewer pixels to fit and therefore a larger scatter in $\theta_{\rm s}$ values recovered.}
\label{fig:fitg_rot_bPA}
\end{figure}

Explanation (i) can, in fact, be ruled out by rotating the images by
some angle prior to fitting. If the problem were intrinsic to the
code, we would expect to see no change in the distribution of
$\theta_{\rm s}$. In fact, when we rotate the images by $90^\circ$, we find that the
distribution of $\theta_{\rm s}$ is shifted by $90^\circ$ (see
Fig.~\ref{fig:fitg_rot_bPA}), indicating that it is some feature of
the images themselves that is causing this problem. The same is true
if we instead rotate galaxies by $45^\circ$. (Note that, in the case
of a $45^\circ$ rotation we crop to the largest square which fits
within the rotated image. As a result, there are fewer pixels
available to fit and therefore larger errors in the fit parameters.) 
Point (ii) is a plausible explanation since the data were taken in
drift-scan mode along the easterly direction which corresponds to
$\theta=0^\circ$. This can lead to small asymmetries in the actual PSF
\cite{bern02}. Since we are using a circularly symmetric PSF in our
photometric model, {\sc Galactica} may try to fit slightly elliptical
bulges with $\theta_{\rm s}\approx 0^\circ$ to match the actual PSF
shape. Note that, as expected, for well-resolved spheroids, the
distribution of $\theta_{\rm s}$ is close to uniform (dashed histogram
in the lower-left panel of Fig.~\ref{fig:fitg_distns}).

\begin{figure*}
 \begin{center}
 \begin{tabular}{cc}
\psfig{file=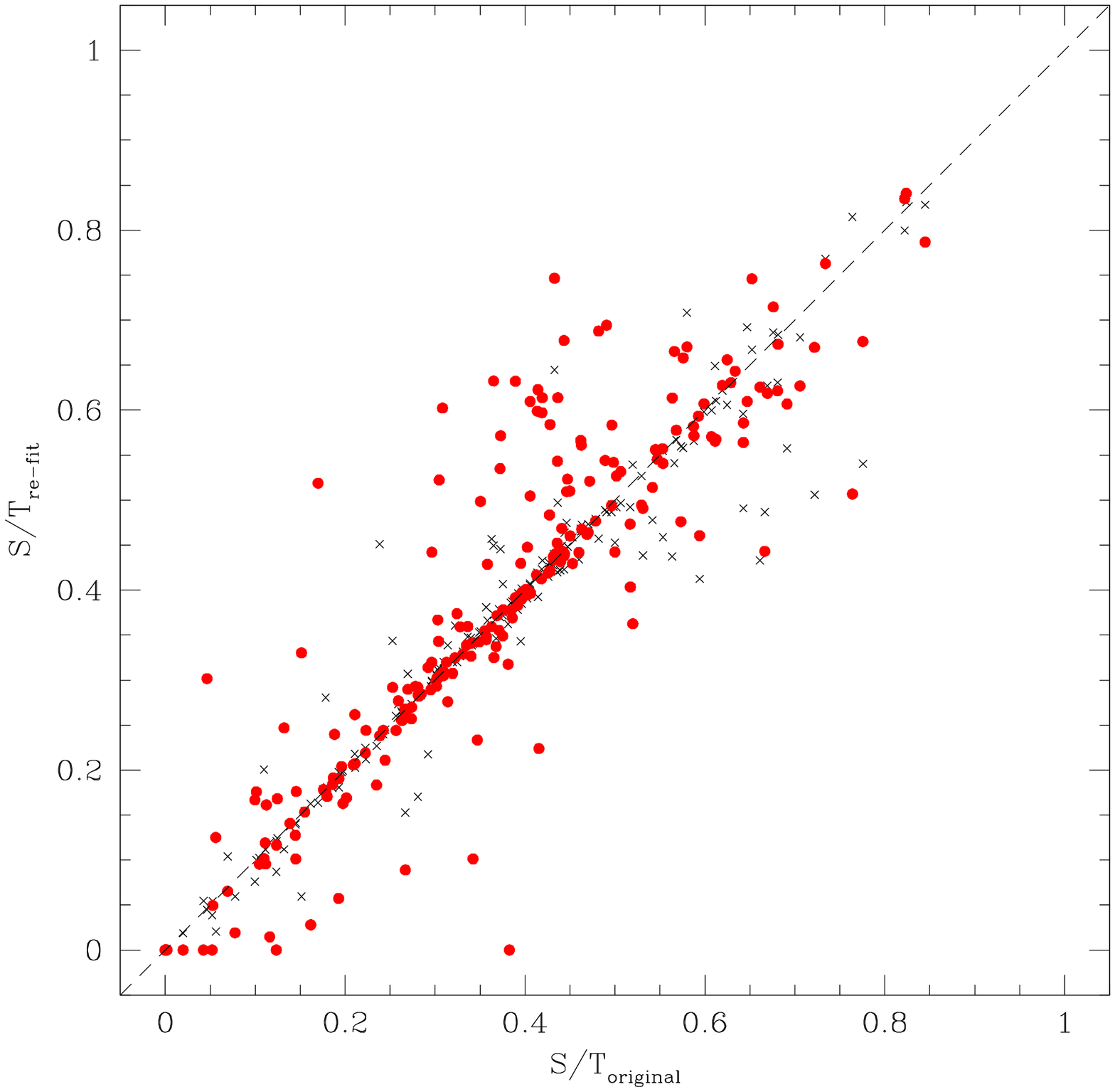,width=8cm} & \psfig{file=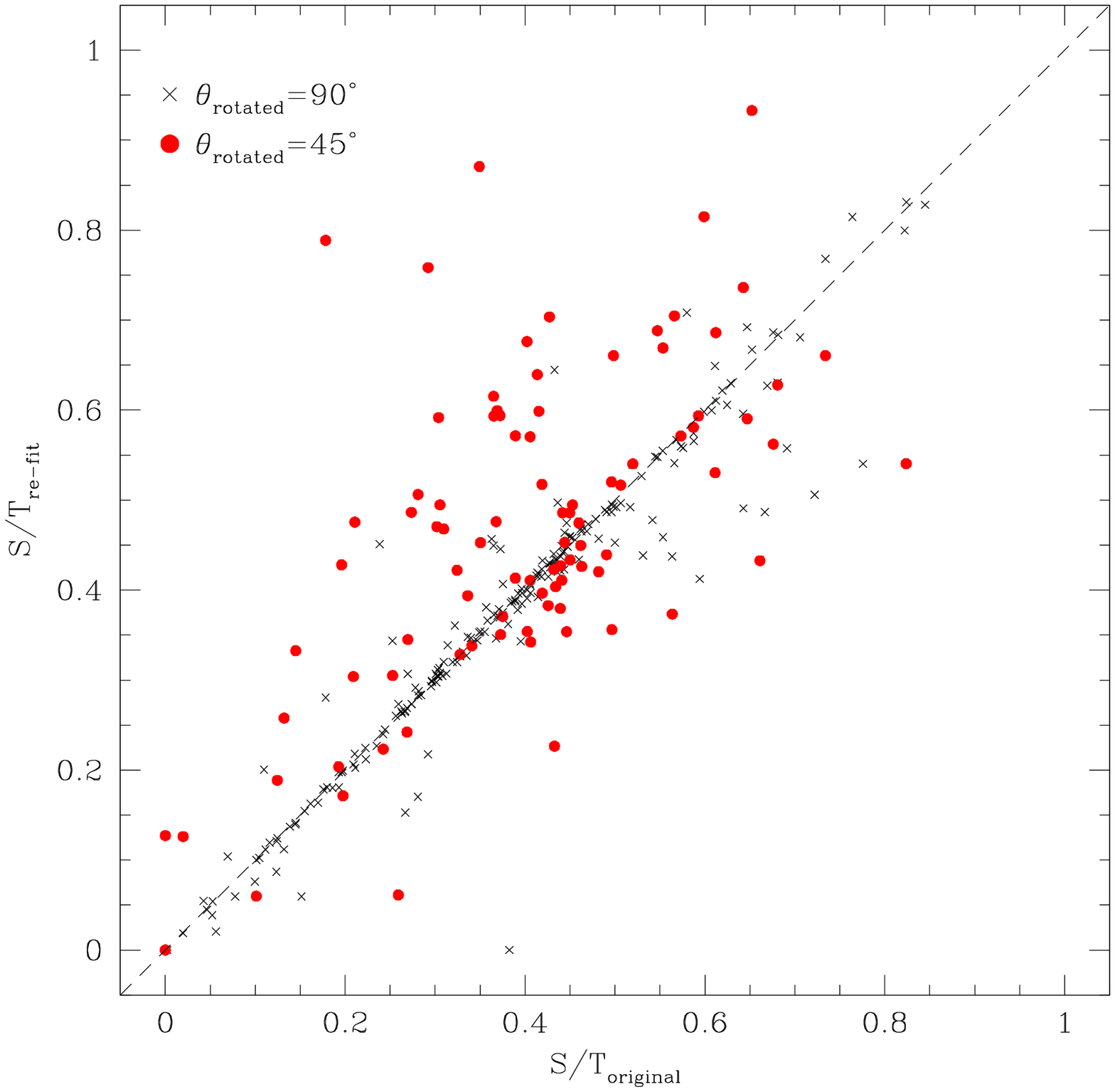,width=8cm} 
\end{tabular}
  \caption{\emph{Left panel:} Comparison of estimates of S/T for a sample of $\sim 100$
SDSS galaxies estimated from the actual image and from the image rotated 
anticlockwise by $90^\circ$ (black crosses). Also shown are the results of re-fitting these images with the spheroid position angle forced to equal the position angle of the disk (red circles).
\emph{Right panel:} Comparison of estimates of S/T for the same sample estimated from the actual image and from the image rotated 
anticlockwise by $90^\circ$ (black crosses) and by $45^\circ$ (red circles). When rotating images by $45^\circ$ we crop them to the largest square which fits entirely within the rotated image. As a result there are fewer pixels to fit and therefore a larger scatter in S/T values recovered.} 
  \label{fig:fitg_rot}
 \end{center}
\end{figure*}

For our purposes, the crucial issue is whether the bias in
$\theta_{\rm s}$ affects the derived S/T. To quantify the effects of
this bias on the recovered S/T ratio, we re-fit a sample of our images
keeping $\theta_{\rm s}$ equal to $\theta_{\rm d}$ (i.e. we did not
allow the spheroid position angle to vary freely). We find that the
S/T ratios recovered correlate extremely well with those found with
$\theta_{\rm s}$ as a free parameter, with scatter consistent with the
fitting uncertainties in S/T (see the left-hand panel of
Fig.~\ref{fig:fitg_rot}). An excellent correlation is also found if we
rotate our images by $45^\circ$ (right hand panel of
Fig.~\ref{fig:fitg_rot}). The larger scatter in this case is caused by
the reduced number of pixels available for fitting in our rotated
images. We conclude that this bias in the distribution of $\theta_{\rm
s}$ does not affect our estimates of S/T. We have further found that
the bias in the distribution of $\theta_{\rm s}$ is strongest for
poorly resolved, low ellipticity spheroids. For larger spheroids,
particularly those which are quite elliptical, there is no apparent
bias.

In conclusion, the bias in $\theta_{\rm s}$ seems to be due to some feature intrinsic to the data, perhaps an asymmetry in the PSF due to the observing method. We do not believe that this bias affects the recovered S/T ratios at any significant level since re-fitting the images with the bulge position angle locked to equal that of the disk (which is essentially unbiased---see Fig.~\ref{fig:fitg_distns}) does not significantly alter the S/T ratio in the vast majority of cases.

\subsubsection{S/T vs. disk inclination}\label{sec:cosibias}

A large number of objects in the sky which are randomly inclined to
the line-of-sight should have a uniform distribution of
$\cos(i)$. Fig.~\ref{fig:fitg_distns} clearly shows that this is not
the case for the inclination angles of the disk components obtained by
decomposing our sample of SDSS galaxies.

To test whether the apparently incorrect recovery of the disk
inclination is an artifact of the fitting procedure, a sample of $200$
mock galaxies was generated using the {\sc Galactica} code (see
Appendix~\ref{Two}). The S/T ratios were chosen at random in the
interval $[0,1]$. The remaining parameters, including the value of
$\cos(i)$, were also chosen at random. Fig.~\ref{fig:fitg_revisit_inc}
demonstrates that the {\sc Galactica} code generally recovers the
$\cos(i)$ distribution for 200 model galaxies quite well. However, a
noticeable feature is a slight deficit at $i=90^\circ$ and a
corresponding at $i \sim 75-80^\circ$.  This
excess reflects the fact that fits avoid the $90^\circ$ limit since this would
correspond to fitting an infinitely thin edge-on disk and, because of
seeing, the disks are never infinitely thin edge-on.
(The feature remains even if the allowed inclination range is increased
from $[0^\circ,90^\circ]$ to $[-180^\circ,180^\circ]$.) The S/T ratios
are not affected by this problem, i.e. model galaxies with input value
$i\sim 90^\circ$ but recovered value $i\sim 85^\circ$ show no bias in
the recovered S/T ratio.

\begin{figure}
\psfig{file=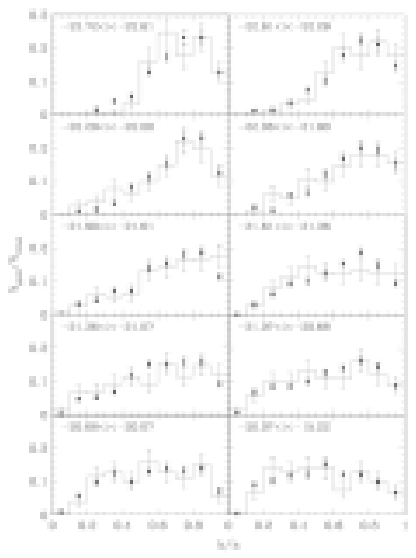,width=80mm}
\caption{The distribution of axial ratio $b/a$ (equivalent to $\cos
i$) as a function of galaxy luminosity. Histograms show results from
\protect\scite{tw} (this figure is a reproduction of their Fig.~10),
while points show results from this work. Both datasets show that the
bias in $\cos i$ occurs primarily for the most luminous galaxies.}
\label{fig:TWcomp} \end{figure}

Such biases in the distribution of $\cos(i)$ have been seen in other
studies employing 2D galaxy decomposition techniques (see, for example, \pcite{sim02,tw}) and can occur because the fitting codes tend to fit disk components to radial variations in axial ratio or position angle in spheroids \cite{sim02}. \scite{tw} used
{\sc Gim2D} to fit the 2D images of galaxies in the SDSS. They found a
biased distribution of $\cos(i)$, with intrinsically brighter galaxies
showing the most biased distribution. Fig.~\ref{fig:TWcomp} reproduces
Fig.~10 of \scite{tw}, with results from this work overlaid. Our
results, using the same dataset but a different galaxy decomposition
code, are in excellent agreement with those of \scite{tw}.

\begin{figure}
\psfig{file=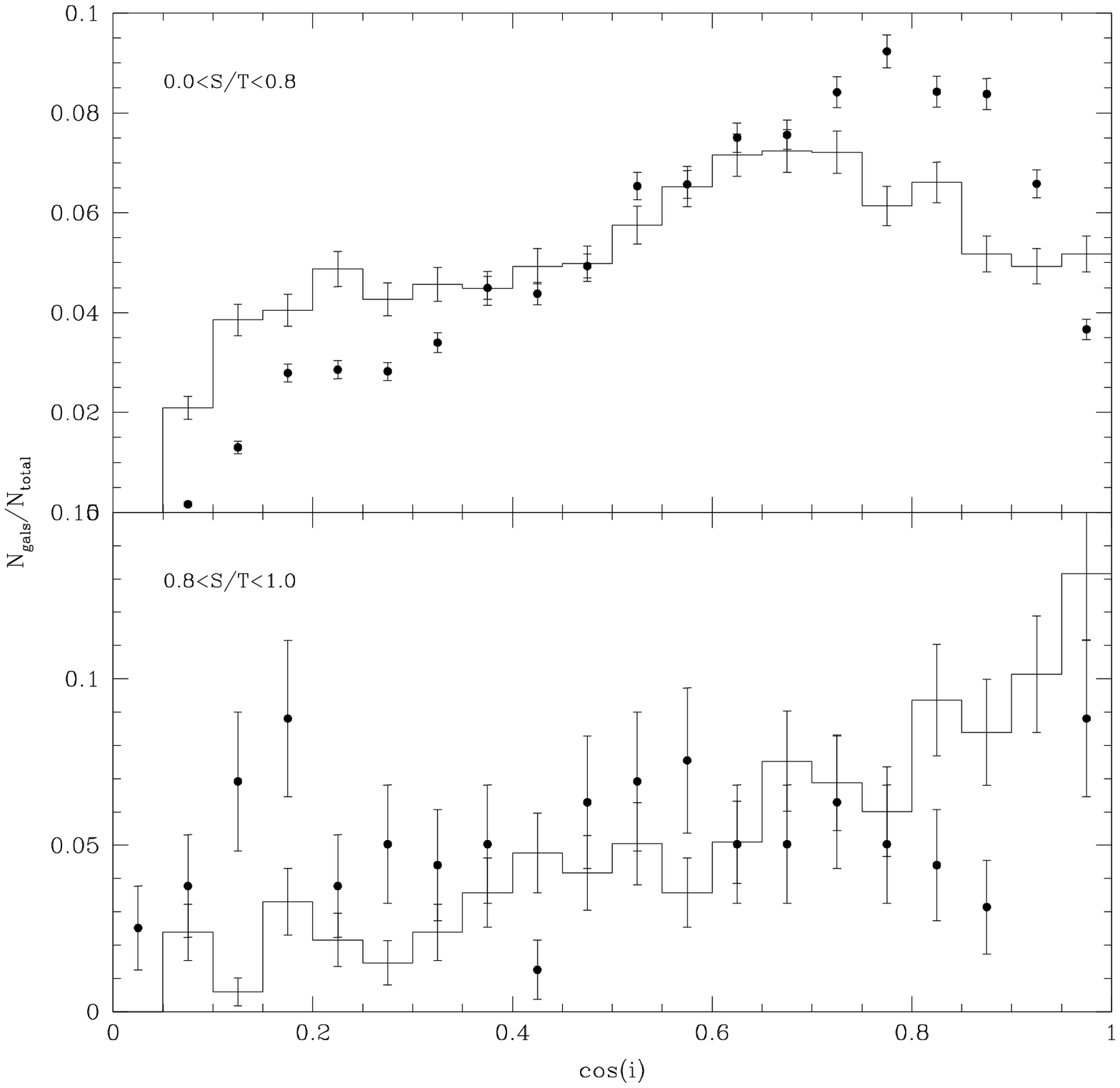,width=80mm}

\caption{The distribution of $\cos i$ split by S/T. Histograms show
results from \protect\scite{allen} (this figure is a reproduction of
their Fig.~9), while points show results from this work. While both
datasets show biased distributions of $\cos i$, the trends with S/T
appear to differ, with the \protect\scite{allen} dataset showing a
more uniform distribution of $\cos i$ for galaxies with low S/T.}
\label{fig:Allencomp} \end{figure}

\begin{figure}
  \begin{center}
 \psfig{file=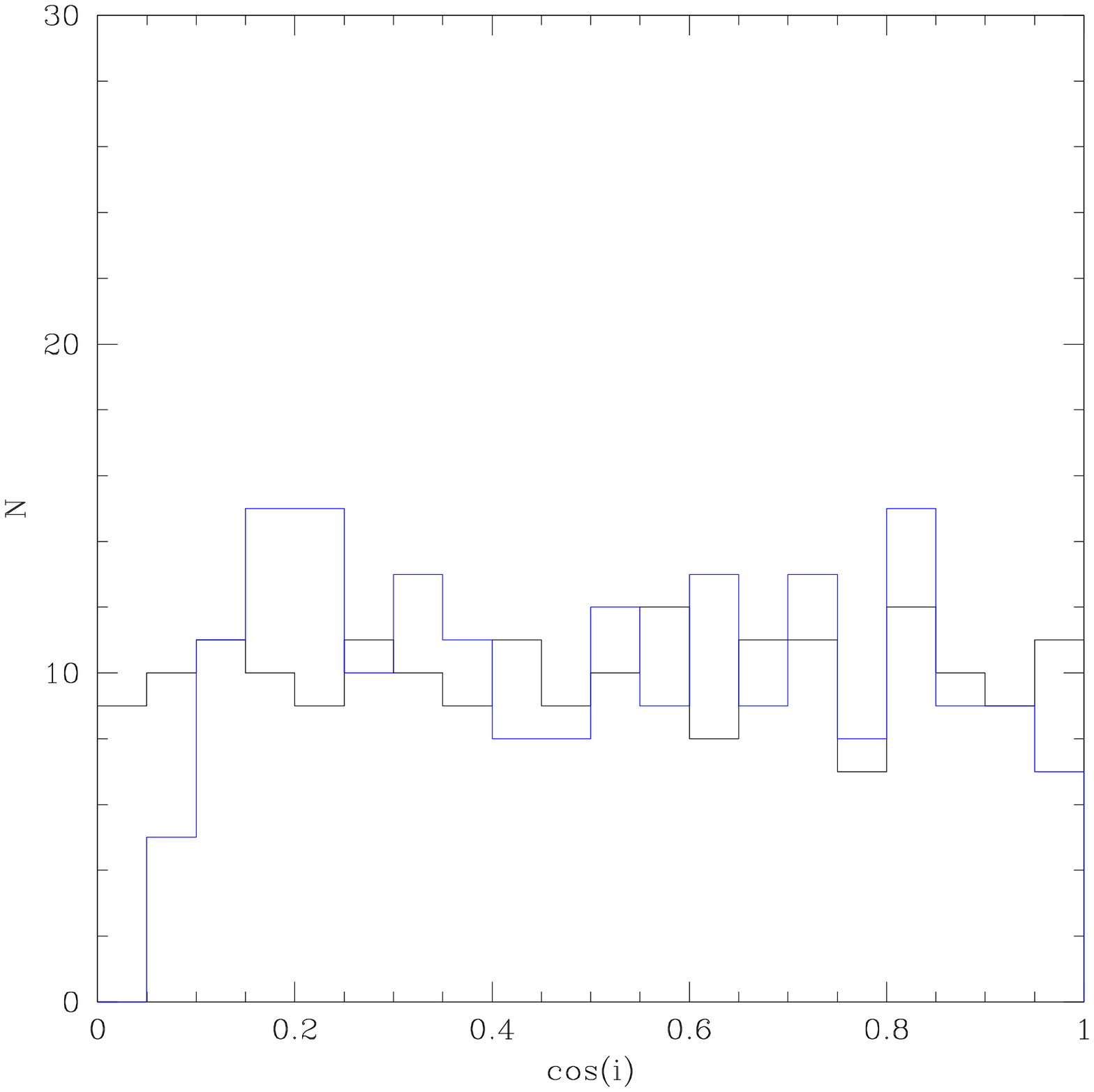,width=8.cm}
   \caption{The input (black) and the recovered (blue) $\cos(i)$ distribution for
 $200$ model galaxies created and decomposed using the 
   {\sc Galactica} code. The figure demonstrates that the non-uniformity in the $\cos(i)$ is not caused by the fitting code. An apparent excess of galaxies with $i\sim 75-80^\circ$ can be seen.}
   \label{fig:fitg_revisit_inc}
  \end{center}
 \end{figure}

\scite{allen} performed 2D galaxy decompositions, also using {\sc
Gim2D}, on galaxies in the Millennium Galaxy Catalogue and found that
disk-dominated galaxies (S/T$<0.8$) had a more uniform (although still
biased) distribution of $\cos(i)$. Fig.~\ref{fig:Allencomp} reproduces
their results, with comparable results from our own work overlayed. In
this case, we find the opposite trend: our $\cos(i)$ distribution is
more uniform for the S/T$>0.8$ sample, although the errors are
large. We find that galaxies must have angular sizes of several times
the seeing half-width at half-maximum in order for the inclination to
be well constrained. From Fig.~1 of \scite{allen}, we would therefore
conclude that a large fraction of their galaxies should have poorly
constrained inclinations.

Assuming that the bias in $\cos i$ arises due to {\sc Galactica} using
the disk component of the photometric model to fit radial variations
in the spheroid, it is possible to make an approximate correction for
this bias. Such a correction was developed by \scite{tw}. In
\S\ref{sec:lfest} we will employ their correction, and a similar yet
more detailed correction to assess the impact of this bias on our
results.

\subsubsection{Effects of fitting S\'ersic index}\label{sec:sersic}

\begin{figure}
 \psfig{file=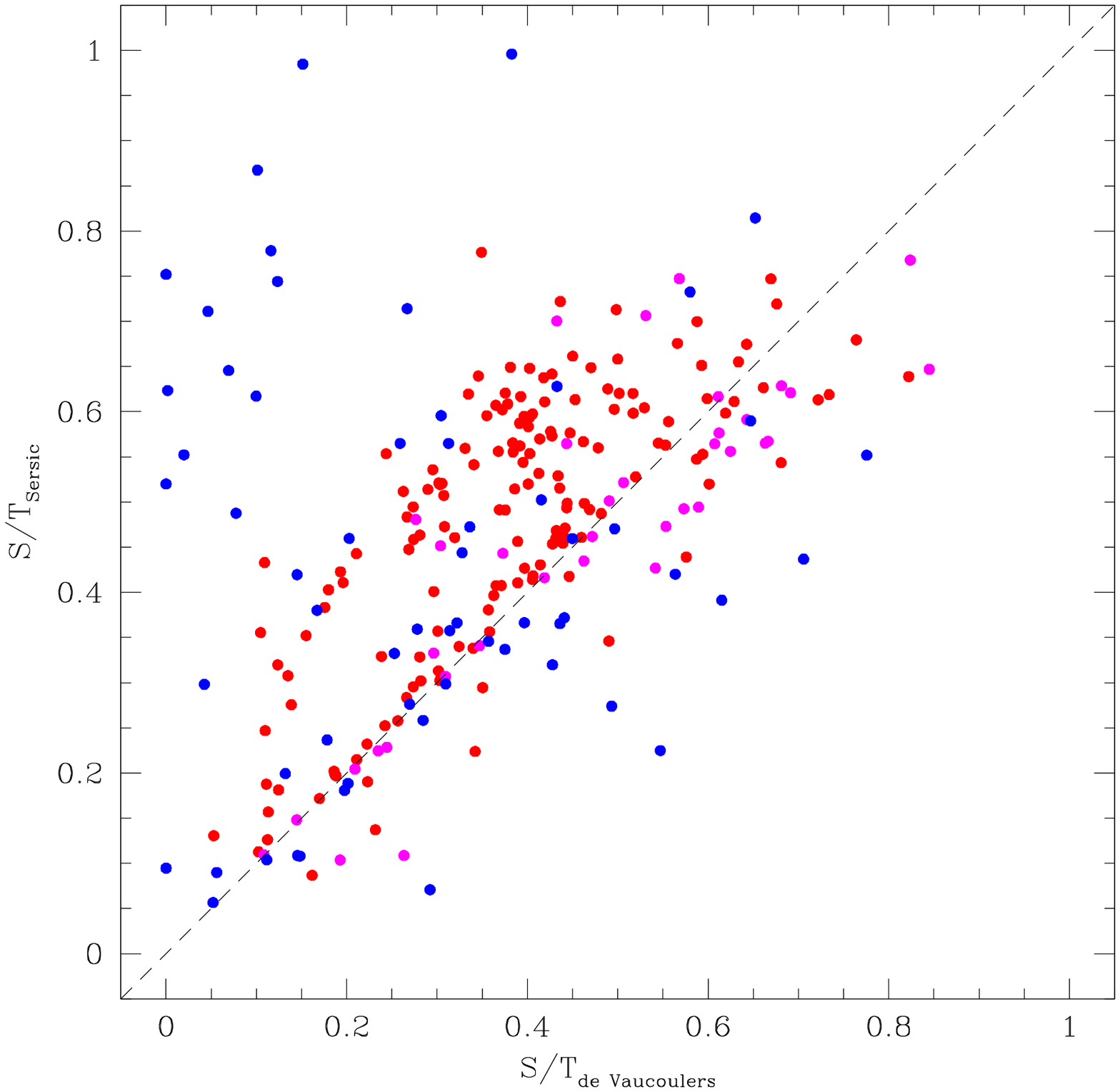,width=80mm}
\caption{The S/T ratios measured for a subsample of our galaxies. On the x-axis we plot the value obtained assuming a de Vaucouler's profile for the spheroid (i.e. a S\'ersic profile with fixed $n=4$) while on the y-axis we show the results of fits in which we allow the S\'ersic index $n$ to vary. Red points show galaxies for which the best fit $n>4$, magenta points show galaxies for which $2.5<n\le4$ while blue points show galaxies with $n\le 2.5$.}
\label{fig:sersic}
\end{figure}

We have chosen not to include the S\'ersic index as a free parameter
in our photometric model, instead holding it fixed at $n=4$
(corresponding to a de Vaucouler's profile). \scite{tw} demonstrated
that $n=4$ provides a good fit to the majority of spheroids in a
magnitude limited sample, and that there is a very good correlation
between the values of S/T obtained using fixed $n=4$ and free $n$
fits. To examine this in our own data we fit a subsample of our
galaxies allowing $n$ to vary. In Fig.~\ref{fig:sersic} we show the
recovered S/T ratios assuming a de Vaucouler's profile (x-axis) and a
S\'ersic profile (y-axis). There is a good correlation between
the results obtained using de Vaucouler's and S\'ersic profiles. This
is particularly true when $n \gsim 2.5$. For lower values of $n$ (blue
points in Fig.~\ref{fig:sersic}) we see some large
discrepancies. These occur for galaxies which had a low S/T in the de
Vaucouler's fit, but are given a high S/T when fit by a S\'ersic
profile. Of course, for $n\approx 1$ there is no difference in our
photometric model between disks and spheroids (except for the fact
that disks may be highly inclined to the line-of-sight while spheroids
are limited in how elliptical they may become). It is not surprising
therefore that {\sc Galactica} mixes disk light between the two model
components in such cases. We find that, when allowing the S\'ersic
index to be fit as a free parameter the fraction of light emitted by
disks (averaged over all galaxies in our sub-sample using a $1/V_{\rm
max}$ weighting) decreases from 60\% to 52\%. This effect is very
similar to that found by \scite{tw}. We consider this to be a lower
limit on the disk light fraction since, as discussed above, for some
galaxies a fraction of the disk light will have been fit by a
spheroidal component with $n\approx 1$. 

\section{Luminosity and mass functions}\label{sec:lfs}

\subsection{Introduction}

The spatial abundance of galaxies is expressed by means of the
luminosity function (LF), defined as:
\begin{equation}
  \d n(M) = \phi(M)\d M, 
\end{equation} 
where $dn$ is the number density of galaxies with absolute magnitude
in the range $M, M+\d M$. The simplest way to calculate the luminosity
function is using
  the $1/V_{\rm max}$ method in which the number of galaxies in each
  individual absolute magnitude bin is divided by the volume of space
  that has been surveyed at that magnitude. Galaxies in any given
  absolute magnitude range are assumed to be uniformly distributed in
  the surveyed volume which is not the case if any local
  overdensities are present. Maximum likelihood techniques circumvent
  this problem and provide more accurate estimates of the
  luminosity function. Here, we will employ the $1/V_{\rm max}$ as
  well as the
  Stepwise Maximum Likelihood (SWML) non-parametric estimator
  \cite{swml} which characterizes the LF as a series of steps. 
We will also employ the STY parametric estimator
  \cite{sty}, assuming a \scite{sch} functional form
\begin{equation}
 \phi(M) = 0.4\ln10\phi_{*}10^{-0.4(M-M_{*})(\alpha+1)}\exp[-10^{-0.4(M-M_{*})}],  
  \label{eqn:sch}
\end{equation}
where $M_*$ is a characteristic magnitude, $\alpha$ is the faint-end
slope and $\phi_*$ is the normalization. Integrating
over the Schechter function provides an estimate of the luminosity
density. This can also be obtained by summing up all the individual
SWML contributions.

Computing the spheroid and disk LFs is more complicated since there is
an additional constraint to be considered \cite{benson02}, namely the
detectability of a spheroid/disk depends both on the component's 
apparent magnitude and on the corresponding S/T. This needs to be
accounted for when constructing the luminosity function. A detailed
discussion of the application of these methods can be found in
\scite{benson02}. We use exactly the same methods as \cite{benson02} to estimate luminosity functions from our present dataset.

\scite{benson02} used a functional form for STY
parametric fits to the spheroid and disk LFs which had a
Schechter$\times$exponential form. We find that the functional form of
Benson et al.  does not provide a good description of our larger
sample of galaxies. We have been unable to find a suitable functional
form which does provide a good description and so have not performed
STY fits to the spheroid and disk luminosity function data.
 
\subsection{SDSS absolute magnitudes and K+E corrections}\label{sec:KE}

In order to estimate  the luminosity function, we require  galaxy
absolute magnitudes. A galaxy at redshift $z$, 
with apparent magnitude $m$, has an absolute magnitude $M$ given by:
\begin{equation}
  m - M = 25 + 5\log_{10}(D_{\rm L}) + {\rm KE}(z) 
\end{equation}
where $D_{\rm L}$ is the luminosity distance in megaparsecs and
KE$(z)$ is the K+E correction. 

K+E corrections for our catalogued galaxies were obtained using a code
kindly provided by Carlton Baugh. It employs the revised isochrone
stellar population synthesis models of \scite{bc93} to determine
present-day galaxy luminosities. The model assumes a stellar initial
mass function (IMF) and a star-formation rate $\psi(t) \propto
\exp(-t/\tau)$, with timescale, $\tau$.  A grid of models was
generated by varying the metallicity and $\tau$.  We assume a Salpeter
(1955) IMF and apply a simple dust extinction law. At every point on
the grid, a table of absolute magnitudes, galaxy colours, K+E
corrections and galaxy stellar mass-to-light ratio is generated.  The model
that best matches the observed $g-r$ and $r-i$ colours of each galaxy
is then used to infer its present-day ($z=0$) $r$-band absolute
luminosity, K+E correction and stellar mass-to-light ratio. The
mass-to-light ratio is used to convert luminosities to stellar masses
in order to estimate stellar mass functions (see \S\ref{sec:MF}). Note
that the K+E corrections are based on the total (i.e. spheroid plus
disk) colour of a galaxy.

\subsection{Luminosity function estimates}\label{sec:lfest}

We estimate luminosity functions using the methods described in detail by \scite{benson02} and employ both the SWML and $1/V_{\rm max}$ estimators (for the total luminosity function we also employ the STY estimator). We estimate the
luminosity functions of spheroids and disks, as well as the total
galaxy luminosity function for our sample of SDSS EDR galaxies.

As noted in Appendix~\ref{app:goodfit}, our requirement that images be
reasonably well fit by {\sc Galactica} (i.e. $\chi_\nu^2<2$) introduces
some bias in both the apparent magnitude and redshift distributions of
our galaxy sample. To correct this bias we make the assumption that
the distribution of S/T for galaxies with $\chi_\nu^2>2$ is the same as
that for galaxies of comparable apparent magnitude and redshift and
with $\chi_\nu^2<2$. Such an assumption may of course not be correct, for
example if disk-dominated galaxies are more likely to be poorly
described by our photometric model. Nevertheless, this assumption
represents the simplest correction that can be made for the
bias. Therefore, for each galaxy with $\chi_\nu^2>2$ we identify all
well-fit galaxies with apparent magnitude differing by less than $0.1$
and redshift differing by less than $0.03$ from the true values for
the poorly-fit galaxy. We then select a galaxy from this sample at
random and adopt its S/T ratio for our poorly-fit galaxy. 

In Appendix~\ref{model_tests} we find that the value of S/T recovered by {\sc Galactica} (and also {\sc
Gim2D}) for mock images are biased. The median bias in
S/T produced by {\sc Galactica} can be approximated by a linear
dependence on the true S/T (see Fig.~\ref{fig:fitg-MKOBJ}). We use
this linear relation to apply a correction to the value of S/T
recovered for each SDSS galaxy in order to obtain an estimate of the
unbiased value. We use these corrected estimates of the S/T when estimating luminosity and mass functions.

We find that there are only small changes in the measured luminosity
functions, the most significant being a small enhancement in the
abundance of bright spheroids. The luminosity density ratio quoted
above varies by less than $0.5\sigma$ after correcting for this bias.

\begin{figure}
 \begin{center}
\psfig{file=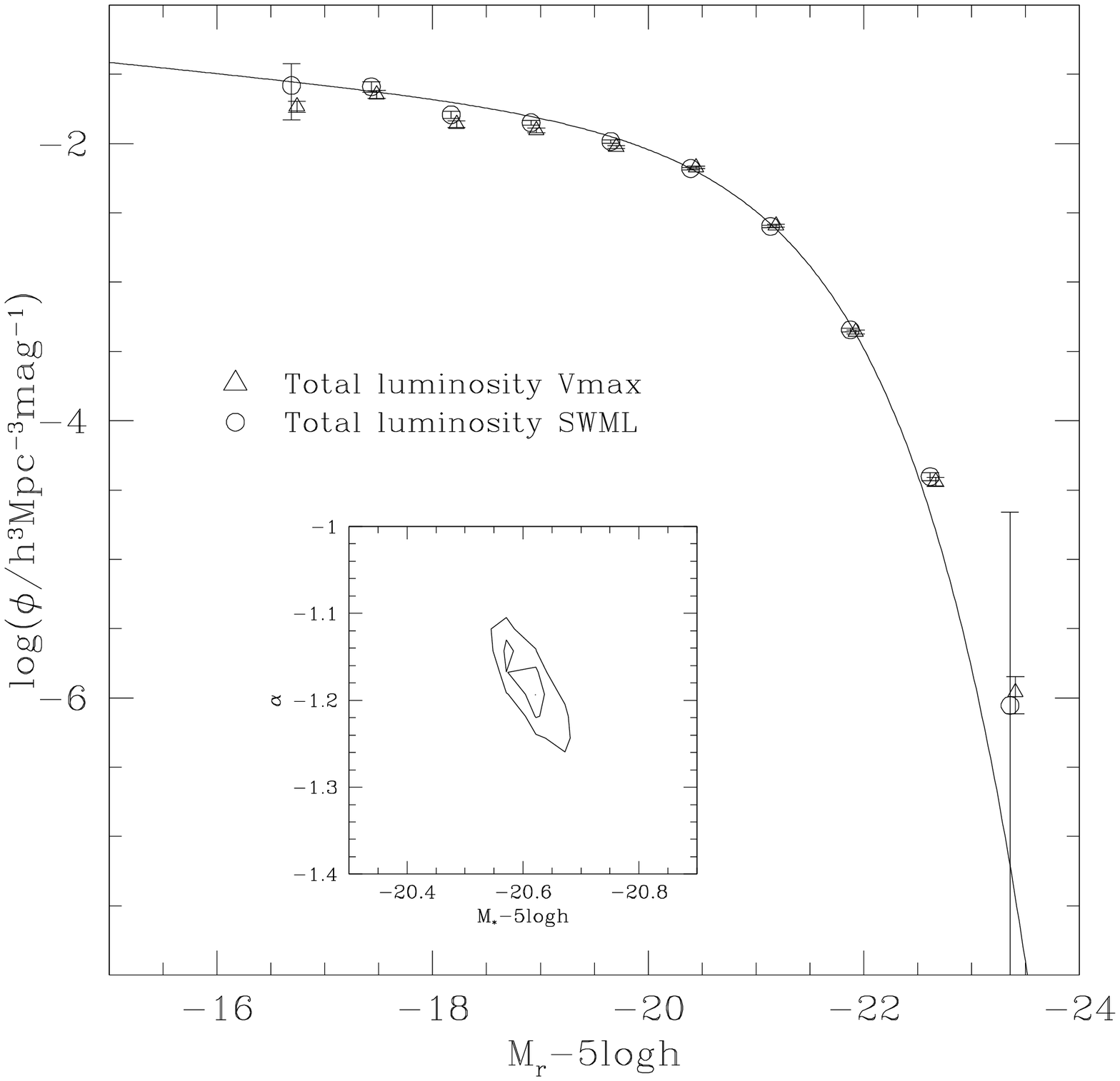,bbllx=0mm,bblly=69mm,bburx=210mm,bbury=245mm,clip=,width=7.88cm}
\psfig{file=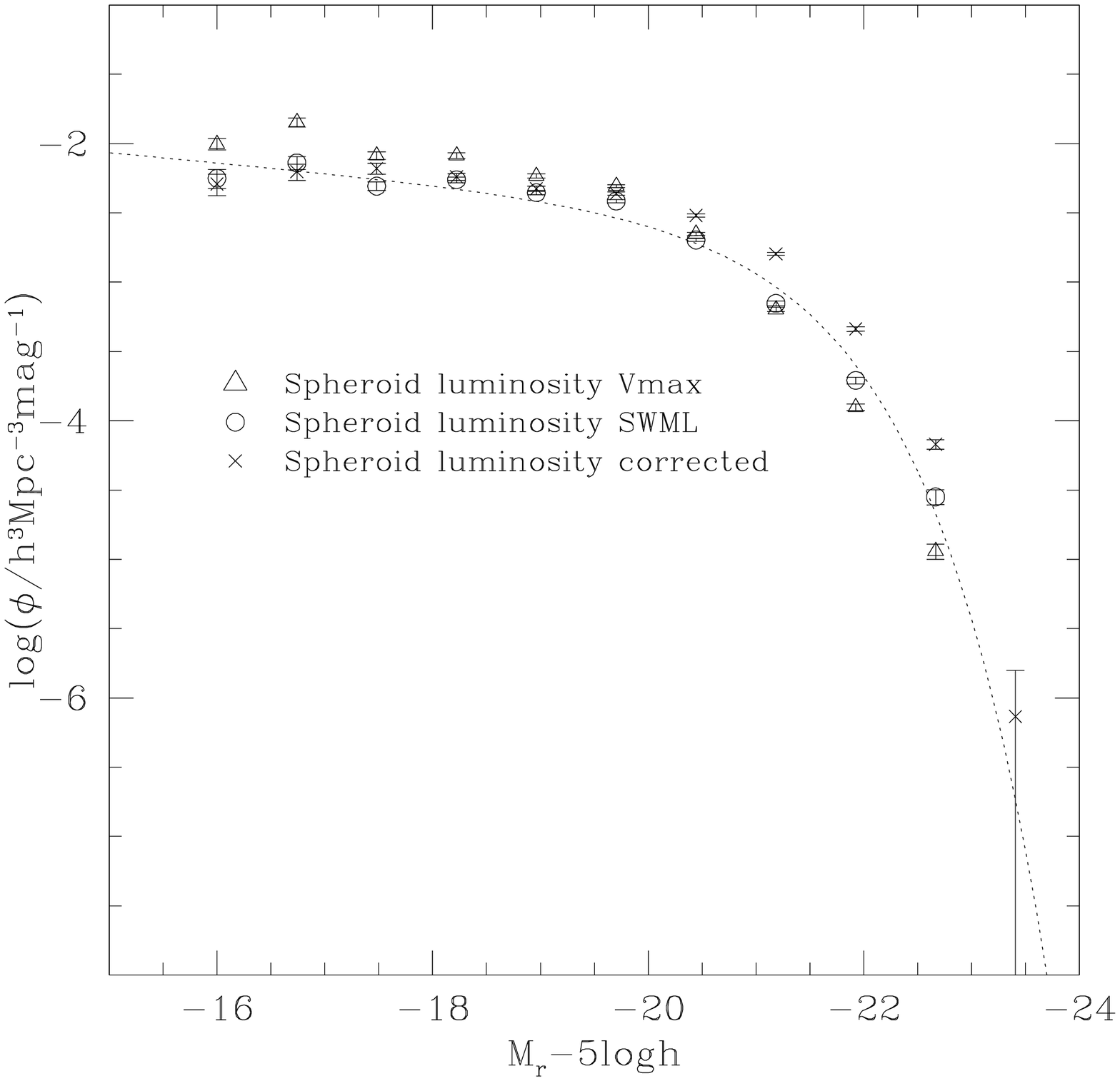,bbllx=0mm,bblly=69mm,bburx=210mm,bbury=241mm,clip=,width=7.88cm}
\psfig{file=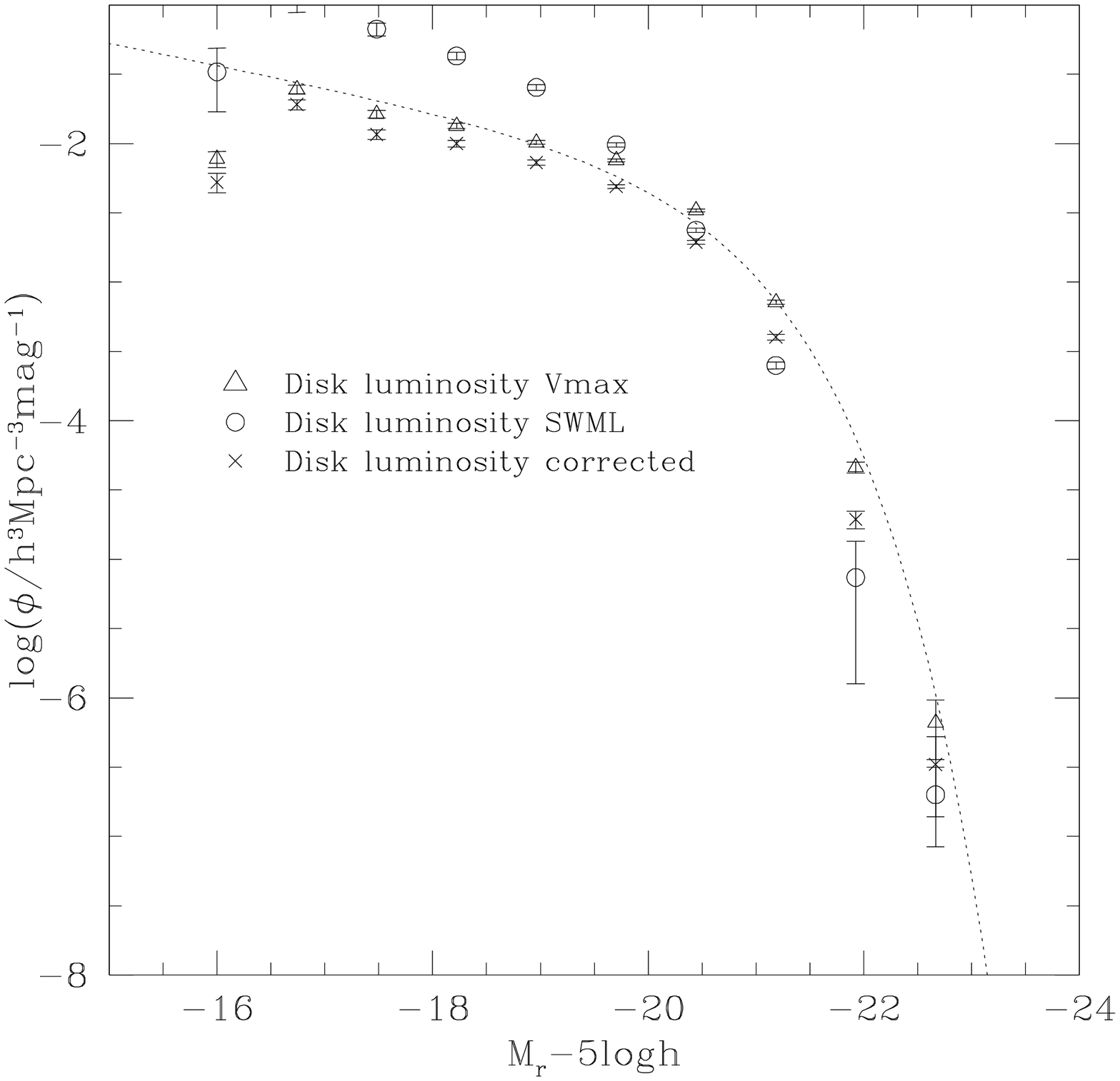,bbllx=0mm,bblly=50mm,bburx=210mm,bbury=241mm,clip=,width=7.88cm}
\caption{Luminosity functions for our sample of $7493$ galaxies
   with $\chi^2_\nu<2.0$. Magnitudes are the total absolute magnitudes of the galaxies in the upper panel, and the absolute magnitudes of spheroid and disk components in the lower panels. Open symbols show the $1/V_{\rm max}$ and SWML
   estimates, while crosses show he SWML estimate after correcting for the biased distribution of $\cos i$; the solid line in the top panel represents the STY
   fit, while dotted lines in the lower panels indicate the best fit Schechter function to the SWML data points. The top panel displays the total galaxy luminosity function,
   the middle panel the luminosity function of spheroids and the lower
   panel, the luminosity function of disks} 
\label{fig:lf_real_all}
\end{center}
\end{figure}

Our results are displayed in Fig.~\ref{fig:lf_real_all}.  For galaxies
whose images are well-fit by our model, we find that the STY method
accurately recovers the parameters of the total luminosity function;
furthermore, the STY fit traces the corresponding SWML points very
well. The values of $M_*$ and $\alpha$ obtained from the STY fit to the total
luminosity function agrees very well with that of \scite{naka03} (SDSS
$r$-band, $z=0$). While we have not been able to find a
parametric form which fits the spheroid and disk 2D luminosity
functions ($\Phi(M,S/T)$) we have determined the parameters of
Schechter functions which fit the SWML data points reasonably
well. These should not be considered good fits in a statistical sense,
merely useful fitting functions. The parameters of the best fit
Schechter functions are given in Table~\ref{tb:schech}. 

\begin{table*}
\caption{Best fitting Schechter function parameters for luminosity
functions of total, spheroid and disk components. For the total
luminosity function the best fit parameters are determined using the
STY method. For the disk and spheroid luminosity function we instead
fit a Schechter function to the non-parametric luminosity function
determined using the SWML method---these should be considered useful
fitting functions only, not good fits in any statistical sense. For
the spheroid and disk luminosity function fits, the maximum deviation
from the SWML data points is given in the final column.} 
\label{tb:schech}
\begin{center}
\begin{tabular}{lcccc}
\hline
{\bf Component} & \boldmath{$M_*-5\log h$} & \boldmath{$\alpha$} & \boldmath{$\phi_0/h^3$}{\bf Mpc}\boldmath{$^{-3}$} & {\bf Max. Dev.}\\
\hline
Total     & -20.62        & -1.19    & 0.00155 & N/A \\
Spheroid  & -20.98        & -1.18    & 0.00348 & 32\% \\
Disk      & -20.40        & -1.39    & 0.00830 & 27\% \\
\hline
\end{tabular}
\end{center}
\end{table*}

We calculate luminosity densities of disks and spheroids by integrating over the SWML
points\footnote{No correction is included for galaxies fainter than the lower limit shown in the figures. Using the best-fit Schechter functions listed in Table~\protect\ref{tb:schech} we estimate that including fainter spheroids/disks would lead to corrections of 1\%/4\% respectively. We regard these corrections as speculative since the Schechter function does not provide a good fit to the spheroid and disk luminosity functions.}. We find the luminosity densities for spheroids and disks to
be: $\rho_{\rm L}=0.611 \pm 0.008 \times 10^8 hL_\odot$ Mpc$^{-3}$ and $\rho_{\rm
L}=1.07 \pm 0.02 \times 10^8 hL_\odot$ Mpc$^{-3}$ respectively.

These values are in
contrast with the findings of the previous study of \scite{benson02}
who found the spheroid and disk luminosity densities to be very nearly
equal. Of course, \scite{benson02} used a very small sample of
galaxies to compute luminosity densities, finding a ratio of disk to
spheroid luminosity density of $1.2\pm0.9$. Our current sample gives a
ratio of $1.75\pm0.04$, which is consistent with that of
\scite{benson02}.

\subsubsection{Corrections for Systematic Effects}\label{sec:syscor}

\begin{figure}
 \psfig{file=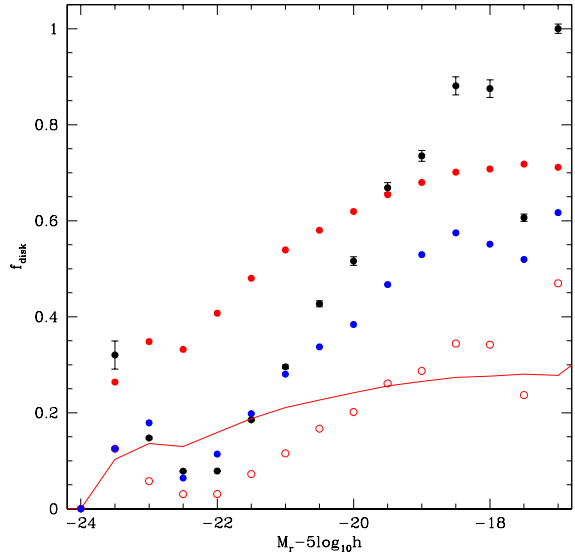,width=8cm}
\caption{The disk light fraction, $f_{\rm disk}$, as a function of
absolute magnitude. Red points show $f_{\rm disk}$ measured from our
image decompositions with no attempt to correct for the non-uniform
distribution of $\cos i$. Black points show the disk fraction
resulting from the correction described by \protect\scite{tw} (error
bars are shown only for these points for clarity---they are similar
for other points). Blue points show the result of applying a more
detailed correction motivated by the assumptions of
\protect\scite{tw}. Red open circles indicate the fraction of light in
disks with $\cos i < 0.5$, while the red line shows the filled red
points reduced by a factor of 2.56.} 
\label{fig:fdisk}
\end{figure}

\scite{tw} propose a method to correct for the bias introduced by the non-uniform $\cos i$ distribution discussed in \S\ref{sec:cosibias}. This method involves using only those galaxies with $\cos i<0.5$ (which \scite{tw} consider to be true disks) to estimate the fraction of light emitted by disks as a function of absolute magnitude. This function, $f_{\rm disk}(M_r)$, is then averaged over the total galaxy luminosity function in order to obtain an estimate of the fraction of light emitted by disks. \scite{tw} include a correction for the inclination-dependent dust extinction experienced by galaxy disks, finding that, at any given magnitude, $f_{\rm disk}$ should be 2.56 times\footnote{This correction would be precisely 2 if there were no dust-extinction of the galaxy disks.} the fraction of light emitted by disks with $\cos i<0.5$. Figure~\ref{fig:fdisk} shows the function $f_{\rm disk}$ for our galaxies. Filled red points are the result of summing the luminosities of disks over all values of $\cos i$, i.e. with no attempt to correct for the biased distribution of $\cos i$. Filled black points show the result after applying the \scite{tw} correction (note that in cases where this correction would imply $f_{\rm disk}>1$ we limit the value to unity). Averaging over the SDSS $r$-band luminosity function of \scite{blanton} we find that $(53\pm 3)$\% of the local luminosity density is contributed by disks. This is consistent with the $(54\pm 2)$\% obtained by \scite{tw}. It should be noted that this result is robust to changes in our decision to include all galaxies with $\chi_\nu^2<2$ in our final sample. Reducing this cut to $\chi_\nu^2<1.2$ for example results in a disk luminosity fraction of $(47\pm 3)$\% ---consistent with the previous result within the quoted errors.

We can attempt to use this same approach to construct disk and
spheroid luminosity functions corrected for the non-uniform $\cos
i$. To do this, we take our catalogue of galaxies and identify those
with $\cos i<0.5$ These galaxies are assumed to have been correctly
fitted (i.e. the disk component of our fit corresponds to a real thin
disk in these galaxies) and are placed into a refined catalogue. Since
we assume that the true $\cos i$ distribution should be uniform we
expect one galaxy with $\cos i>0.5$ for each galaxy with $\cos i
<0.5$. Therefore, for each galaxy in our $\cos i<0.5$ sample we search
for a galaxy with similar spheroid and face-on disk absolute
magnitudes but with $\cos i > 0.5$. The most similar galaxy is added
to our refined catalogue. At the end of this procedure what remains is
a sample of galaxies with $\cos i>0.5$ for which there are no $\cos
i<0.5$ counterparts. We assume that in these cases the disk component
has been used to fit some feature in the spheroid. Therefore, we set
the S/T ratio of these remaining galaxies to 1 and include them in our
refined catalogue.

This procedure should give a conservative lower limit to the disk luminosity fraction. The disk fraction obtained via this method is shown by the blue points in Fig.~\ref{fig:fdisk}. Note that this matches the \scite{tw} method for bright galaxies, but falls below it at faint magnitudes. The reason for this is simple: the \scite{tw} method assumes that the total disk luminosity in any bin of absolute magnitude is 2.56 times that of disks with $\cos i<0.5$ in that bin, \emph{even if that exceeds the total fitted disk luminosity of all galaxies in that bin of absolute magnitude}. Thus, the \scite{tw} method can create additional disk light in some bins, contrary to the assumption that the image decomposition code has added in extra disk light to fit details of the spheroid component. The open red circles in Fig.~\ref{fig:fdisk} show the fraction of light from disks with $\cos i<0.5$ while the solid red line indicates the total disk luminosity reduced by a factor $2.56$. Where the open red circles lie above the red line the \scite{tw} method must create additional disk light. In our more detailed method, disk light can never be created, and so the blue points always lie below the red points.

The problem just discussed illustrates the limitations of the
\scite{tw} method, and indicates that the reality here is
significantly more complicated than the simple assumption adopted by
\scite{tw}. Nevertheless, our more detailed implementation of their 
method should still give a good lower limit on the disk luminosity
fraction. The resulting disk and spheroid luminosity functions are
shown as crosses in Fig.~\ref{fig:lf_real_all}. We find a disk
luminosity fraction of $(43\pm 1)$\%.

In short, the \scite{tw} method works provided all objects with $\cos
i<0.5$ are correctly fit (i.e. the model disk is fit to a true
disk). \emph{If} this assumption is correct, then our data imply that
{\sc Galactica} and {\sc Gim2D} must be systematically failing to fit
the disk components of equivalent galaxies with $\cos i>0.5$,
assigning some of the disk light to a spheroid component. This could
occur, for example, if in face on galaxies the codes use the spheroid
component to fit a bar feature in the disk.

To summarize, our results suggest that stars in disks contribute
between 43 and 64\% of the local luminosity density. \scite{tw} found
a disk contribution of 54$\pm$2\% which is entirely consistent with
this range. Furthermore, if we apply Tasca \& White's correction for
the bias in $\cos i$ precisely as they did we find a disk fraction of
$(53\pm3)$\%, in excellent agreement with their result. However, as we
have shown above, it is not clear that the \scite{tw} correction is
entirely valid and hence we prefer to quote the range above which we feel is a very conservative estimate of the disk contribution to the luminosity density.

Finally, as noted in \S\ref{sec:ellip}, a we suspect that {\sc Galactica} frequently uses a highly elliptical spheroid component to fit bar-like features in galaxy disks. If we assume that all spheroids at the upper limit of allowed ellipticities (i.e. those in the final bin in Fig.~\ref{fig:fitg_distns}) are in fact bars, and therefore count their light as originating from the disk we find that our estimate of the disk luminosity density is increase by 16\% while that of the spheroid luminosity density is decreased by 10\%. Consequently, this correction would adjust the disk luminosity density fraction from 43\% up to 50\%.

\subsection{Comparison with theoretical predictions}

\begin{figure}
\psfig{file=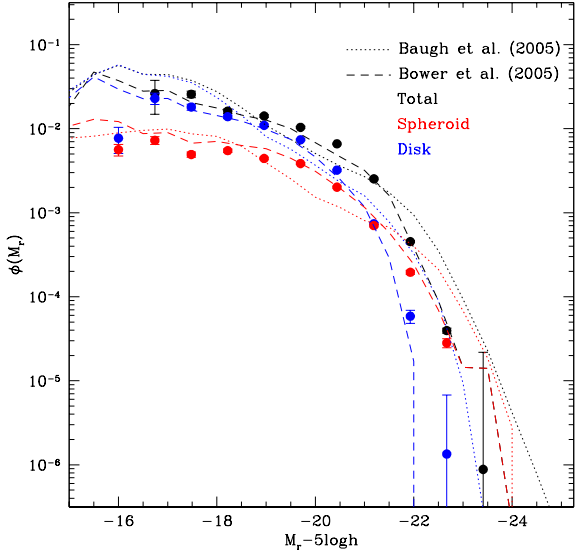,width=80mm}
\caption{Luminosity functions disk and spheroid and total light
(indicated by colour; see legend for details). The symbols show our
estimates for SDSS galaxies in this work and the lines two different
implementations of the {\sc galform} semi-analytic model:
\protect\scite{baugh05} (dotted lines) \protect\scite{bower06} (dashed lines).} 
\label{fig:GFcompare}
\end{figure}

In Fig.~\ref{fig:GFcompare} we compare our estimate of the disk and
spheroid luminosity functions with predictions from the \scite{baugh05} and
\scite{bower06} implementations of the {\sc galform} semi-analytic
model of galaxy formation. These two models differ in a number of
important respects. For example, in \scite{bower06} feedback 
from the emission of active galactic nuclei plays a role in quenching
cooling flows in clusters; in the \scite{baugh05} model, a top-heavy
IMF is assumed for stars that form in starbursts. The two models,
however, assume similar mechanisms for the formation of disks and
spheroids: disks form when spinning gas cools in a halo while
spheroids form either by major mergers or by instabilities in the
disks. Although both models generally provide a reasonable description
of many galaxy properties, they have different strengths and
weaknesses. Neither of them has been previously applied to the
study of galaxy morphology, although the parameter $f_{\rm ellip}$ (first introduced into semi-analytic models by \scite{kwg93}), which controls the mass ratio at which a galaxy merger is deemed to destroy any pre-existing disks and create a spheroid, was constrained to produce a good match to morphological fractions as a function of absolute magnitude and galaxy colours.

Fig.~\ref{fig:GFcompare} shows that both {\sc galform} models
reproduce the main trends seen in the SDSS luminosity functions. At
faint magnitudes, the luminosity function is dominated by disks while
at bright magnitudes disks and spheroids make comparable
contributions. The \scite{bower06} model in particular provides a good
match to the SDSS luminosity functions.

\subsection{Stellar mass functions}\label{sec:MF}

As noted in \S\ref{sec:KE}, our procedure for determining K+E
corrections also provides an estimate of the stellar mass of each
galaxy. Using these stellar masses we have constructed total, spheroid
and disk stellar mass functions using the SWML and STY (for total mass
only) methods. For the spheroid and disk mass functions we also derive
the Schechter function which best fits the SWML data points. It should be noted that
we implicitly assume that the mass-to-light ratio, $\Upsilon$, 
determined for each by our K+E correction procedure is the same for
both the disk and spheroid components. In reality, the recovered value
of $\Upsilon$ reflects some weighted average of the $\Upsilon$ of each
component. To improve upon this situation would require a more
advanced procedure in which the $\Upsilon$ (and K+E correction) of
disk and spheroid components were estimated separately using
measurements of the disk and spheroid colours. This would require
performing spheroid-disk decompositions in multiple wavebands. 

\begin{table*}
\caption{Best fitting Schechter function parameters for stellar mass
functions of total, spheroid and disk components. For the total
stellar mass function the best fit parameters are determined using the
STY method. For the disk and spheroid stellar mass functions we
instead fit a Schechter function to the non-parametric stellar mass
function determined using the SWML method---these should be considered
useful fitting functions only, not good fits in any statistical
sense. For the SMBH mass function we find that a generalized Schechter
function (see eqn.~\protect\ref{eq:genschech}) provides a better fit
to the data. The $\gamma$-parameter of this function is given in the
final column.} 
\label{tb:mschech}
\begin{center}
\begin{tabular}{lcccc}
\hline
{\bf Component} & \boldmath{$\log_{10}(M_*/h^{-2}M_\odot)$} & \boldmath{$\alpha$} & \boldmath{$\phi_0/h^3$}{\bf Mpc}\boldmath{$^{-3}$} & \boldmath{$\gamma$} \\
\hline
Total     & 10.82   & -1.57 & 0.0035$\pm$0.0002 & 1 \\
Spheroid  & 10.87   & -0.79 & 0.0019          & 1 \\
Disk      & 10.64   & -0.78 & 0.0035          & 1 \\
SMBH      &  7.61   & -0.65 & 0.0029          & 0.6 \\
\hline
\end{tabular}
\end{center}
\end{table*}

\begin{figure}
\begin{tabular}{c}
\psfig{file=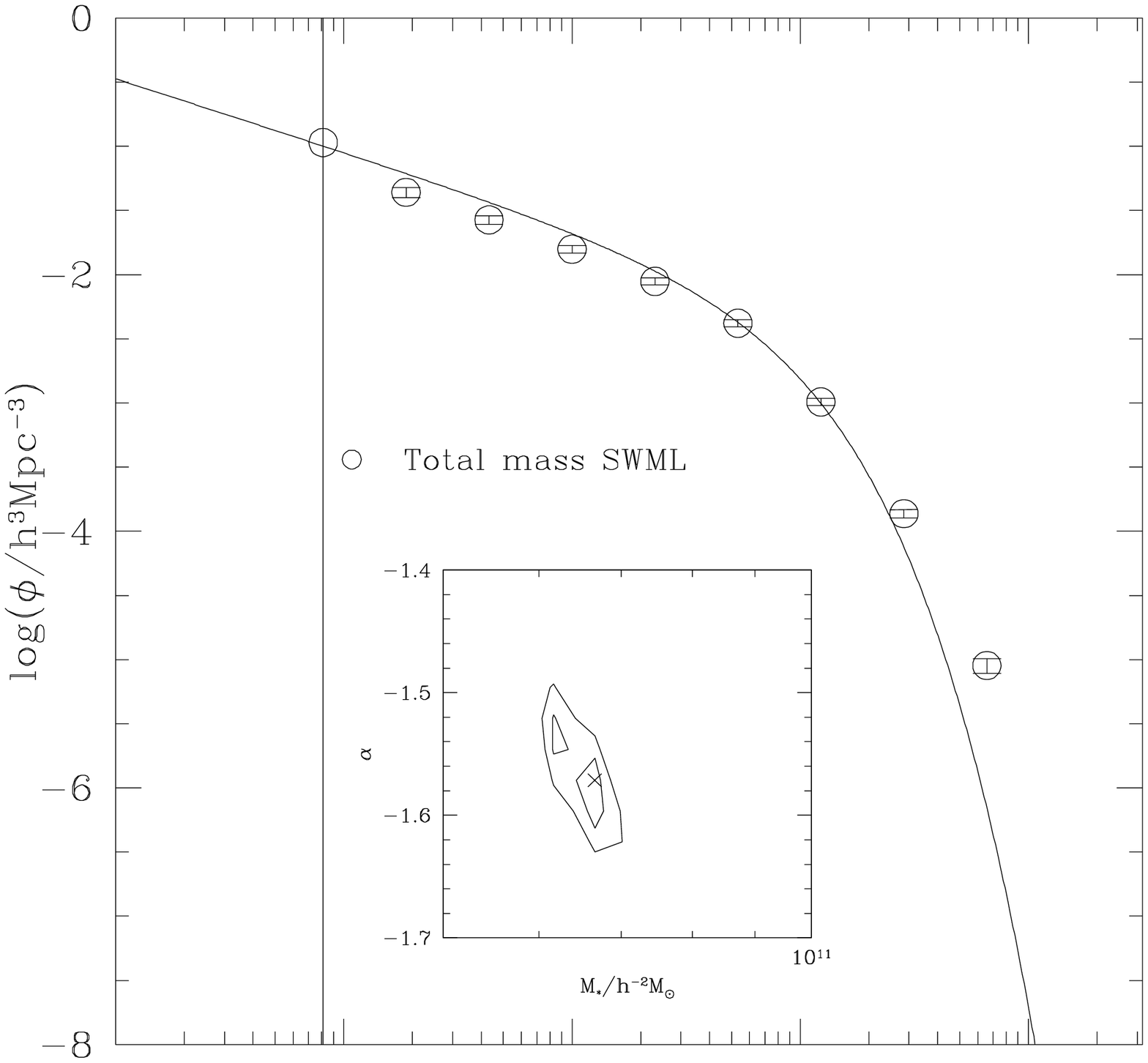,width=80mm,bbllx=0mm,bblly=67mm,bburx=210mm,bbury=244mm,clip=}\\ 
\psfig{file=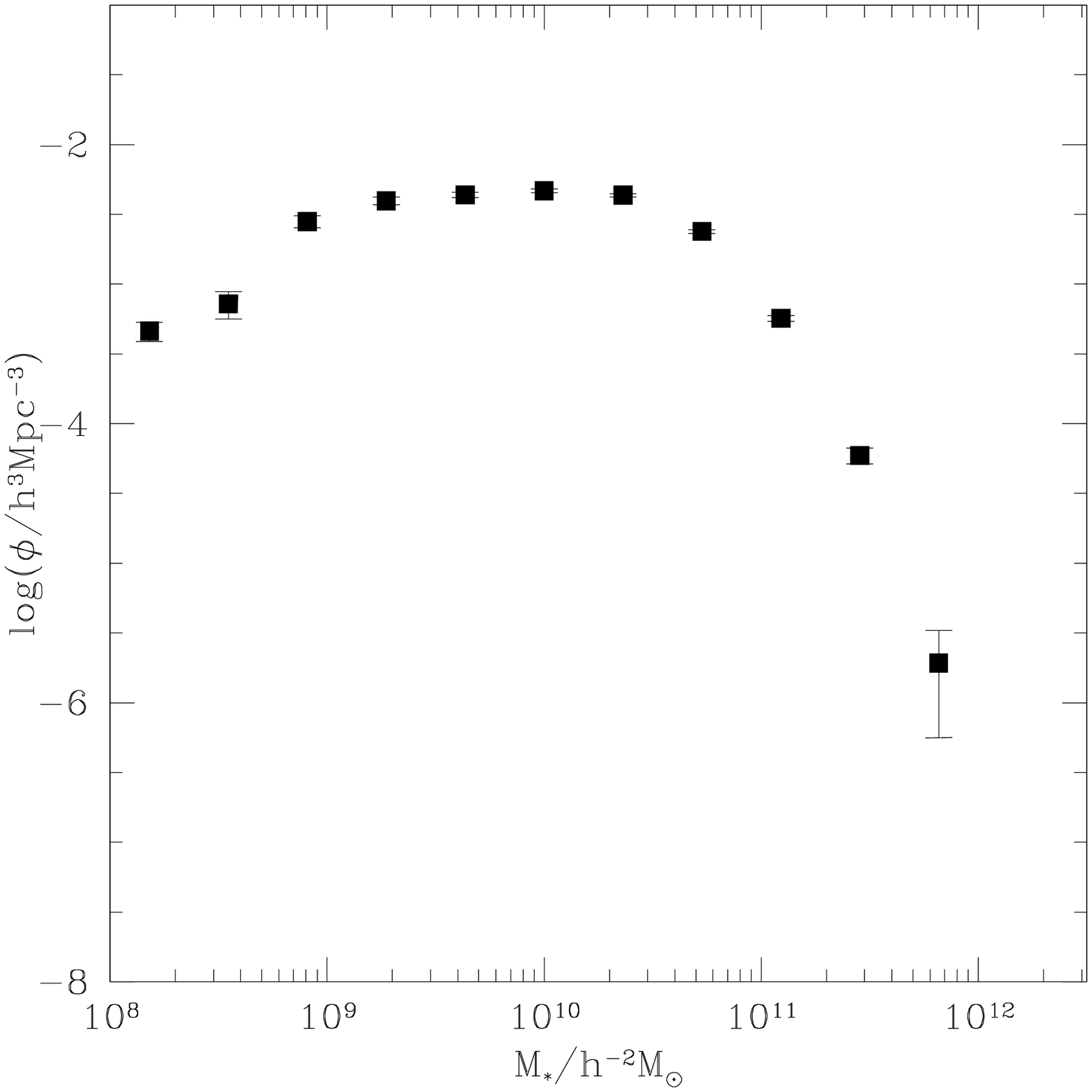,width=80mm,bbllx=0mm,bblly=50mm,bburx=210mm,bbury=244mm,clip=}
\end{tabular}
\caption{Stellar mass functions for galaxies as a whole and their disk
and spheroid components (upper and lower panels) obtained using
the SWML estimator. For the total mass function we also plot the
Schechter function derived using the STY method (solid line), with the
constraints on the parameters $M_*$ and $\alpha$ shown in the
inset. For the disk and spheroid mass functions, the dotted lines show
the Schechter function which best fits the SWML data points.} 
\label{fig:mf}
\end{figure}

A quantity of interest is the ratio of stellar mass in disks to that
in spheroids, averaged over the entire galaxy population. Integrating
the stellar mass densities we obtain
the average density of stars in disks and spheroids in units of the
critical density. We find $\Omega_{\rm stars,disks} = (0.486 \pm
0.004)h^{-1}10^{-3}$ and $\Omega_{\rm stars,spheroids} = (0.465 \pm
0.006) h^{-1}10^{-3}$. These results are in good agreement with those
of \scite{benson02} who found $\Omega_{\rm stars,disks} = (0.51 \pm
0.08)h^{-1}10^{-3}$ and $\Omega_{\rm stars,spheroids} = (0.39 \pm
0.06) h^{-1}10^{-3}$. We conclude that the fraction of stellar mass
found in disks today is $51\pm1$\%.

If we adopt the same correction for the biased distribution of $\cos i$ as we used in \S\ref{sec:syscor} we can construct stellar mass functions for disks and spheroids. The results are shown in Fig.~\ref{fig:mf}, with parameters of Schechter
function fits given in Table~\ref{tb:mschech}. After applying this correction we find stellar mass densities of $\Omega_{\rm stars,disks} = (0.330 \pm
0.004)h^{-1}10^{-3}$ and $\Omega_{\rm stars,spheroids} = (0.622 \pm
0.010) h^{-1}10^{-3}$ so that $(35\pm 1)$\% of stellar mass at $z=0$ is found in disks.

As discussed above, a difficulty in converting from light to stellar mass is that we expect that disks and spheroids should have rather different mass-to-light ratios. In the above, we have used a mean mass-to-light ratio, estimated from our K+E method, to convert disk and spheroid light to disk and spheroid stellar mass. To examine the consequences of this, we perform the following simple experiment. We use our dataset to find the mean mass-to-light ratios as a function of redshift of systems identified as pure disks and pure spheroids (technically we identify systems which are at least 90\% disk or 90\% spheroid respectively). We then assume that in composite systems (i.e. galaxies with comparable fractions of light in their disk and spheroid) the mass-to-light ratios of the individual components are given by these mean values for pure systems. We can then estimate the stellar mass of the disks and spheroids using the total luminosity, measured S/T and the estimated mass-to-light ratios for disk stars and spheroid stars separately. We then compute stellar mass densities using these revised masses. We find that this changes our results by less than the errors quoted above. This approach represents only an approximate method for finding the mass-to-light ratios of individual components. Nevertheless, it suggests that such corrections will be small.

\subsection{Black hole mass function}

In the last few years, it has been conclusively demonstrated that many
galaxies posses central supermassive black holes and that their mass
is strongly correlated with the properties of the galaxy's spheroid
such as its luminosity, stellar mass and velocity dispersion
\cite{kr95,mag98,mf01,mh03,harrix}. Although there is only direct evidence
for these black holes in bright galaxies, it seems quite plausible
that galaxies of all sizes have a central black hole (e.g.
\pcite{malbon06}). 

From the mass function of galactic spheroids determined in
\S\ref{sec:MF}, assuming that all spheroids contain a supermassive
black hole at their centre, we can estimate the mass function of
supermassive black holes in the local Universe. We assume that the
black hole mass is given by $M_\bullet/M_\odot =1.6\times 10^8 [M_{\rm spheroid}/10^{11}M_\odot]^{1.12}$ \cite{harrix} and ignore any scatter in this relation since
an accurate determination of the black hole mass function would first
require a deconvolution of the (uncharacterized) error distribution of
spheroid masses.

\begin{figure}
\psfig{file=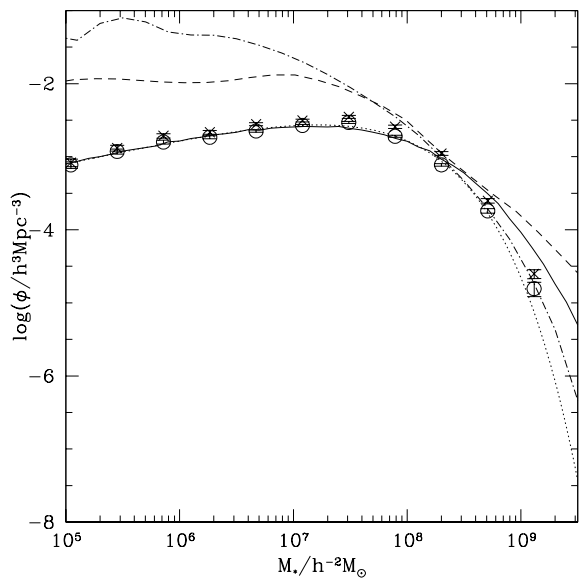,width=80mm} \\
\caption{The mass function of supermassive black holes in galactic
spheroids. Symbols show the black hole mass function implied by our
observationally determined spheroid stellar mass function assuming
that $M_\bullet/M_\odot = 1.6\times 10^8 [M_{\rm spheroid}/10^{11}M_\odot]^{1.12}$
\protect\cite{harrix}. The dotted line shows the Schechter function
which best fits the SWML data points. Other lines show results from the galaxy formation
model of \protect\scite{malbon06} when using the parameters specified
by Baugh et al. (2005; dot-dashed line) and Bower et al. (2006; dashed
line).} 
\label{fig:mfSMBH}
\end{figure}

The resulting black hole mass function is shown in
Fig.~\ref{fig:mfSMBH}. Integrating this mass function
gives a total black hole mass density in the local universe of
$\rho_{\bullet}=(2.8
\pm 0.7) \times 10^5 h  M_{\odot} {\rm Mpc}^{-3}$ where we have included a scatter of 0.3 dex in the spheroid-SMBH mass relation \cite{harrix} and have
included the error in the zero-point of the \cite{harrix} relation in
our error budget. This result is in agreement with previous
determinations
\cite{yt02,allerrichstone02,marconi04,md04,shankar04} based on much
smaller samples of galaxies.

Applying our correction for the biased distribution of $\cos i$ results in an SMBH mass function shown by the crosses in Fig.~\ref{fig:mfSMBH}. We find that a generalized Schechter function of the form
\begin{equation}
 \phi(M_\bullet) = \phi_0 \left({M_\bullet \over M_*}\right)^\alpha \exp\left[-\left({M_\bullet \over M_*}\right)^\gamma\right]
\label{eq:genschech}
\end{equation}
provides a better fit to this SMBH mass function. Parameters of the
generalized Schechter function which best fits the SWML data points
are given in Table~\ref{tb:mschech}. 
The SMBH mass density after applying this correction is $\rho_{\bullet}=(3.77 \pm 0.97) \times 10^5 h  M_{\odot} {\rm
Mpc}^{-3}$, which is consistent with previous determinations.

For comparison with our inferred black hole mass function, we show
results from the recent model of
\scite{malbon06} who incorporate a calculation of supermassive black
hole growth into the {\sc Galform} semi-analytic model of galaxy
formation in a $\Lambda$CDM universe using methods similar to those first described by \scite{kh}. The lines in Fig.~\ref{fig:mfSMBH}
show their results for two different specific galaxy formation
models. While the calculation based upon the parameters of
\scite{baugh05} seems to match the abundance of the high-mass black
holes quite well, neither model is able to reproduce the inferred low
abundance of less massive black holes. 

Before our results can be used to constrain such models strongly, it
will be necessary to achieve a significantly better understanding of
the uncertainties in the measured spheroid mass, and to perform the
conversion from luminosity to stellar mass using a technique which
accounts for the different stellar populations in the spheroid and
disk. 

\subsection{The distribution of S/T}\label{sec:distST}

\begin{figure*}
 \begin{tabular}{cc}
  \psfig{file=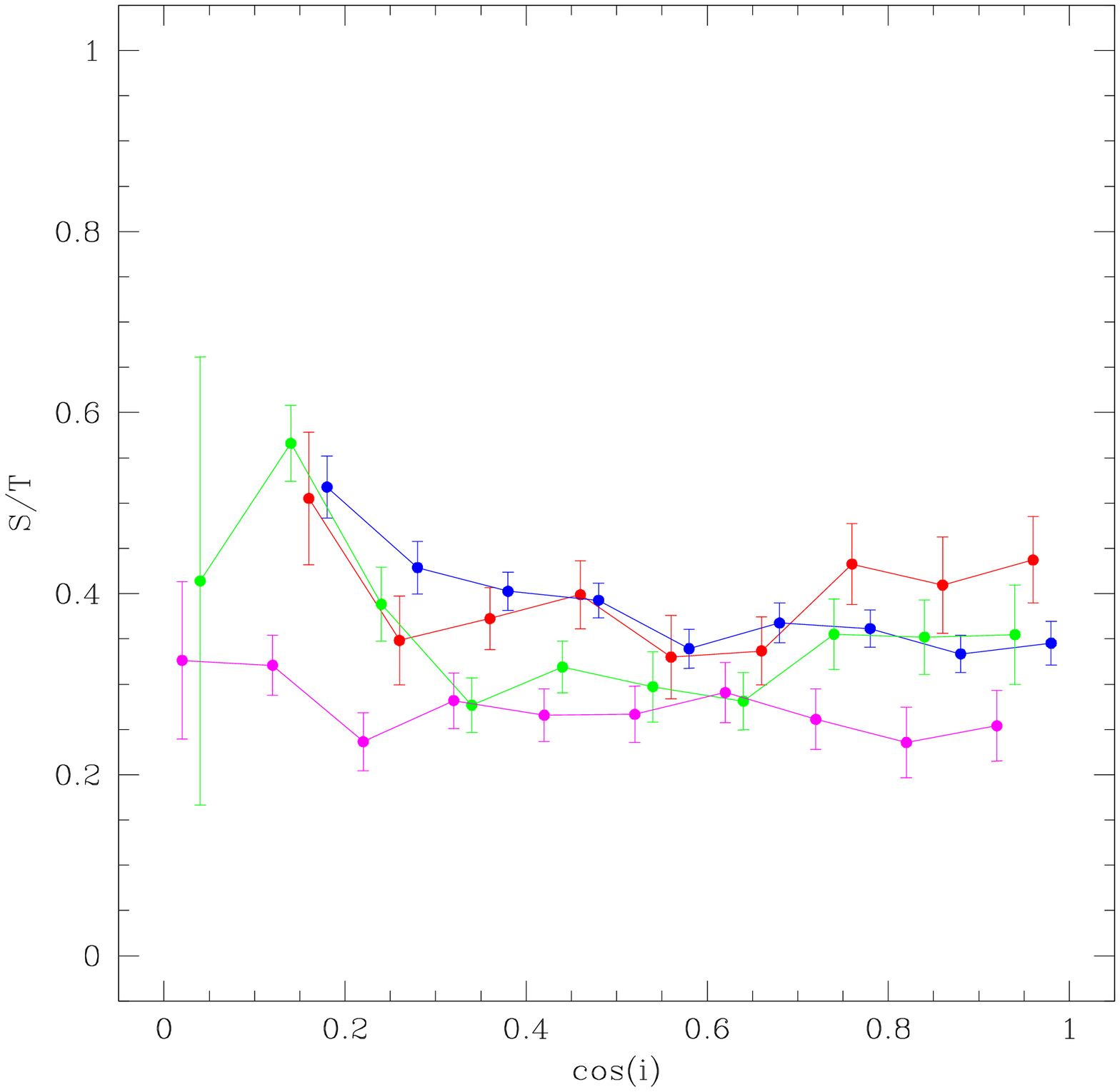,width=80mm} & \psfig{file=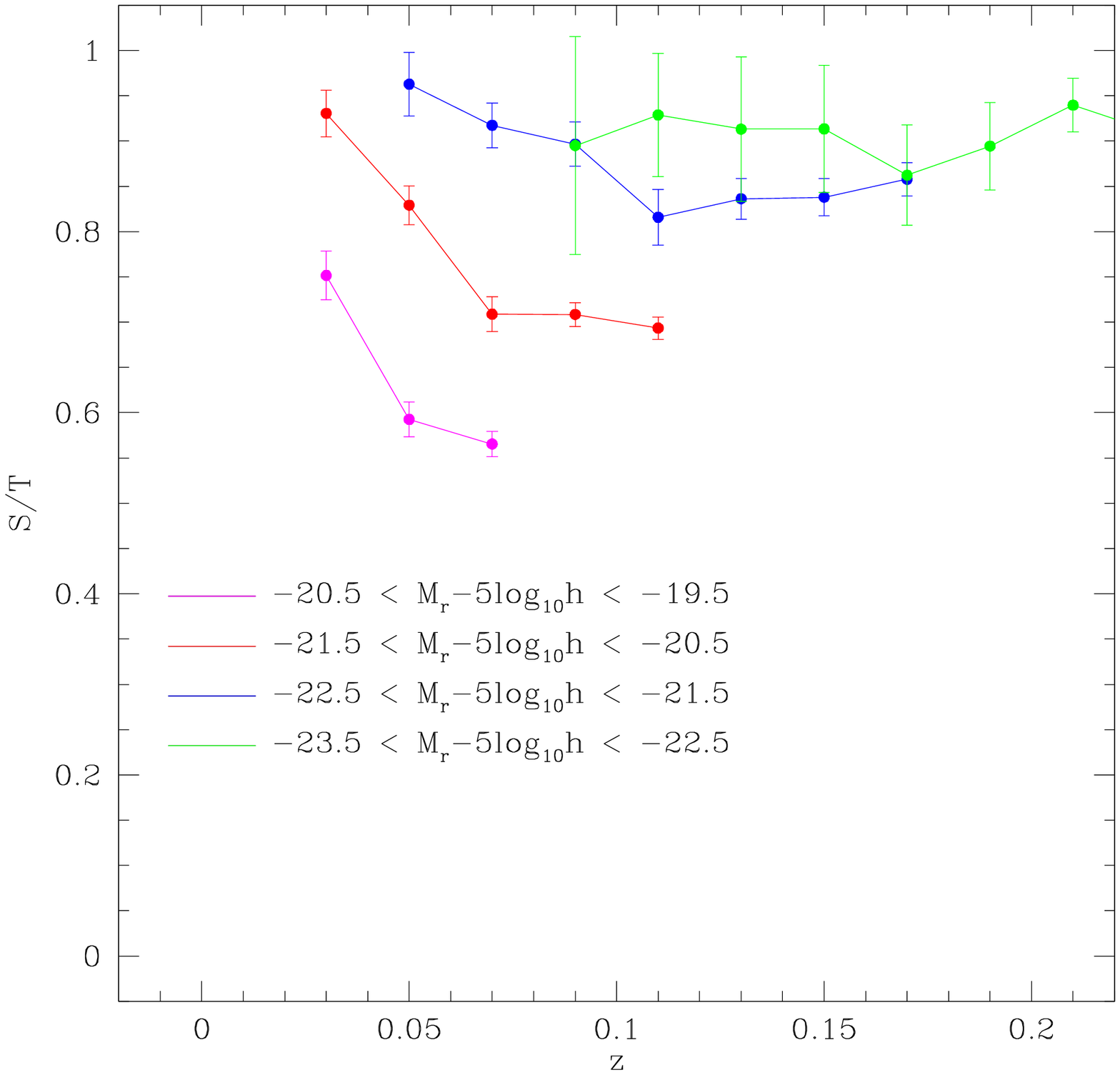,width=80mm}
 \end{tabular}
 \caption{The mean, $1/V_{\rm max}$ weighted S/T ratio as a function of disk inclination (left-hand panel) and redshift (right-hand panel). Points show the mean S/T in each bin, while error bars show the $1\sigma$ error on the mean. In the left-hand panel, galaxies are split into four groups by luminosity and redshift at $M_r-5\log_{10}h<-20.57$ and $z<0.085$. The bright, low-redshift sample is shown by red points, the bright high-redshift sample by blue points, the faint low-redshift sample by magenta points and the faint high-redshift sample by green points. Galaxies for which the disk component of the model is thought to have been used to fit features of the observed spheroid have been excluded. In the right-hand panel galaxies are split by luminosity as indicated by the labels in the figure.}
 \label{fig:S2Tdists}
\end{figure*}

Finally, we examine the distributions of S/T obtained after applying
the corrections for systematic effects described previously in this
section. Not only are these distributions of interest in their own
right, but they can also serve as valuable checks for additional
systematic biases in our fitting procedure. We expect that the
distribution of S/T (and therefore the mean S/T) should be independent
of disk inclination and of redshift (at least for a sufficiently
shallow sample that evolution can be ignored). In
Fig.~\ref{fig:S2Tdists} we show the mean S/T (weighted by $1/V_{\rm
max}$) as a function of these two quantities, split by galaxy
luminosity and redshift. The mean S/T seems to be reasonably
independent of disk inclination, although there is some evidence for a
rise at low $\cos(i)$. Note that we have excluded all galaxies for
which we believe that the disk component of the photometric model has
been used to fit some aspects of the true spheroid.

When we consider the mean S/T as a function of redshift we see a
significant increase in S/T at low-redshifts for the fainter
samples. Our sample is not sufficiently large to be unaffected by
large-scale structure and, in fact, shows clear evidence of peaks in
the redshift distribution presumably caused by large scale structure
(see Fig.~\ref{fig:sdss_distns}). This could create a redshift dependence in the mean S/T if,
for example, a cluster of galaxies (likely to contain a substantial
population of elliptical galaxies) is present in the sample at low
redshifts. Larger samples, unaffected by large scale structure, would
be needed to address this issue further. For now, we simply note that
for our brightest cut, the mean S/T seems quite independent of
redshift.

\begin{figure}
 \psfig{file=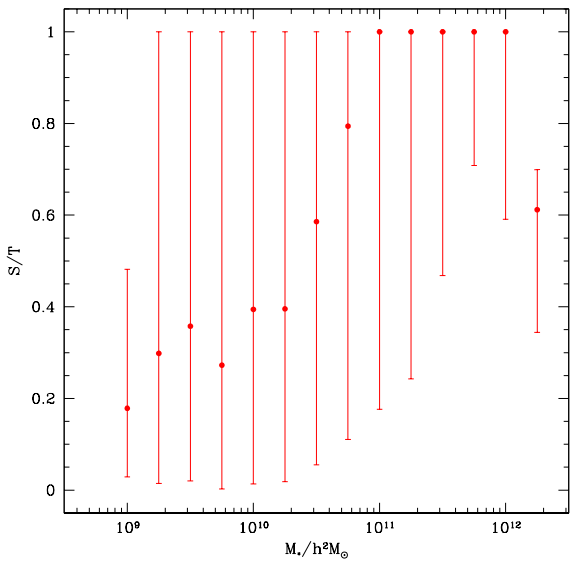,width=80mm}
\caption{The distribution of S/T as a function of galaxy stellar
mass. Galaxies are weighted by $1/V_{\rm max}$. Note that many galaxies are assigned S/T$=1$
during our process of correcting for cases where the disk in the
photometric model has been fit to a true spheroid component. Red
points with error bars show the median S/T as a function of stellar
mass together with the $10^{\rm th}$ and $90^{\rm th}$ percentiles of the distribution.} 
\label{fig:grey}
\end{figure}

Finally, we show in Fig.~\ref{fig:grey} the median S/T (and $10^{\rm th}$ and $90^{\rm th}$ percentiles) as a
function of stellar mass. There is a strong trend for increasing S/T
with stellar mass---the most massive galaxies are ellipticals. Interestingly, the median S/T is fairly constant at $\sim 0.3$ below around $3\times 10^{10}h^2M_\odot$, after which it rises rapidly to become close to unity. This is similar to the $3\times 10^{10}M_\odot$ found by \scite{Kauffmann03} to mark the division between galaxies with young stellar populations, low surface densities and low concentrations and those which are older, higher density and more concentrated.

\section{Summary and conclusions}
\label{sec:conc}

We have used a sample of $\sim9000$ galaxies extracted from the Sloan
Digital Sky Survey to estimate the spheroid and disk luminosity and
stellar mass functions in the local universe using the {\sc Galactica}
code of \scite{benson02}. The 2D model fits to the surface brightness
have revealed a bias in the recovered disk inclination angle arising
from the lack of strong constraints on this parameter for most
galaxies.

We find that at faint r-band luminosities, the light is dominated by
disks whereas at bright luminosities, it is dominated by spheroids,
with the changeover occurring at around the characteristic luminosity
$L_*$. Integrating the luminosity functions, we find the total r-band
luminosity densities in spheroids and disks to be $\rho_{\rm L}=0.611 \pm 0.008 \times 10^8 hL_\odot$ Mpc$^{-3}$ and $\rho_{\rm
L}=1.07 \pm 0.02 \times 10^8 hL_\odot$ Mpc$^{-3}$ respectively. Thus,
the disks contribute approximately two thirds of the total luminosity
density. This is in contrast with the 
findings of previous studies \cite{sd87,benson02}, based upon galaxy
samples over 40 times smaller, which found the spheroid and disk
luminosities to be very nearly equal.

Due to the fact that real galaxies do not always look like our idealized models we find a biased distribution of disk inclinations $\cos i$. This bias has been noted before by \scite{sim02}, \scite{tw} and \scite{allen}. The bias found here is identical to that found by \scite{tw}. Attempting to correct for this bias leads us to a conservative estimate for the disk contribution to the local luminosity density of between 43 to 64\%. Applying the correction suggested by \scite{tw} we find a disk contribution of $(53\pm3)$\% in excellent agreement with their result of $(54\pm2)$\%.

Current {\it a priori} galaxy
formation models are able to reproduce the disk and spheroid luminosity
functions reasonably well---at least to the extent of predicting the
correct trends of abundance with luminosity. 

Using an approximate conversion of r-band light to stellar mass, we
derive $\Omega_{\rm stars,disks} = (0.486 \pm 0.004)h^{-1}10^{-3}$ and
$\Omega_{\rm stars,spheroids} = (0.465 \pm 0.006) h^{-1}10^{-3}$, in
excellent agreement with the earlier work of \scite{benson02}. Correcting for the bias in the $\cos i$ distribution leads to revised values of $\Omega_{\rm stars,disks} = (0.458 \pm 0.005)h^{-1}10^{-3}$ and
$\Omega_{\rm stars,spheroids} = (0.622 \pm 0.010) h^{-1}10^{-3}$. Thus, stars in disks contribute between 35 and 51\% of the local stellar mass density.
 \scite{bell} claim that between 50 and 75\% of the stellar mass density comes from late-type galaxies. This range is in excellent agreement with our results. It should be noted that the morphological selection chosen by \scite{bell} would essentially select all galaxies with a spheroid-fraction of 0.5 or greater (and a large number of galaxies with smaller spheroid fractions) according to our calculations (see Fig.~\ref{fig:bt_conc}).

From
the inferred spheroid mass function and the observed relation between
central supermassive black hole mass and spheroid stellar mass, and
assuming that all spheroids harbour a central black hole, we infer the
supermassive black hole mass function. The associated black hole mass
density in the local universe is $\rho_{\bullet}=(3.77
\pm 0.97)
\times 10^5 h M_{\odot} {\rm Mpc}^{-3}$, consistent with previous estimates. Improvements in the characterization of the errors
in the spheroid mass function and in the stellar population modelling
will enable a better estimate of the spheroid (and black hole) mass
function.

We conclude that the local Universe contains around roughly comparable amounts of
stars, by mass, in disks and in spheroids. This fundamental ratio is
the outcome of the physical processes at play in the formation of the
galaxy population. 

\section*{Acknowledgments}

We are extremely grateful to Chris Miller for providing us with all of
the SDSS EDR imaging data and Robert Nichol and Tomotsugu Goto for the
encouragement given in the early days of this work. We also wish to
thank Carlton Baugh for providing his code for computing K+E
corrections, Rowena Malbon for providing theoretical predictions for
black hole mass functions from the {\sc Galform} semi-analytic model
({\tt www.galform.org}) and Hans Walter Rix and the anonymous referee
for valuable discussions. We are particularly grateful to Simon White
whose insightful comments on an early draft of the paper led us to
revise our initial analysis and conclusions substantially. AJB
acknowledges support from the Gordon \& Betty Moore Foundation, and
from a Royal Society University Research Fellowship during part of
this work. CSF is the holder of a Royal Society Wolfson Research Merit
award

Funding for the creation and distribution of the SDSS Archive has been
provided by the Alfred P. Sloan Foundation, the 
Participating Institutions, the National Aeronautics and Space
Administration, the National Science Foundation, the 
U.S. Department of Energy, the Japanese Monbukagakusho, and the Max
Planck Society. The SDSS Web site is {\tt http://www.sdss.org/}.

The SDSS is managed by the Astrophysical Research Consortium (ARC) for
the Participating Institutions. The Participating Institutions are The
University of Chicago, Fermilab, the Institute for Advanced Study, the
Japan Participation Group, The Johns Hopkins University, Los Alamos
National Laboratory, the Max-Planck-Institute for Astronomy (MPIA),
the Max-Planck-Institute for Astrophysics (MPA), New Mexico State
University, University of Pittsburgh, Princeton University, the United
States Naval Observatory, and the University of Washington.

\appendix

\section{Methods for quantitative galaxy morphology}\label{Two}

In this Appendix, two independent methods for spheroid-to-disk
decompositions of galaxies are described and compared. A number of
tests are performed to reveal and estimate any potential biases in
the decomposition codes.  

\subsection{Introduction to spheroid-to-disk decomposition}  
  
To first approximation, the surface brightness of a galaxy can be
expressed as the sum of a highly concentrated central component, the
spheroid, and an extended disk. Empirically, the surface brightness
profiles of spheroids and disks are well represented by the following
functions, 
\begin{equation}
 I_{\rm s}=I_{\rm e}\exp(-7.67[(r/r_{\rm e})^{1/4}-1]), 
 \label{eqn:devauc}
\end{equation}
an $r^{1/4}$-law, for the spheroid \cite{dev61}, where $r_{\rm e}$ is
the half-light radius and $I_{\rm e}$ is the surface brightness at
$r_{\rm e}$, and, 
\begin{equation}
 I_{\rm d}=I_{0}\exp(-r/r_{\rm d}),
 \label{eqn:expdisk}
\end{equation}
an exponential-law for a face-on disk, where $r_{\rm d}$ is the
exponential disk scale-length and $I_{0}$ is the central intensity. 

Equations \ref{eqn:devauc} and \ref{eqn:expdisk} can be used to
construct model images of the galaxy. Comparison with the surface
brightness distribution of each galaxy, including the effects of
inclination, enables the fitting parameters to be determined.

\scite{apb95} used this technique to fit the spheroid components of a
sample of morphologically selected galaxies with types ranging from S0
to Sbc. They assumed a more general type
of profile,
\begin{equation}
I_{\rm s}=I_{\rm e}\exp(-b_n[(r/r_{e})^{1/n}-1],
 \label{eqn:sersic}
\end{equation}
first proposed by \scite{sersic}, where $n$ is often referred to as
the S\'{e}rsic index and determines the `peakiness' of the
profile and $b_n$ is a constant dependent on the value of $n$. Andredakis et al. found that the value of $n$ varied systematically
from $1$ for late-type spheroids to $6$ for early-type
spheroids. \scite{deJ96} also suggested that the spheroids of field
spirals are better fit using a pure exponential (i.e. $n=1$) 
profile. 

\subsection{Methods for 2-dimensional spheroid-to-disk decomposition} 

\subsubsection{Fitting parameters revisited}\label{fitparam}

On a Cartesian grid $(x,y)$ the effective $r$ in
eqns.~(\ref{eqn:devauc}) and (\ref{eqn:expdisk}) become
\begin{eqnarray}
  r^2(x,y)&=&\frac{1}{e_{\rm s}}[x\cos(\theta_{\rm
s})-y\sin(\theta_{\rm s})]^2+e_{\rm s}[x\sin(\theta_{\rm s}) \nonumber\\
   & & +y\cos(\theta_{\rm s})]^2\label{eqn:PA_b}\\
\end{eqnarray}
and
\begin{eqnarray}
  r^2(x,y)&=&[x\cos(\theta_{\rm d})-y\sin(\theta_{\rm
d})]^2+\frac{1}{\cos(i)^2}[x\sin(\theta_{\rm d}) \nonumber \\  
   & & +y\cos(\theta_{\rm d})]^2\label{eqn:PA_d}
\end{eqnarray}
respectively. Here $\theta_{\rm s}$ and $\theta_{\rm d}$ are the
spheroid and disk position angles, where a position angle is defined
as the angle of orientation of the galaxy's main axis with respect to
some coordinate system, and $e_{\rm s}$ is the spheroid ellipticity
used to describe the deviation from circularity of the spheroid
component

In terms of a $r^{1/n}$ spheroid and an exponential disk, and including
the sky background, 2D decomposition usually requires a total of 13
free parameters: 
\begin{itemize}
 \item{total flux in the galaxy;}

 \item{S/T: ratio of the amount of light in the spheroid to the total amount of light;}

 \item{$r_{\rm e}$: effective radius of the spheroid;}

 \item{$e_{\rm s}$: spheroid ellipticity;}

 \item{$\theta_{\rm s}$: spheroid position angle;}

 \item{$r_{\rm d}$: scale length of the disk;}

 \item{$i$: inclination angle of the disk;}

 \item{$\theta_{\rm d}$: disk position angle;}

 \item{$x_{\rm c},y_{\rm c}$: subpixel offset of the galaxy centre;}

 \item{residual sky background level;}

 \item{FWHM of the point spread function (PSF);}

 \item{$n$: S\'{e}rsic index.}
\end{itemize}

In order to make the decomposition procedure as accurate and fast as
possible, the following points must be taken into account:
\begin{itemize}
 \item The PSF smooths the galaxy image and so to achieve accurate fits,
 it must be modelled accurately and included in the mock images.

 \item The fit is carried out using small ``thumbnail'' regions around
 each galaxy. The size of the thumbnail must be small enough to enable
 a fast fit to the image, but large enough to include all regions
 of the galaxy with significant signal-to-noise.

 \item The mean sky background level should be $\sim 0$ since the
 decomposition codes are designed to work with no (or very little)
 background; any excess sky light can be mistaken for galaxy light and
 lead to incorrect parameter estimation.
\end{itemize}

We now explore the similarities and differences of two independent
multi-dimensional fitting codes, the {\sc Gim2D} code of \scite{sim02}
and the {\sc Galactica} code of \scite{benson02}.

\subsubsection{Galaxy image 2D decomposition (Gim2D)}

{\sc Gim2D} is a publicly available code written by \scite{sim02} and
widely used for automated spheroid-to-disk decompositions of galaxy
light profiles \cite{balogh02a,nelson02}. This code was purposely
written for imaging with the Hubble Space Telescope Wide Field and
Planetary Camera which has a very well modelled PSF
\cite{krist95}. The code can also be used for ground-based
imaging data but, in this case, special attention must be given to the
much larger and less well defined PSF.\\

\noindent\emph{Object Detection}\\
To extract a postage-stamp image around each galaxy, {\sc Gim2D} relies
upon {\sc SExtractor} \cite{ber96}. {\sc SExtractor} determines the
galaxy centroid position and the area at the faintest detected
isophote to be obtained and also measures the mean level of the sky
background for each galaxy (a $3\sigma$ threshold is usually
sufficient to discriminate between the object and the
background). {\sc Gim2D} extracts a postage-stamp of size
equal to a multiple of a galaxy isophotal area. A value of $15
\times${\tt iso\_area} was found to be optimum. The sky-background is
not recommended to be treated as a free fitting parameter in {\sc
Gim2D} because the underlying sky is not well known and can
potentially bias the output \cite{sim02}. However, before the
decomposition procedure is initiated, {\sc Gim2D} uses the pixels
flagged by {\sc SExtractor} as belonging to the background (flag value
0) to recompute the background value, therefore ensuring that the
mean sky level is close to zero. All the background pixels and also
pixels flagged as `bad' (flag value -2) by {\sc SExtractor} are
subsequently excluded from the fitting altogether.\\

\noindent\emph{Point Spread Function}\\
During the minimization in {\sc Gim2D}, the PSF is kept fixed. For
ground-based imaging, the PSF is obtained from a bright unsaturated
stellar image; for HST data, an analytic PSF modelled using the
Tiny Tim software is used \cite{krist95}.\\

\noindent\emph{Minimization Technique: Metropolis Algorithm}\\
{\sc Gim2D} allows up to 12 parameters to be fit\footnote{Sersic index is held fixed at 4.} and uses the
Metropolis algorithm \cite{metro53} to search for the minimum
$\chi^{2}$ in this multi-dimensional parameter space. Before starting
the Metropolis algorithm, {\sc Gim2D} works in the Initial Condition
Filter (ICF) mode, i.e. it creates a user-specified number of models
between the limits of the user-specified multi-dimensional parameter
space. The ICF computes the model likelihoods and sets the sampling
origin to the parameters of the best model, making it a subvolume to
be exploited by the Metropolis Algorithm.\\

\noindent\emph{{\sc Gim2D} Outputs}\\
After finding the model that corresponds to the highest likelihood,
{\sc Gim2D} produces a residual (object - model) map and calculates
the value of the corresponding $\chi_{\nu}^{2}$. If
$\chi_{\nu}^{2}\sim1$ and the residual map has no remaining
galaxy structure, the best-fit model is accepted. 

\subsubsection{{\sc Galactica}}\label{fitg}

\noindent\emph{Introduction}\\
The 2-dimensional decomposition code described here is based on a
technique proposed by \scite{wad99}. {\sc Galactica} was developed by
\scite{benson02} and assumes the standard empirical formalisms for the
2-dimensional surface brightnesses of a galaxy spheroid and disk
components respectively (eqns.~\ref{eqn:devauc} \&
\ref{eqn:expdisk}).\\ 

\noindent\emph{Object detection}\\
To locate and extract a postage-stamp image around every galaxy, {\sc
Galactica} employs the same method as {\sc Gim2D}. {\sc SExtractor}
is also used to measure the mean level of the sky background for each
galaxy (a $3\sigma$ threshold is usually sufficient to discriminate
between the object and the background) and this value is subtracted
from the corresponding galaxy image. To mask any overlapping objects
within the extracted postage-stamp {\sc Galactica} relies upon an
in-built masking algorithm which finds any objects that contaminate
the galaxy of interest and masks them out. The galaxy itself is also
detected by the algorithm using a $5\sigma$ threshold above the sky
background. Pixels which have not been flagged as belonging to any of
the detected objects are used in the sky background fitting.\\ 

\noindent\emph{Point spread function: Moffat profile}\\
To correct for the effect of seeing, the {\sc Galactica} code
generates a Moffat profile star image \cite{moff69} of a given
full-width half-maximum (FWHM) expressed in terms of $\sigma_{\rm
PSF}={\rm PSF}_{\rm FWHM}/2.35$. This analytic profile is defined by 
\begin{equation} 
 {\rm PSF}(r)=\hbox{const}/[1+(r/\alpha)^2]^{\beta}
 \label{eqn:moff_psf}
\end{equation} 
and is thought to represent the overall PSF shape better than a pure
Gaussian which only approximates the core regions. Here $\alpha$
represents the width of the PSF and is related to the FWHM$=2\alpha
\sqrt{2^{1/\beta}-1}$ \cite{tru01}. $\beta$ governs how ``peaky'' the
PSF profile is (the larger $\beta$ is, the more Gaussian-like the
profile becomes). A value of $\beta=4.5$ is used throughout. The
$\alpha$ parameter can be fine-tuned to a particular dataset using
the average FWHM for the data. {\sc Galactica} lets $\sigma_{\rm
PSF}$ be a free fitting parameter to allow for any small changes in
the PSF between the position of the stars in the image and the galaxy position.\\ 

\noindent\emph{Minimization technique: Powell's method}\\
The code requires explicit initialization of the fitting
parameters. The initial value of the S/T ratio is always fixed at 0.5,
although starting with a value of S/T randomly distributed in the
range 0 to 1 has no effect on the recovered S/T distribution. The
starting position angles of the disk and spheroid components, their
characteristic radii and the disk inclination angle are calculated
directly from the image. The ranges over which parameters are allowed
to vary during the fitting are specified and fitting outside these
limits is not possible.

$\chi^{2}$ is minimized in a 12-parameter space; the S\'{e}rsic index
is set to $n=4$ and the FWHM of the PSF and the residual sky
background level are additional fitting parameters not included in
{\sc Gim2D}. Note that in fitting the total flux in the galaxy {\sc Galactica} we find a typical variation of only 5\% around the value estimated from SDSS photometry. The minimization routine is also rather different from
the one used by {\sc Gim2D}. In {\sc Gim2D} every parameter is varied
at each step according to the `temperature' of the fit. In {\sc
Galactica} one parameter is minimized at a time, i.e. all but one
parameters are `frozen' until a minimum for this parameter is found
and the process is repeated for the entire set of parameters until the
global minimum is found ---the essence of Powell's method (see
\pcite{numrec} for further details). This method is good for finding a
global extremum but is typically slower than the Metropolis algorithm
employed by {\sc Gim2D}.\\

\noindent\emph{{\sc Galactica} outputs and error estimation}\\
After convergence is achieved, the best-fit parameters are output
along with the best-fit model image and the residual map obtained by
subtracting the model galaxy from the real image. The value of
$\chi^2$ per degree of freedom, $\chi^2_\nu$, is then
calculated. Errors on the fitted parameters are obtained using a Monte
Carlo method: we create 30 realizations of the best fitting model for
each galaxy by adding random noise, drawn from a Poisson 
distribution, to the model image. The
distribution of the best-fitting parameters of the model realizations
is then used to estimate the uncertainty in the fit. This method
allows the uncertainties in the image parameters to be obtained
without any assumptions about their distribution.\\

\subsection{{\sc Gim2D} vs. {\sc Galactica} comparison}

\begin{figure}
  \begin{center}
\epsfig{file=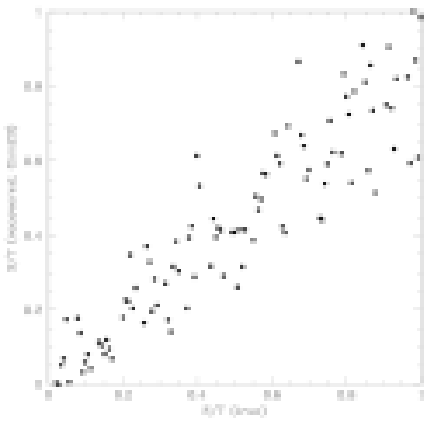,width=8.cm}
         \caption{Correlation between the input S/T ratios for 100 model
galaxies and the best-fit S/T ratios recovered using {\sc Gim2D}. The
scatter around the mean is \protect{$\sigma_{\rm rms}=0.10$}. Note a
small systematic underestimate of S/T, accompanied by increased
scatter, for large input values. The recovered characteristic radii of
these galaxies are the largest and the sky background recalculated by
{\sc Gim2D} for these galaxies is high. The resulting confusion
between the background and an extended surface brightness profile
accounts for this effect. }
\label{fig:Gim2D-mock} 
\end{center}
\end{figure}

\begin{figure}
  \begin{center}
\epsfig{file=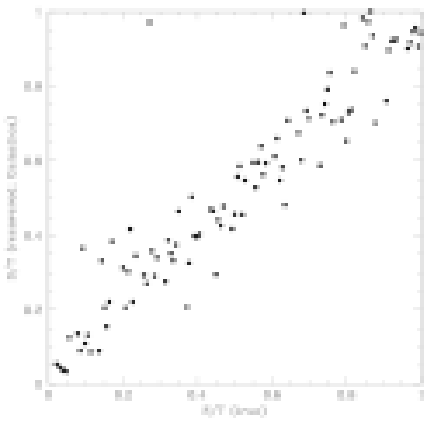,width=8.cm}
      \caption{Correlation between the input S/T ratios for 100 model
      galaxies and the best-fit S/T ratios recovered using {\sc
      Galactica}. The scatter around the mean is
      \protect{$\sigma_{\rm rms}=0.11$}.}
 \label{fig:fitg-mock}
      \end{center}
\end{figure}

\subsubsection{Introduction}

The previous section described two independent codes for estimating
basic galaxy structural parameters. Although these codes assume the
same analytic surface brightness profiles to fit the spheroid and disk
components, the differences in the number of fitting parameters and in
the minimization techniques are sufficiently large to make a
comparison interesting and important. \\

We first note the following points relevant to the comparison:
\begin{itemize}
 \item both codes assume a fixed value of $n=4$;
 \item the sky background is always kept fixed by {\sc Gim2D}
although the code is allowed to recompute and correct the
background level before the minimization procedure starts; 
 \item {\sc Galactica} always treats the sky background as a free parameter;
 \item the ellipticity is defined differently: {\sc Gim2D} fits $e=1-b/a$
while {\sc Galactica} fits $a/b$; 
 \item the seeing is fixed in {\sc Gim2D} but in {\sc Galactica} it
is allowed to fluctuate between $\pm5\%$ of the specified $\sigma_{\rm PSF}$; 
 \item the position angles in {\sc Gim2D} are defined with respect to
the $y$-axis of a Cartesian system while in {\sc Galactica} they are 
defined clockwise from the $x$-axis. (The position angles of the 
spheroid and disk are allowed to vary in both codes; a
large difference between these can be a signature of a barred
structure; \pcite{sim02}). 
\end{itemize}

To quantify the performance of the codes, a series of tests were
conducted as we now describe.

\subsubsection{Tests using model galaxies}\label{model_tests}

A useful in-built feature of both {\sc Galactica} and {\sc Gim2D} is
the ability to create model galaxies. The initial tests and code
comparisons described below were carried out on model galaxies
generated ``internally'' by {\sc Galactica}. \\

Model galaxies were constructed adopting parameter values chosen at
random between realistic limits (Table \ref{tab:limits}) and matching
the total counts measured in a typical real galaxy. Poisson noise was
added to the model galaxy after its image was convolved with an
analytic Moffat PSF corresponding to a typical value of the
seeing. This PSF is subsequently used as the {\sc Gim2D} PSF. Model
galaxies were analyzed with both codes using exactly the same
procedures as for real galaxies.

\begin{table}
 \begin{center}
  \caption{Parameter ranges used for constructing mock galaxies.}  
 \begin{tabular}[htbp]{lll} \hline
{\bf Parameter}   & {\bf Low limit} & {\bf High limit} \\\hline
S/T         & $0.0$ & $1.0$    \\
$r_{\rm e,d}$ (pixels)  & $1$ & $12$\\
$e_{\rm s}$       & $0.0$ & $0.8$\\
$i$ (degrees) & $0.0$ & $90.0$ \\
$\theta_{\rm s,d}$ (degrees) & $0.0$ & $180.0$ \\
FWHM  ('')      & $1.4$ &  $1.4$\\ \hline 
 \end{tabular}
\label{tab:limits}
 \end{center}
\end{table}    

\begin{figure*}
  \begin{center}
  \begin{tabular}{cc}  
  \epsfig{file=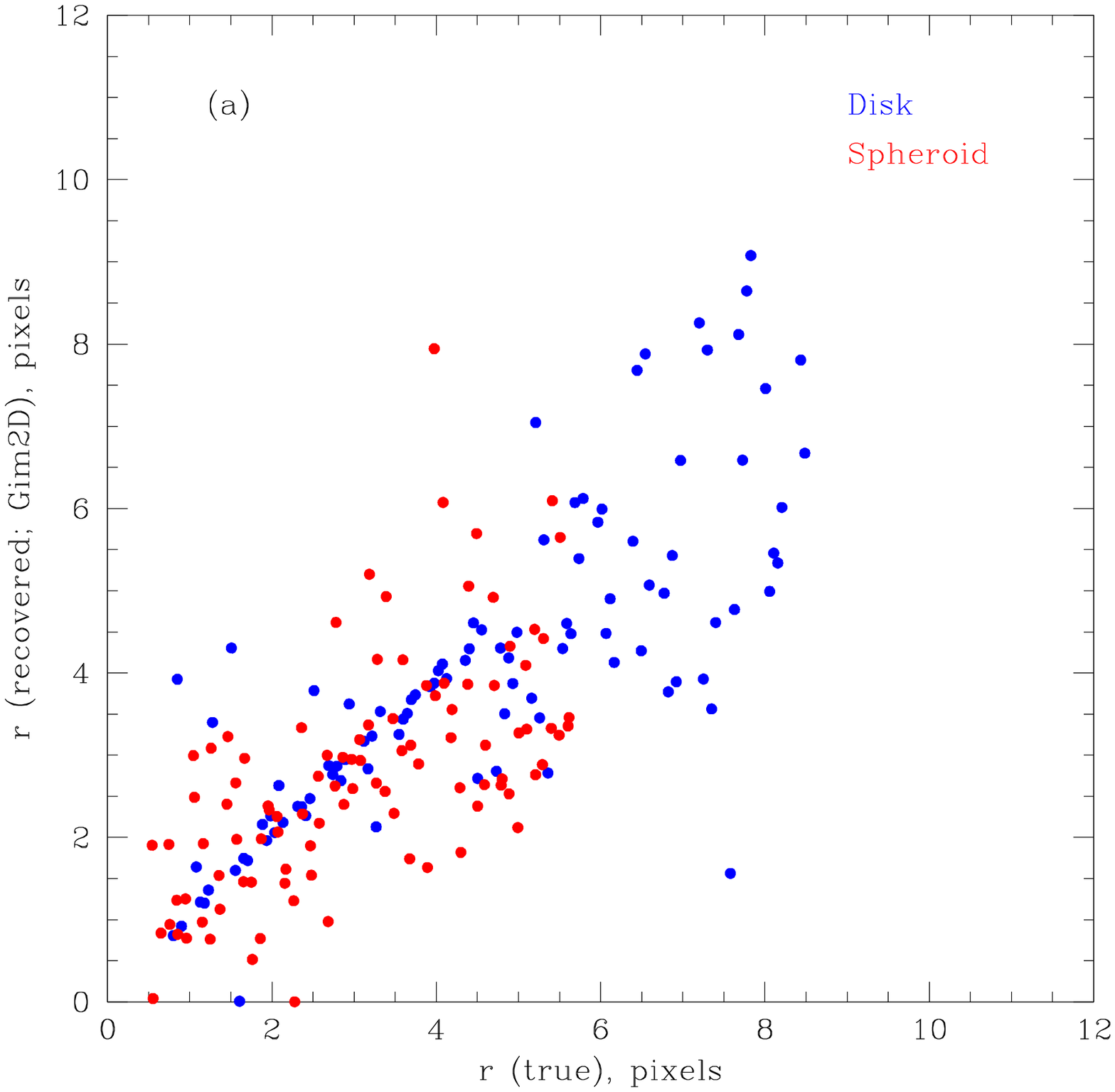,width=8.cm} & \epsfig{file=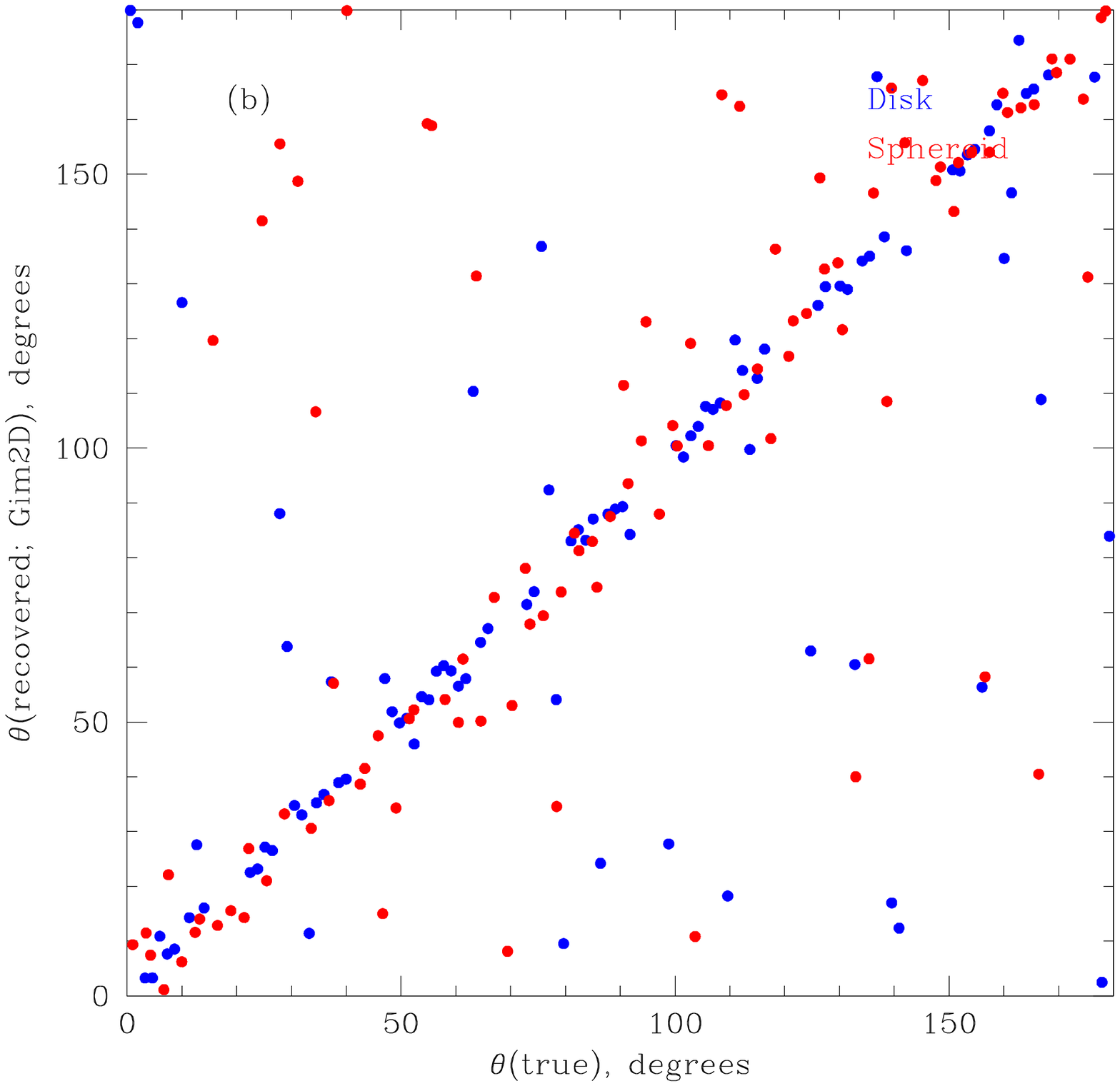,width=8.cm}\\
  \epsfig{file=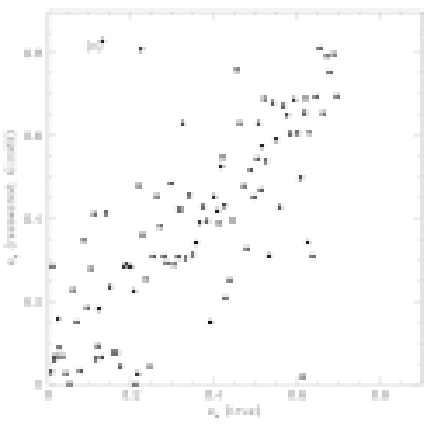,width=8.cm} & \epsfig{file=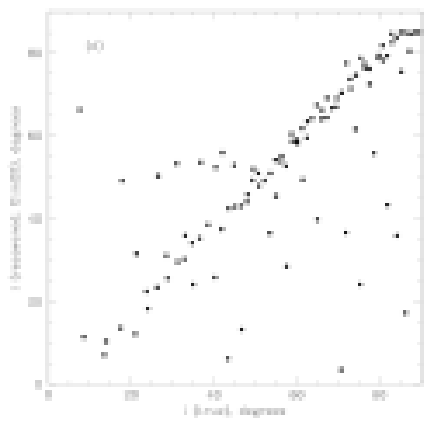,width=8.cm}
  \end{tabular}
        \caption{Correlations between input and recovered values of 
the characteristic radius, position angle, ellipticity and inclination for
$100$ model galaxies created using the {\sc Galactica} code and
decomposed using {\sc Gim2D}. a) Disk and spheroid radii. b) Disk and
spheroid position angles. c) Spheroid ellipticity. d) Disk
inclination. There is a saturation at $i=85^\circ$ which is
the upper limit that {\sc Gim2D} allows for the disk inclination. All
other parameters correlate well although significant scatter is
seen.}
\label{fig:Gim2D-mock1}
\end{center}
\end{figure*}

\begin{figure*}
  \begin{center}
  \begin{tabular}{cc}
  \epsfig{file=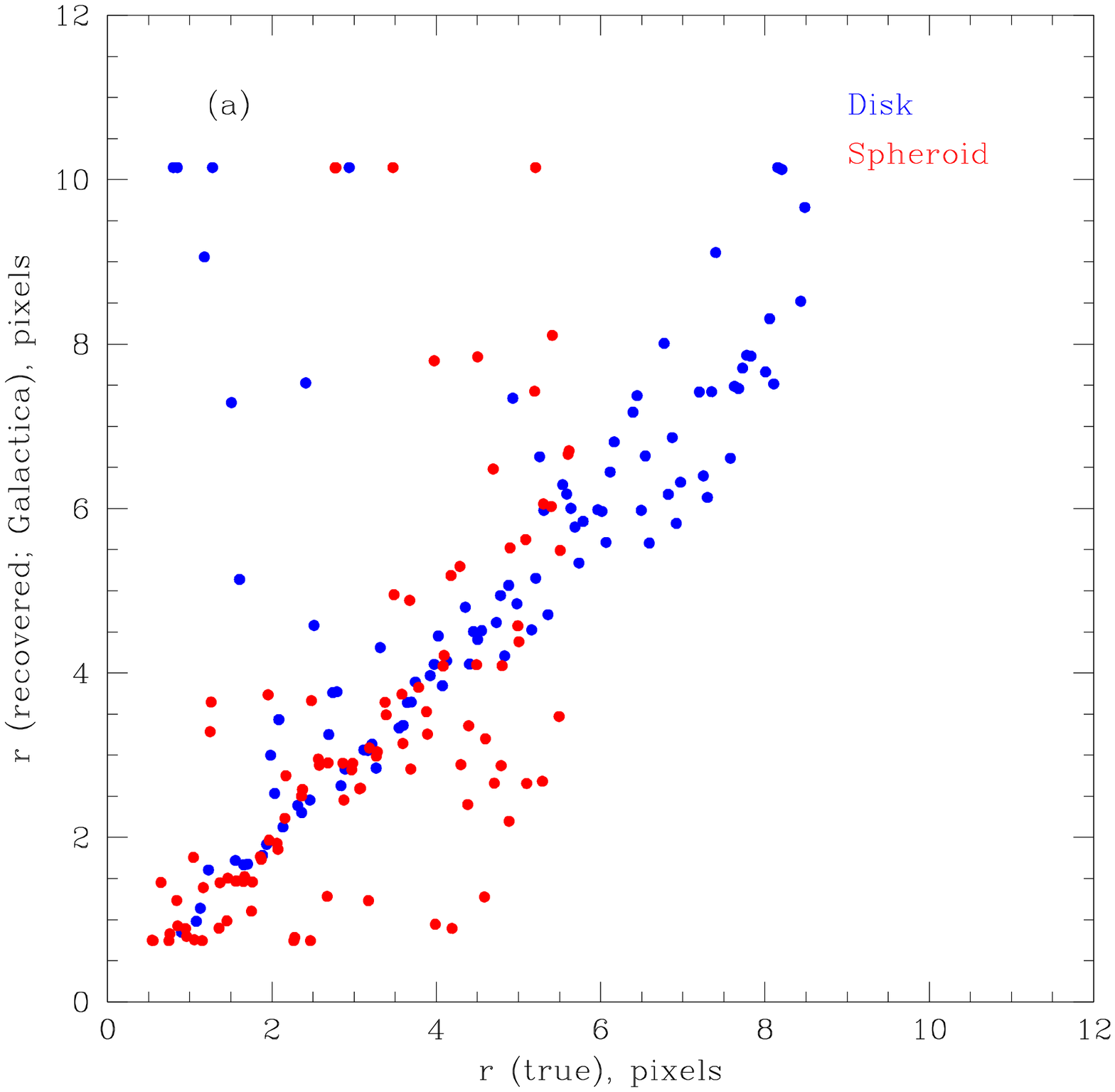,width=8.cm} & \epsfig{file=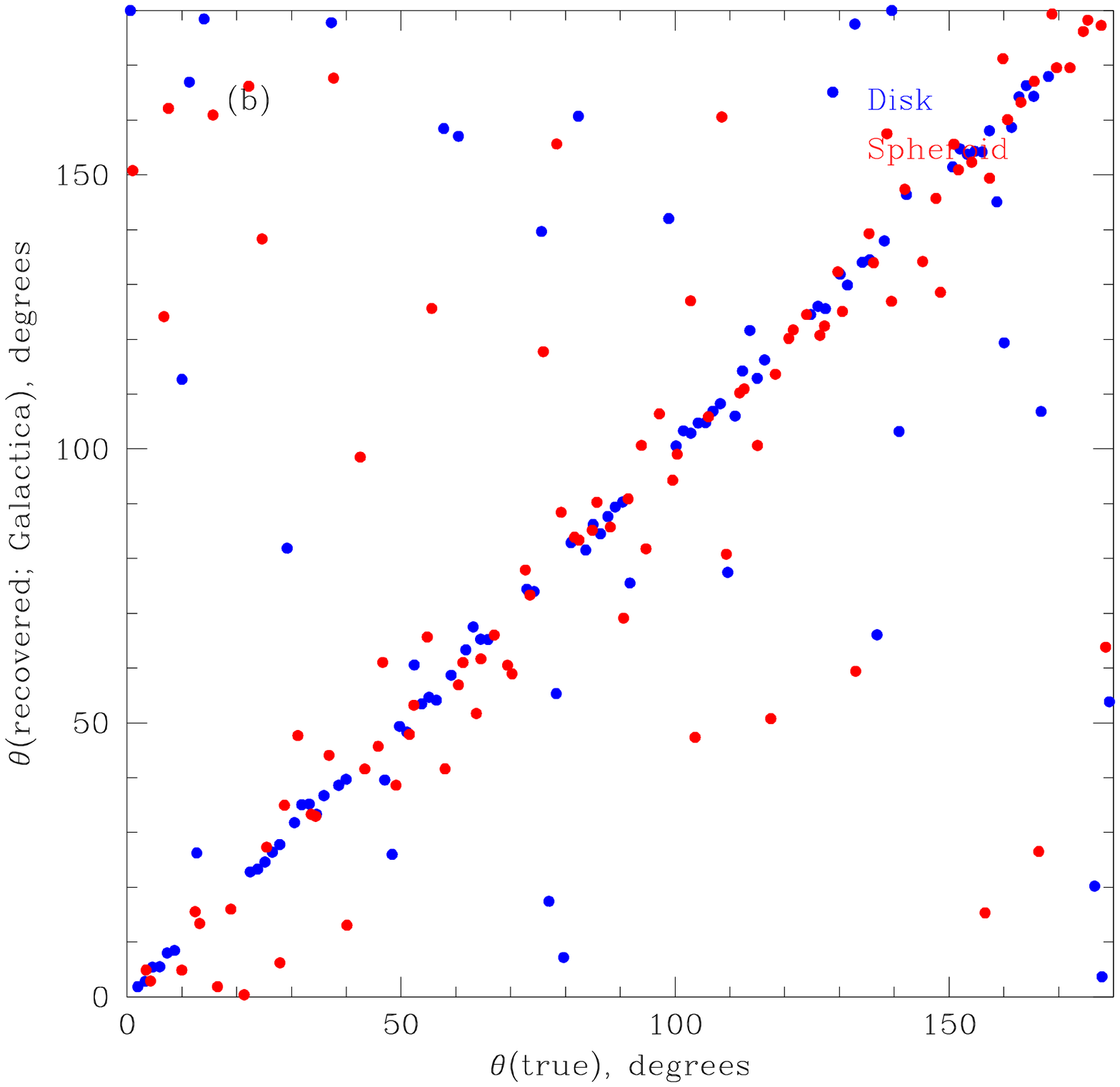,width=8.cm}\\
  \epsfig{file=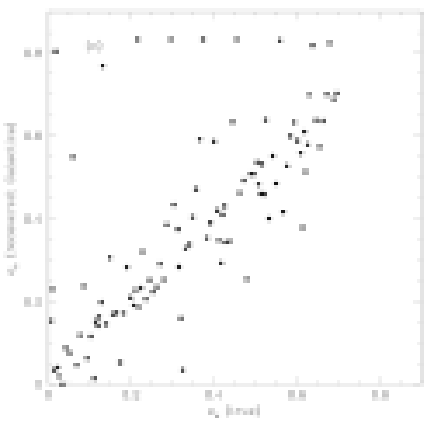,width=8.cm} & \epsfig{file=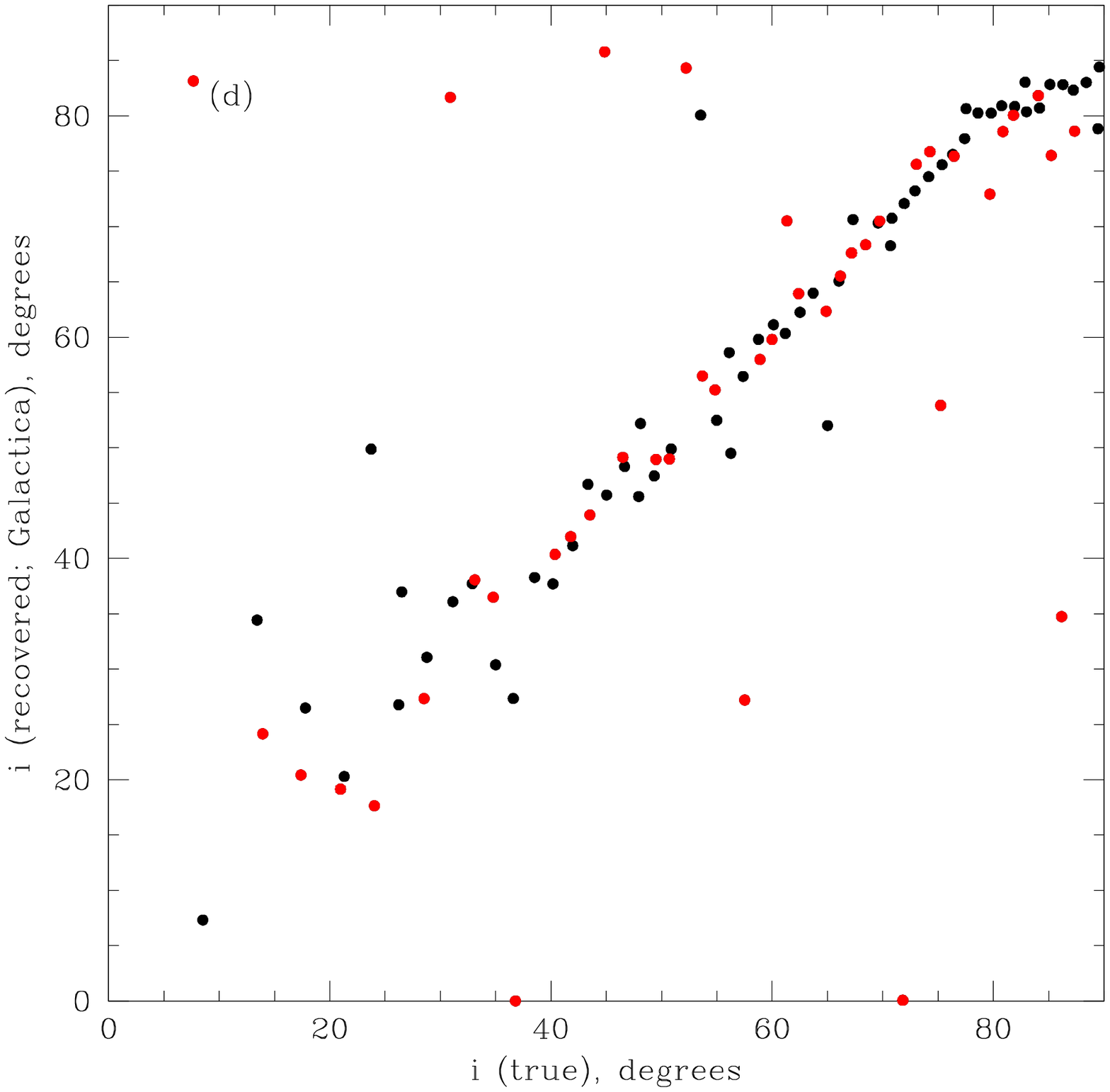,width=8.cm}
  \end{tabular}

\caption{Correlations between input and recovered values of 
the characteristic radius, position angle, ellipticity and inclination
for 100 model galaxies created and decomposed using the {\sc
Galactica} code. a) Disk and spheroid radii. Note that the
reconstruction seems to hit the upper limit on the characteristic
radii when the input radii are very small or very large. b) Disk and
spheroid position angles. c) Spheroid ellipticity. Note that a number
of very elliptical spheroids are found. These are mostly galaxies
which have very small spheroids. d) Disk inclination. Note an apparent
saturation at $i\sim85^\circ$. Even though the {\sc Galactica} code
allows the disk to be fully inclined ($i=90^\circ$), the
reconstruction avoids this upper limit.}
\label{fig:fitg-mock1}
\end{center}
\end{figure*}

Comparison of the known input S/T values and the values output by {\sc
Gim2D} and {\sc Galactica} for 100 model galaxies are shown in
Figs.~\ref{fig:Gim2D-mock} and~\ref{fig:fitg-mock} respectively. In
both cases the codes recover the input S/T very well. The scatter in
the recovered S/T ratios is $\sigma_{\rm rms} \sim 0.10$. The
remaining parameter recoveries are shown in
Figs.~\ref{fig:Gim2D-mock1} and~\ref{fig:fitg-mock1}.

The results shown in Figs.~\ref{fig:Gim2D-mock} to
\ref{fig:fitg-mock1} demonstrate that both {\sc Gim2D} and {\sc
Galactica} produce, on the whole, reliable spheroid/disk decompositions
for a set of artificial galaxies generated according to the model
assumed by the code. A more stringent test of the codes is to apply
them to model galaxies generated independently of the codes
themselves. Model galaxies were therefore externally created using the
{\sc iraf} task {\tt MKOBJ}. 

The model parameters were taken from Table~\ref{tab:limits}; a model
galaxy is created for given values of the size, orientation and
ellipticity (in this case defined as $b/a$) and the image is convolved
with a specified seeing. A useful feature of this approach is that a
real science frame can be fed into {\tt MKOBJ} and the model galaxy
added to a blank patch of the sky on this science frame. By matching,
on average, the counts in a real galaxy, this procedure ensures that
the artificial image has similar noise characteristics to the real
data.

To generate the model galaxies several SDSS r'-band frames were
extracted, each typically containing $\sim 5$ SDSS catalogued
galaxies. Each frame was taken from a different patch of the
sky. Counts associated with the SDSS galaxies were measured using the
{\sc SExtractor} {\tt flux\_best} parameter. In order to test the
decomposition algorithm over a realistic range of galaxy properties,
this procedure was applied to galaxies spanning a range of apparent
magnitude and apparent shape and size. The model galaxies were
inserted across the blank regions of the sky in the original
frames. The postage-stamps for these galaxies were extracted using
{\sc SExtractor} and the decomposition codes run treating the
extracted model galaxies just as the real ones.\\

\begin{figure}
  \begin{center}
  \epsfig{file=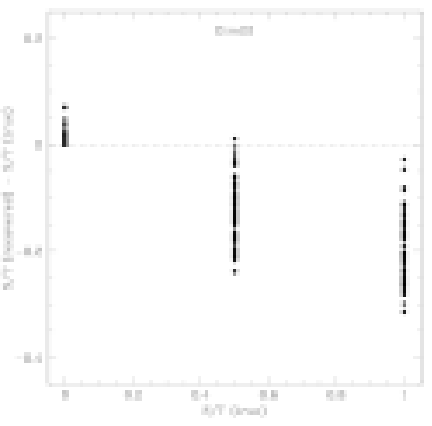,width=8.cm}
     \caption{\small The difference between input and recovered values of
S/T using {\sc Gim2D} for a set of $250$ model galaxies. The mean
offset in the recovered values are $\Delta$S/T$=0.02$ (for S/T$=0.0$),
$\Delta$S/T$=0.13$ (for S/T$=0.5$) and $\Delta$S/T$=0.20$ (for S/T$=1.0$) Model
galaxies span a range of apparent magnitudes, sizes and orientations.}
      \label{fig:Gim2D-MKOBJ}
  \end{center}
\end{figure}

The results of the {\sc Gim2D} decompositions of the model galaxies
are shown in Fig~\ref{fig:Gim2D-MKOBJ}. The agreement between the
input and output S/T ratios for the pure exponential disks
(S/T$=0$) is excellent. However, the recovered S/T ratio for S/T$=0.5$
is biased by $\Delta$S/T$=0.1$ and for S/T$=1.0$ it is biased by
$\Delta$S/T$=0.2$. The tendency is always to underestimate the amount
of spheroid or, equivalently, overestimate the amount of disk. {\sc
Gim2D} can be fine-tuned to recover the input S/T ratios with
$\Delta$S/T$\simeq 0.1$ across the full S/T range. For this, {\sc
Gim2D} requires that the size of the zone around the lowest {\sc
SExtractor} isophote used in the re-calibration of the sky background
should be set to $\sim 30$ pixels (the default value is $10$
pixels). This ensures that any faint galaxy flux does not contribute
to the re-calibrated background flux thus minimizing any bias in S/T.

\begin{figure}
  \begin{center}
  \epsfig{file=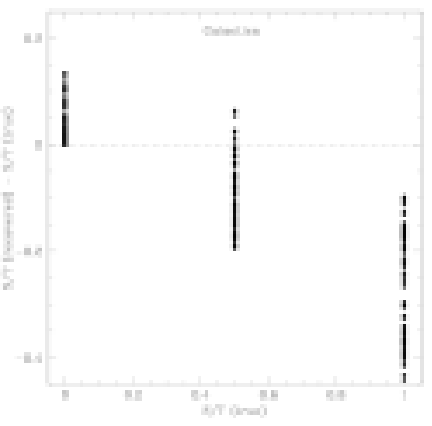,width=8.cm}
     \caption{\small The difference between input and recovered values of
S/T using {\sc Galactica} for a set of $250$ model galaxies. The mean
offset in the recovered values are $\Delta$S/T$=0.05$ (for S/T$=0.0$),
$\Delta$S/T$=0.11$ (for S/T$=0.5$) and $\Delta$S/T$=0.24$ (for S/T$=1.0$) Model
galaxies span a range of apparent magnitudes, sizes and
orientations. The median $\Delta$S/T is well fit by the relation
$\Delta$S/T$=0.02-0.26$S/T$_{\rm true}$.} 
      \label{fig:fitg-MKOBJ}
  \end{center}
\end{figure}

The results of the {\sc Galactica} decompositions of the model
galaxies are shown in Fig.~\ref{fig:fitg-MKOBJ}. For the pure
exponential disks the recovered S/T ratios are, again, very good. As
before, for larger S/T, this ratio is underestimated and peaks at
S/T$= 1.0$, implying that many pure $r^{1/4}$ galaxies have acquired a
fictitious disk component. There appears to be no correlation between
the size of the underestimation and the magnitude or scale radius of
the input galaxy but a weak correlation with the minor/major axis
ratios: the S/T deviation is largest for the most elliptical
profiles. The most prominent correlation, however, is between the
recovered S/T ratio and the sky background. As discussed earlier, the
{\sc Galactica} code allows the sky background to fluctuate a little
to allow for uncertainties in the estimated background. The fact that
the deviation between the input and the output S/T ratios is largest
when the `fitted' background is smallest implies that an extraneous
disk component is found where, in fact, the extra counts are due to
the sky background. Since there is no sharp cut-off in either the
spheroid or the disk, at the galaxy edges the galaxy surface
brightness profile and the sky background are indistinguishable.

{\sc Gim2D} performs marginally better than {\sc Galactica} in the
recovery of the S/T ratios. However, both codes show similar biases in
the decompositions. This is despite the fact that we allow the background and PSF to be fit by {\sc Galactica} but not by {\sc Gim2D} indicating that this choice does not significantly bias our results. The relative performance of these two codes on a set of real
galaxies is compared next. This allows a more realistic comparison of
the codes but, of course, there is no \emph{a priori} correct answer.

\subsubsection{Tests using real galaxies}

\begin{figure}
  \begin{center}
  \epsfig{file=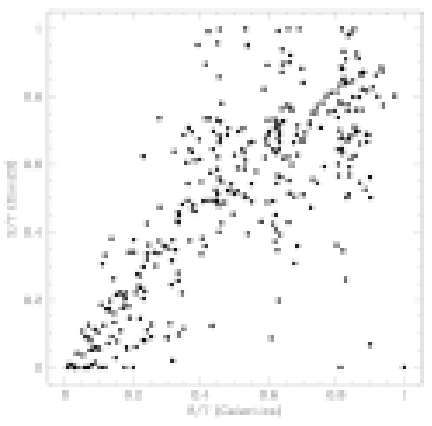,width=8.cm}
      \caption{The correlation between the S/T ratios of a sample of 350
SDSS galaxies inferred using {\sc Gim2D} and {\sc Galactica}. The
Spearman rank correlation coefficient of $0.74$ indicates a
significant correlation. Note that almost no pure disks
(i.e. S/T$=0.0$) are detected by {\sc Galactica} but a few are
detected by {\sc Gim2D}. Most of these galaxies have spheroid
characteristic radii of less than $2$ pixels (as does the single {\sc
Galactica} detection at S/T$=1.0$). {\sc Gim2D} finds larger
characteristic radii for these galaxies. The overall scatter is
$\sigma_{\rm rms}=0.19$, but there is a marked increase in the scatter for
larger S/T ratios.}
\label{fig:gim-fitg-sdss} 
\end{center}
\end{figure}

\begin{figure*}
  \begin{center}
  \begin{tabular}{cc}
  \epsfig{file=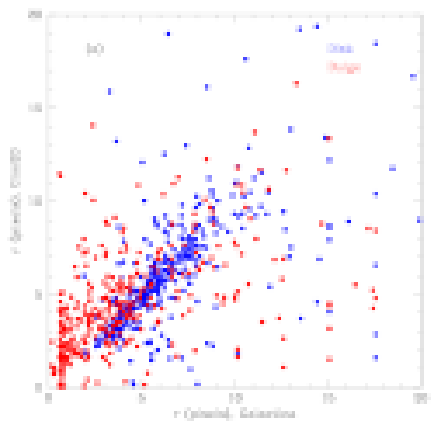,width=8.cm} & \epsfig{file=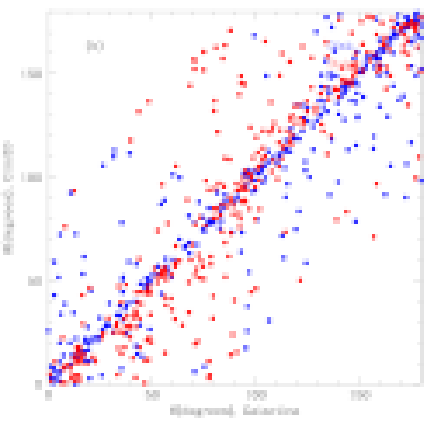,width=8.cm}\\
\epsfig{file=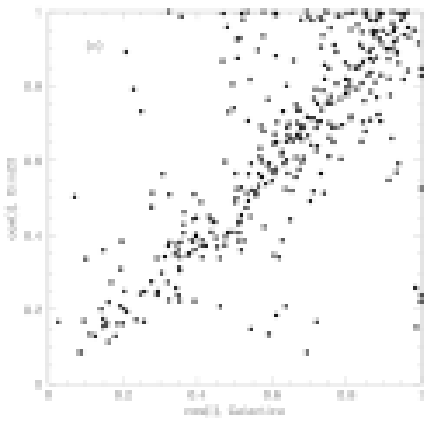,width=8.cm} & \epsfig{file=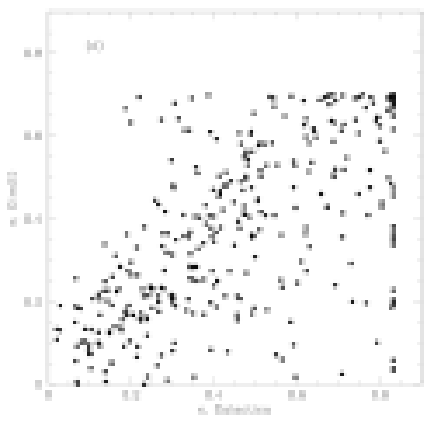,width=8.cm}
      \end{tabular}
    \caption{Correlation between various parameters 
inferred using {\sc Gim2D} and {\sc Galactica} for a sample of 350
SDSS galaxies. No systematic differences in the recovered parameters
are apparent although the scatter can be quite large.}
      \label{fig:gim-fitg-sdss1}
  \end{center}
\end{figure*}

To ensure a uniform sampling of the [S/T, apparent magnitude] space,
the comparison was carried out using a subsample of SDSS galaxies
selected in bins of 0.5 in apparent magnitude and 0.2 in S/T ratio (as
determined by {\sc Galactica}).  Unsaturated stellar images with high
S/N were extracted from the SDSS galaxy frames and used in the {\sc
Gim2D} PSF deconvolution. The {\sc Galactica} Moffat PSF was
fine-tuned to fit the SDSS data well. Fig.~\ref{fig:gim-fitg-sdss}
demonstrates a significant correlation (Spearman rank correlation
coefficient of $0.74$) between the S/T ratios for $\sim 350$ SDSS
galaxies inferred using {\sc Gim2D} and {\sc Galactica}. There are no
systematic differences between the results of the two codes. The
correlations between other parameters are displayed in
Fig.~\ref{fig:gim-fitg-sdss1}.\\

\section{{\sc Galactica} methodology}\label{sec:tech}

\begin{figure}
 \begin{center}
\begin{tabular}{c}
\psfig{file=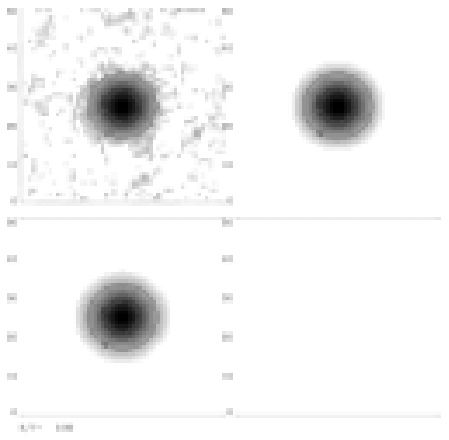,width=8.cm,bbllx=20mm,bblly=142mm,bburx=220mm,bbury=330mm,clip=} \\
\psfig{file=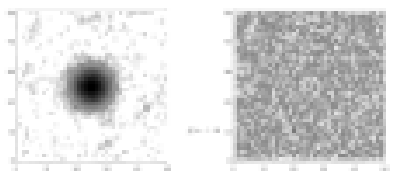,width=8.cm}\\
S/T$=0.00$
\end{tabular}
\caption{Fit to a pure exponential model galaxy after the original
$101\times101$ model image was binned $2\times2$. The top row shows
the real (left) and model (right) images. The middle row shows the
disk (left) and spheroidal (right) components. The bottom row shows
the real (left) and residual (right) images.  The recovered S/T$=0$,
corresponding to a pure exponential. The good fit is evident from both
the $\chi^2_\nu \sim 1$ and the noise-dominated residual image.}
\label{fig:sdss_bin_exp} 
\end{center}
\end{figure}     

\begin{figure}
 \begin{center}
\begin{tabular}{c}
\psfig{file=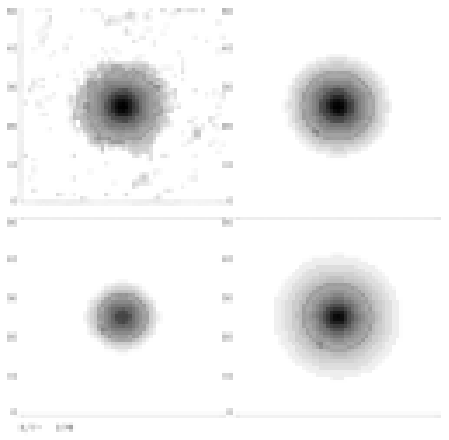,width=8.cm,bbllx=20mm,bblly=142mm,bburx=220mm,bbury=330mm,clip=} \\
\psfig{file=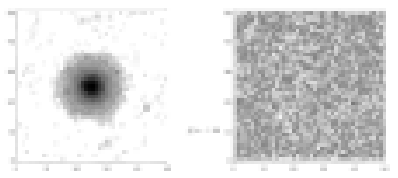,width=8.cm}
\end{tabular}
S/T$=0.79$ \caption{Fit to a pure $r^{1/4}$ model galaxy after the
original $101\times101$ model image was binned $2\times2$. The top row
shows the real (left) and model (right) images. The middle row shows
the disk (left) and spheroidal (right) components. The bottom row
shows the real (left) and residual (right) images.  The recovered
$S/T=0.8$, showing once more that the {\sc Galactica} code returns a
biased estimate of S/T. The good fit is inferred from both $\chi^2_\nu
\sim 1$ and the noise-dominated residual image.}
\label{fig:sdss_bin_dev} 
\end{center}
\end{figure}     

\subsection{Code speed limitations and galaxy binning}
\label{sec:bins}

The SDSS galaxy sample contains less than $200$ galaxies whose
postage-stamp size exceeds $91$ pixels on a side. To reduce the
processing time, these large postage-stamps were binned
$2\times2$. This has the added advantage that large nearby bright
galaxies are sampled at a resolution comparable to that of more
distant objects. To make sure that the binning procedure does not bias
the recovery of the galaxy S/T ratios, we carried out a series of
tests.

Several model galaxies were created using our standard procedure (see
Appendix~\ref{model_tests}) and binned using the {\sc iraf} task {\tt
BLKAVG}. {\sc Galactica} was then used to perform the fitting ensuring
that the pixel for the binned image is set to $2\times$ the normal
pixel size ($2\times0.396''$) and that the noise properties in this
`super-pixel' are changed accordingly. Figs.~~\ref{fig:sdss_bin_exp}
and~\ref{fig:sdss_bin_dev} show fits to model galaxies consisting
either of a pure exponential disk or a pure $r^{1/4}$ spheroid after
the original model images were binned by $2\times2$. In the case of
the pure exponential disk, the fit to the model is perfect. The fit to the
pure $r^{1/4}$ galaxy, however, shows a similar bias to that seen for
the unbinned data (Appendix~\ref{Two}), as indicated by the low value of
the recovered S/T$=0.8$. This shows that the binning in itself is not
responsible for the observed S/T bias. The $2\times2$ binning was
therefore applied to all the SDSS galaxies whose postage-stamps are
greater than $91\times91$ pixels.

\begin{figure}
 \begin{center}
\begin{tabular}{c}
\psfig{file=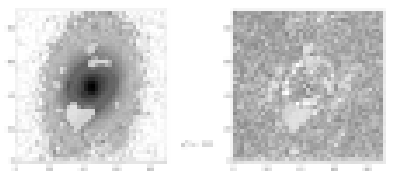,width=8.cm} \\
\psfig{file=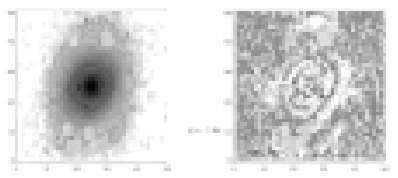,width=8.cm}
\end{tabular}
    \caption{Fits to a galaxy that exhibits internal structure. The top
images show the unbinned galaxy postage-stamp (left) and the
corresponding residual image (right). The bottom images show the galaxy
and the residual after the galaxy is binned $2\times2$. In both cases
the $\chi^2_\nu$ is poor ($\chi^2_\nu>2.0$) and the residuals are not
noise-dominated. This supports the conclusion that galaxies with
significant structure give poor fits irrespective of whether
they are binned or not.} 
   \label{fig:sdss_bin_real1}
 \end{center}
\end{figure}

\begin{figure}
 \begin{center}
\begin{tabular}{c}
\psfig{file=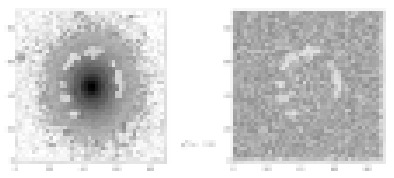,width=8.cm} \\
\psfig{file=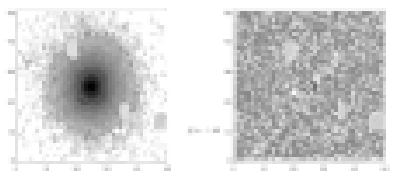,width=8.cm}
\end{tabular}
   \caption{Fits to a galaxy that does not exhibit internal
structure. The top images show the unbinned galaxy postage-stamp
(left) and the corresponding residual image (right). The bottom images
show the galaxy and the residual after the galaxy is binned
$2\times2$. In both cases, the $\chi^2_\nu$ is good ($\chi^2_\nu<2.0$)
and the residuals are noise-dominated. This supports the conclusion
that galaxies without significant structure result in acceptable fits
irrespective of whether they are binned or not.}
   \label{fig:sdss_bin_real2}
 \end{center}
\end{figure}

The binning works very well if the binned galaxy does not exhibit much
internal structure (as in the model galaxies). However, for a galaxy
which exhibits significant internal structure, a fit with
$\chi^2_\nu>2.0$ is more typical. Whether decomposing such galaxies
even without binning would lead to a good fit is unclear as
demonstrated for 2 SDSS galaxies in Fig.~\ref{fig:sdss_bin_real1} and
\ref{fig:sdss_bin_real2}. The top images in both figures show the
postage stamp and the residual map for the unbinned galaxy which has
size $101\times101$ pixels. (To speed up the calculation, the postage
stamp was trimmed by $5$ pixels on either side.) The bottom panels
show the corresponding images for the binned versions of the same
galaxy.

The galaxy in Fig.~\ref{fig:sdss_bin_real1} exhibits much more
internal structure than the galaxy in Fig.~\ref{fig:sdss_bin_real2}, as
is clear from both the value of $\chi^2_\nu$ and the residual
image. This supports the conclusion that galaxies which exhibit
internal structure are poorly fit irrespective of whether they are
binned or not. The recovered S/T ratios for the unbinned and binned
data, although different, are consistent with the typical errors in
the S/T ratio. We conclude that the S/T distribution of the final SDSS
galaxy sample is not biased by binning these large, bright nearby
objects (most of which contribute to the faint-end of the luminosity
function -- see Section \ref{sec:lfs}).

\subsection{SDSS data and the goodness-of-fit}\label{app:goodfit}

\begin{figure*}
  \begin{center}
\begin{tabular}{cc}
\psfig{file=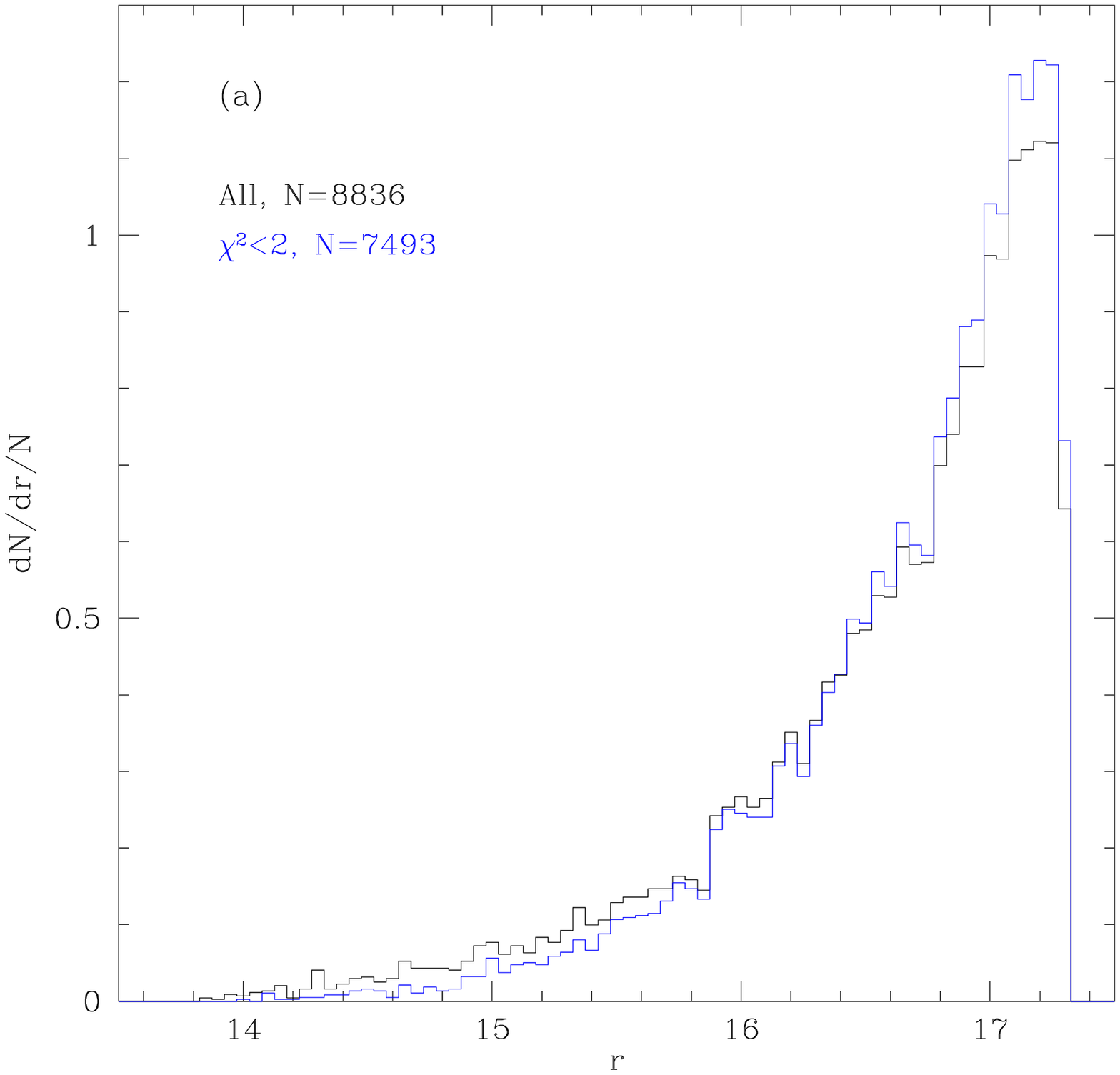,width=8.cm} & \psfig{file=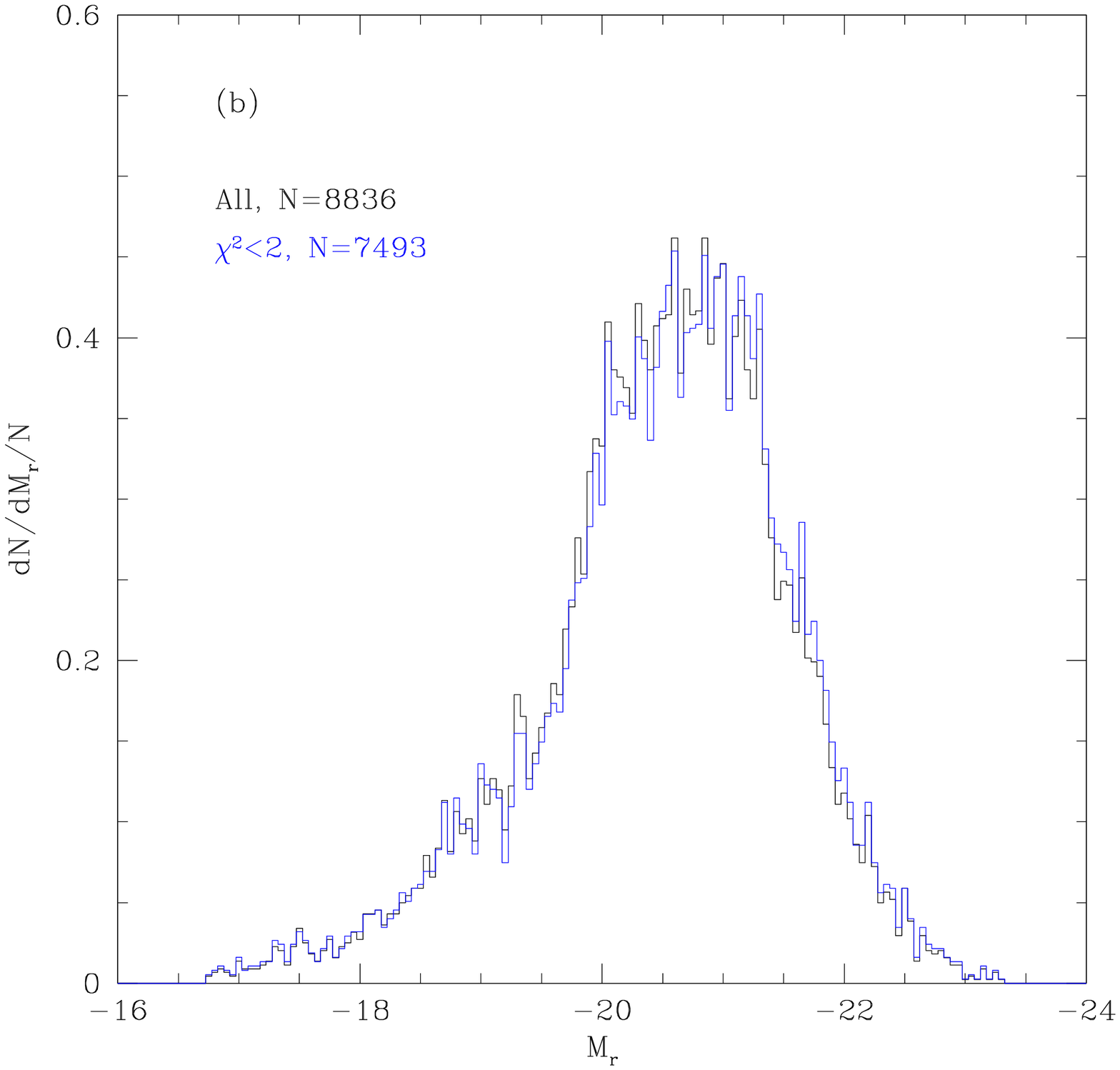,width=8.cm} \\
\psfig{file=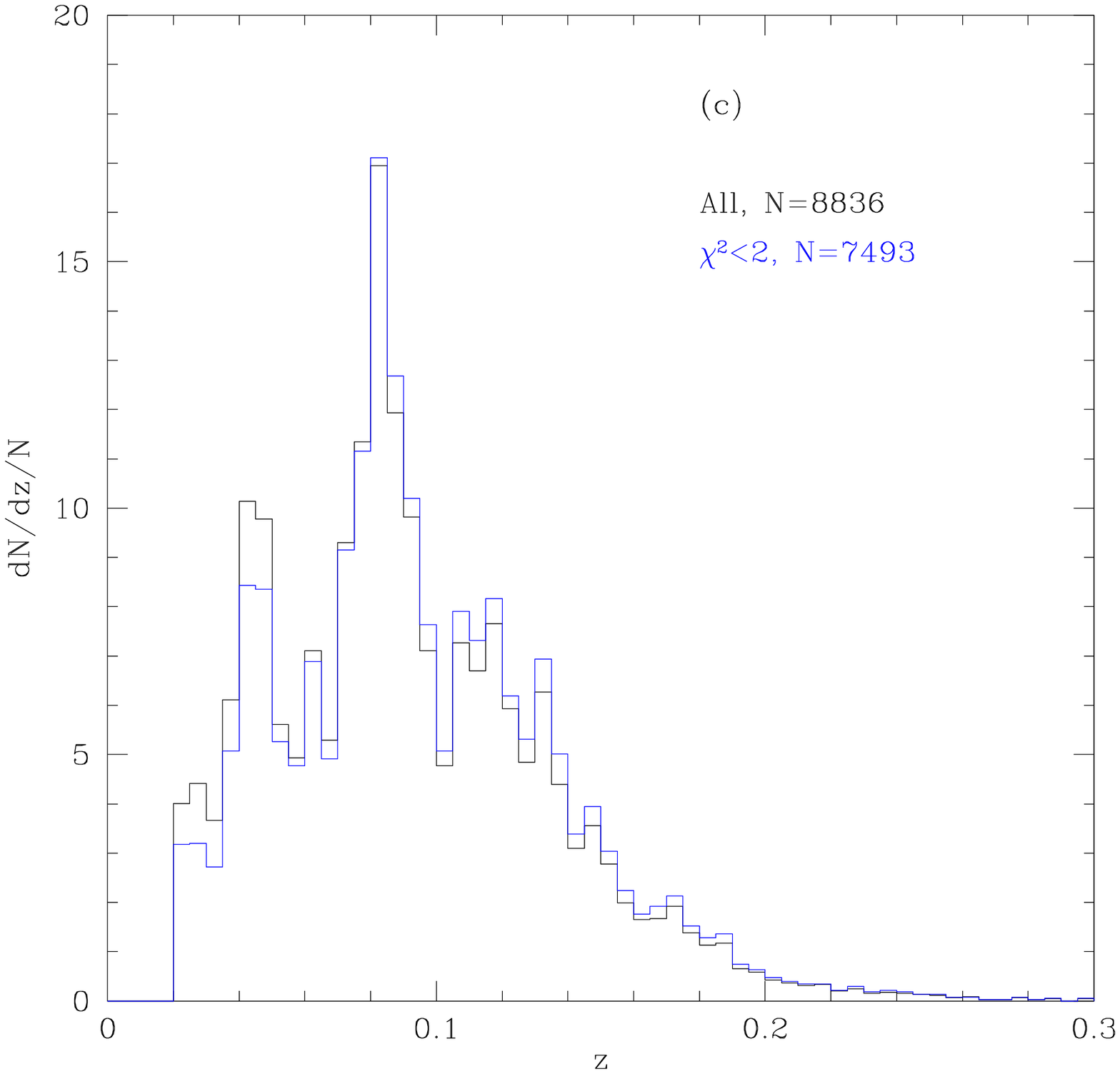,width=8.cm} & \psfig{file=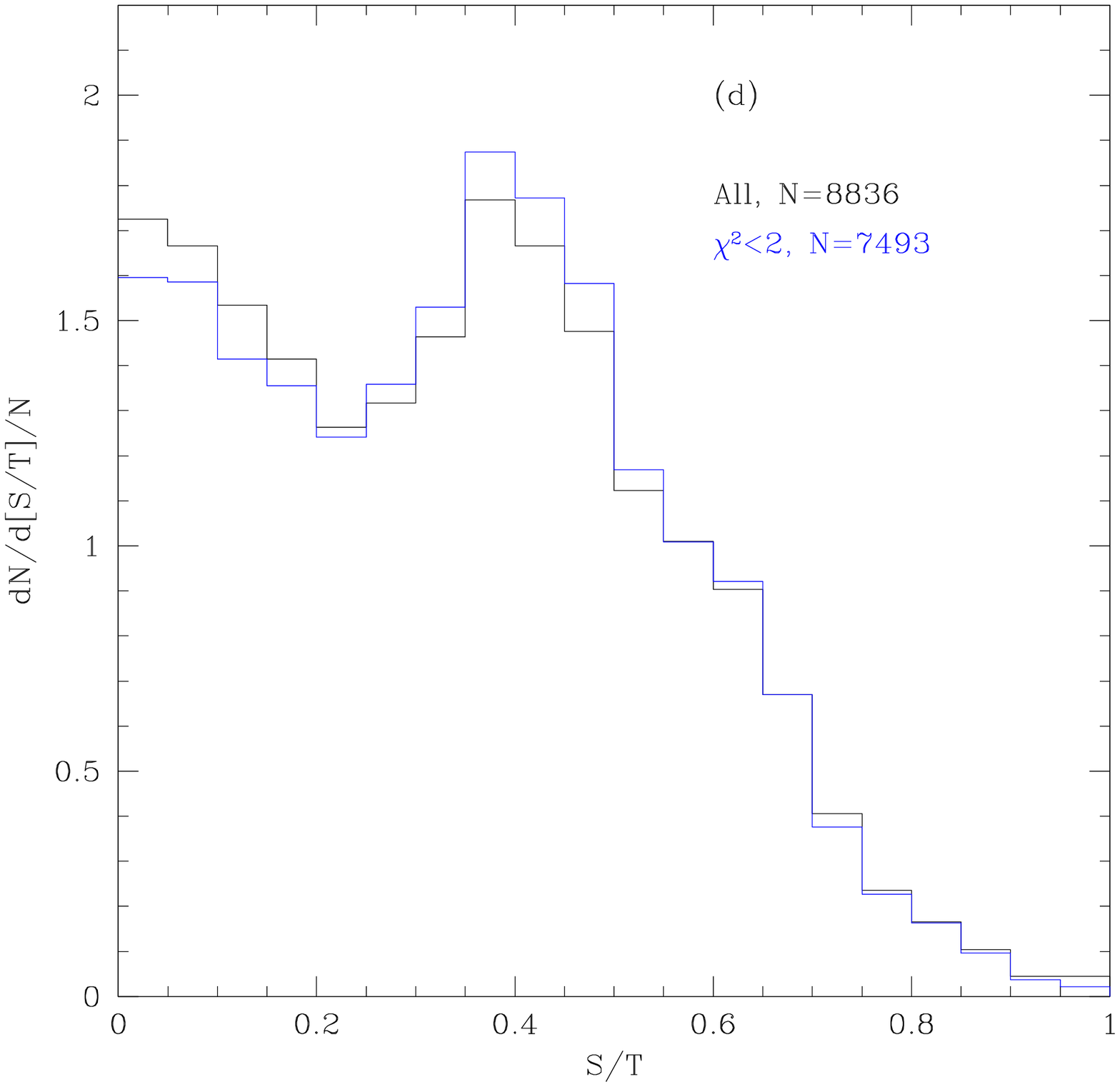,width=8.cm}
\end{tabular}
    \caption{Distributions of (a) apparent magnitude, (b) absolute
magnitude, (c) redshift and (d) recovered S/T ratios. Results for the total
sample are shown by the black histograms and for the sample with 
$\chi^{2}_{\nu}<2.0$ by the blue histograms. The two distributions are
similar, indicating that excluding poorly fit galaxies does not
introduce any obvious biases in the sample.}
    \label{fig:sdss_distns}
  \end{center}
\end{figure*}

The selected SDSS sample of $8839$ galaxies is too large for each of
the residual images to be inspected by eye to ensure a satisfactory
decomposition as suggested by the $\chi^{2}_{\nu}<2.0$. However, a
randomly selected sample of residuals was examined by eye to ensure
that they were indeed predominately noise dominated. The
$\chi^{2}_{\nu}<2.0$ criterion was therefore adopted to define a `well
fit' dataset of $7493$ galaxies. To test for any selection biases were
introduced by the rejection of galaxies with $\chi^2_\nu>2.0$, we
compare the distributions of some basic properties of these galaxies
and of the well-fit subset in Fig.~\ref{fig:sdss_distns}. There are
small but noticeable biases introduced in the distributions of
apparent magnitudes and redshifts. These biases are taken into account
when estimating galaxy luminosity functions (see \S\ref{sec:lfest})
The figure also shows a deficit of objects with $S/T>0.7$. This is
most likely due to the bias in the {\sc Galactica} code discussed in
Appendix~\ref{Two}. The significance of this bias and its influence on
the final results is discussed in the main body of the paper.
 
\begin{figure}
 \psfig{file=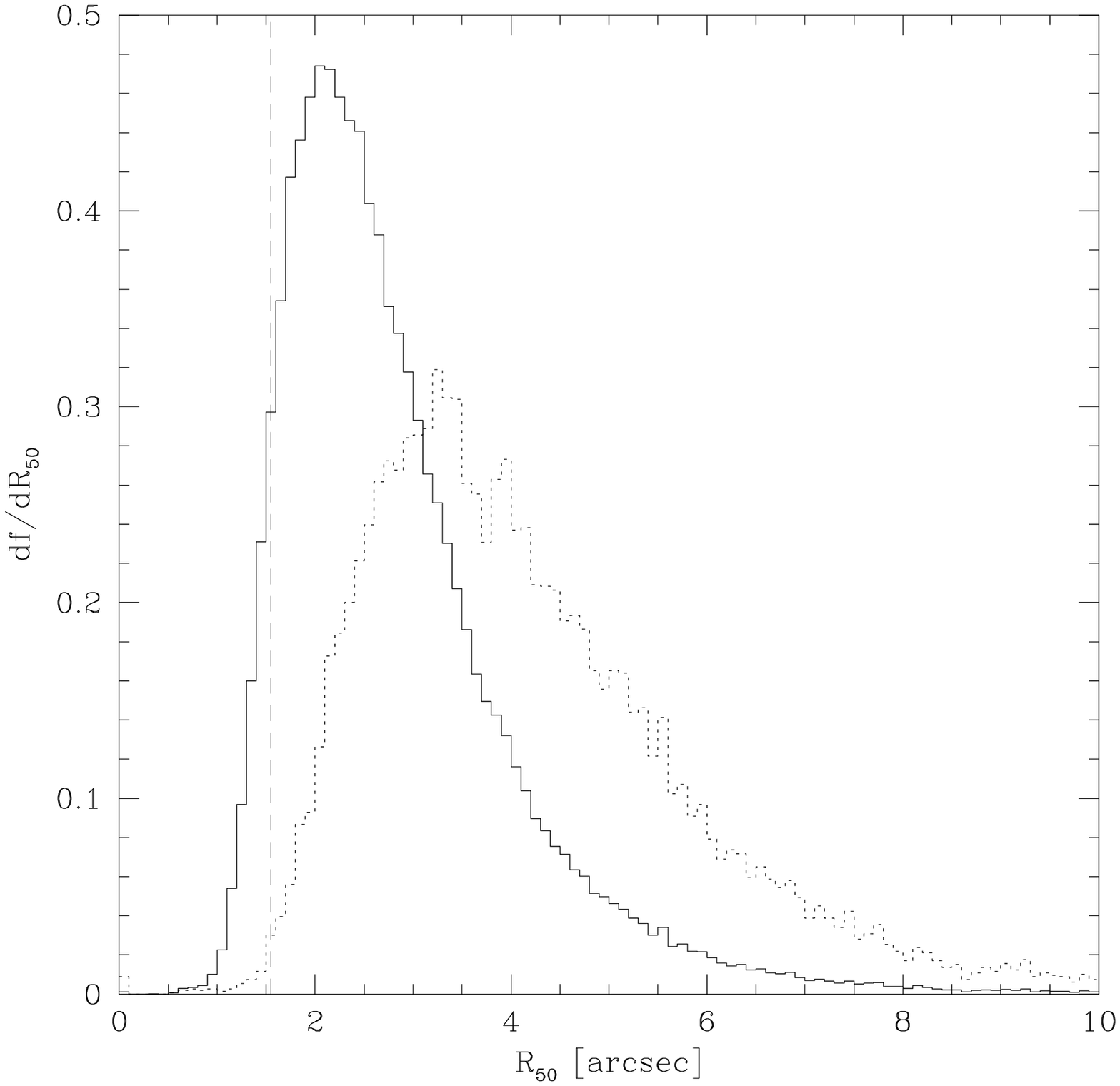,width=80mm}
\caption{The distribution of $R_{50}$ (the radius enclosing 50\% of the Petrosian flux) for galaxies in the SDSS which match our selection criteria (solid histogram). Results are also shown for the sample used by \protect\scite{tw} (dotted histogram). For comparison, we indicate the maximum allowable seeing for our sample by the vertical dashed line.}
\label{fig:r50}
\end{figure}

Finally, in Fig.~\ref{fig:r50} we plot the distribution of $R_{50}$ (the radius enclosing 50\% of the Petrosian flux) for galaxies meeting the selection criteria of this work and that of \scite{tw}. Our galaxies are typically 40\% smaller than those of \scite{tw}.

\subsection{S/T error estimates}

Benson et al. (2002) developed a Monte Carlo approach to estimate the
errors on the fitted parameters in {\sc Galactica}. This method is
very time consuming, requiring several CPU days for a typical
galaxy. A full Monte Carlo analysis is therefore impractical for a
large dataset such as the one in this paper. Instead, to obtain
representative error estimates, we split the sample into bins of
apparent magnitude of width $0.5$, and selected from each of these
bins $5$ galaxies from each of three further bins in S/T
($0.0<$S/T$<0.3$, $0.3<$S/T$<0.6$, $0.6<$S/T$<1.0$). The full Monte
Carlo analysis was performed on the selected subsample and the medians
of the derived errors taken to be representative for galaxies in each
$[r_{\rm mag}, S/T]$ bin.

\section{Spheroid-to-disk ratios and galaxy morphologies}\label{sec:bt_morph}

\subsection{Morphological classification using colour}

It has long been known that galaxy colour is a useful indicator of
whether a galaxy is elliptical (old, red) or spiral (young, blue)
\cite{dev61} since the dominant stellar populations are reflected in
the galaxy colours. Investigating the colour-magnitude and
colour-colour diagrams, \scite{stra01} have shown that the ($u-r$)
colour distribution of SDSS galaxies has two maxima which are
separated by a well-defined minimum at $(u-r)=2.2$ and that $98\%$ of
galaxies spectroscopically classified as `early' types have
$(u-r)>2.2$ whilst $73\%$ of spectroscopically classified `late' types
have $(u-r)<2.2$. \scite{stra01} have also shown that this separator
also applies for a subsample of visually classified morphological
types where $80\%$ of galaxies visually classified as E, S0 or Sa have
colours redder than $(u-r)=2.2$ and $66\%$ of galaxies visually
classified as Sb, Sc and Irr have colours bluer than $(u-r)=2.2$. The
$(u-r)$ separator has already been used to study morphological
properties of galaxies in the SDSS sample as a function of environment
\cite{goto02,balogh04b}.

\begin{figure}
  \begin{center}
    \psfig{file=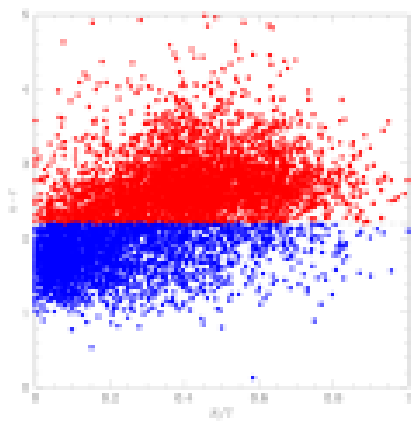,width=8.cm}
\caption{{\sc Galactica} S/T ratio vs. ($u-r$) colour for the
sample of $7493$ SDSS galaxies studied here. Most galaxies
colour-classified as late types are predominately disk-dominated
systems while most galaxies colour-classified as early type have
$S/T>0.4$.}
	
\label{fig:bt_col}
  \end{center}
\end{figure}

Our derived S/T ratios are plotted against u$-$r colour in
Fig.~\ref{fig:bt_col}. The B/T distributions of red galaxies and blue
galaxies are significantly displaced relative to each other: most
galaxies classified by colour as early types ($u-r>2.2$) have
$S/T>0.4$ whereas most galaxies classified by colour as late types
($u-r<2.2$) have $S/T<0.2$.  However, the red galaxies in particular
span a large range in S/T ratio. This suggests that blue star-forming
galaxies are predominantly disk-dominated but that disk-dominated
galaxies include both star-forming (blue) and passive (red)
galaxies\footnote{The red disk population may also include galaxies
with heavily obscured star formation.}

\subsection{Morphological classification using concentration index}

Galaxies can also be classified according to how `peaky' their light
distribution is by using the concentration index \cite{abr94}. The
surface brightness distribution of ellipticals and S0s is considerably
more centrally concentrated than that of spirals and
irregulars. \scite{shima01} defined the (inverse) concentration index
for SDSS galaxies as the ratio of the half- to the $90\%$ light radii
and define an optimum division between late and early types to be at
$C=0.33$ (with $15-20\%$ contamination from opposite types). This
separator has also been used to investigate the morphological
properties of SDSS galaxies \cite{goto02,naka03}.

\begin{figure}
 \begin{center}
\psfig{file=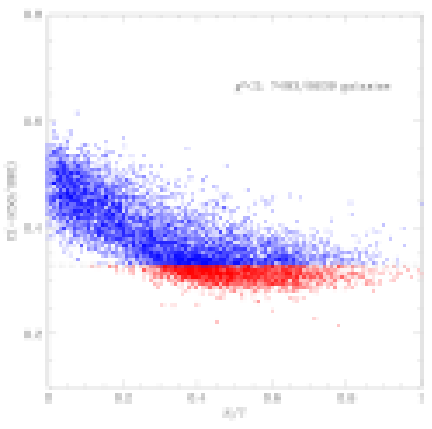,width=8.cm}
  \caption{{\sc Galactica} S/T ratio vs. (inverse) concentration index,
$C=R_{50}/R_{90}$, for our sample of $7493$ SDSS galaxies.  More
centrally concentrated galaxies ($C<0.33$) predominately have higher
S/T ratios.}
   \label{fig:bt_conc}
 \end{center}
\end{figure}

Fig.~\ref{fig:bt_conc} shows that there is a good correlation between
the S/T ratios and the (inverse) concentration index, $C$, for
galaxies with small $S/T \lsim 0.3$. Although spheroid-dominated
galaxies tend to be more centrally concentrated than disk-dominated
galaxies, there is little correlation of concentration with $S/T$ for
$S/T\gsim 0.3$. There is, however, considerable scatter in $C$ for a
given $S/T$.

\subsection{Morphological classification: S/T vs eye-morphology}

\begin{figure}
 \begin{center}
\psfig{file=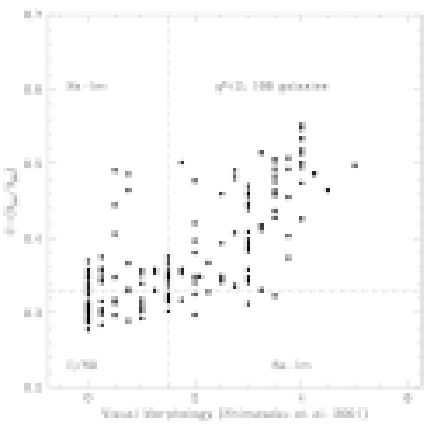,width=8.cm}
   \caption{Concentration index vs. eye morphology for $166$ galaxies in
common between our sample and that of \protect \scite{shima01}.}
  \label{fig:conc_morph}
 \end{center}
\end{figure}

\begin{figure}
 \begin{center}
\psfig{file=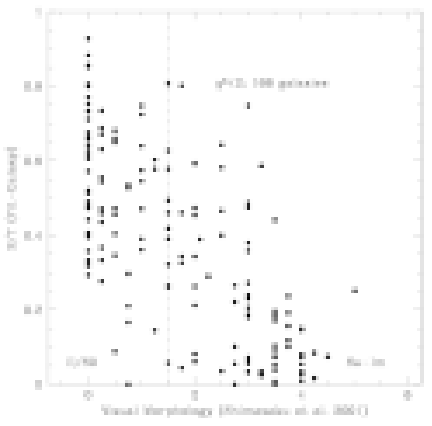,width=8.cm}
  \caption{{\sc Galactica} S/T ratio vs. the visual morphology
determined by \protect \scite{shima01} for the $166$ galaxies in
common in the two sample. There is a general trend for the S/T ratio
to increase along the S-S0-E morphological sequence but the scatter is
large.}
  \label{fig:bt_morph}
 \end{center}
\end{figure}

\scite{shima01} used a sample of $456$ bright SDSS galaxies
($g'<16.0$) visually classified into seven morphological types (Hubble
types E, S0, Sa, Sb, Sc, Sdm and Im) to investigate correlations
between galaxy colours, effective sizes and concentrations. The
(inverse) concentration index was found to correlate well with the
visual estimates of morphology. \scite{shima01} have kindly provided
us with their visual morphologies in order to compare them with the
{\sc Galactica} S/T ratios. The are $166$ galaxies in common in the
two samples which have $\chi^2_{\nu}<2.0$. Fig.~\ref{fig:conc_morph}
shows that there is a fair correlation between the (inverse)
concentration index and the visual morphology for these $166$
galaxies, confirming the conclusions of \scite{shima01}. Plotted in
Fig.~\ref{fig:bt_morph} is the correlation between our derived S/T
ratios and the visual morphology for these galaxies. Although the
scatter is large, there is a clear 
trend for the earlier types to have larger S/T ratios.

\begin{thebibliography}{} 
\bibitem[Abraham et al. <1994>]{abr94}
Abraham R.G., Valdes F., Yee H.K.C. \& van den Bergh S., 1994, ApJ, 432, 75-90. 

\bibitem[Allen et al. <2006>]{allen}Allen~P.~D., Driver~S.~P.,
Graham~A.~W., Cameron~E., Liske~J., de Propris~R., 2006, MNRAS, 371, 2

\bibitem[Aller \& Richstone <2002>]{allerrichstone02}
Aller M.~C., Richstone D., 2002, AJ, 124, 3035

\bibitem[Andredakis et al. <1995>]{apb95}
Andredakis Y.C., Peletier R.F. \& Balcells M., 1995, 275, 874A.

\bibitem[Barnes \& Hernquist <1992>]{bar92} 
Barnes J.E. \& Hernquist L., 1992, ARA\&A, 30, 705-742.

\bibitem[Baugh et al. <1996a>]{baugh96a} 
Baugh C.M., Cole S. \& Frenk C.S., 1996a, MNRAS, 282, L27-L32.

\bibitem[Baugh et al. <1996b>]{baugh96b} 
Baugh C.M., Cole S. \& Frenk C.S., 1996b, MNRAS, 283, 1361-1378.

\bibitem[Baugh et al. <2005>]{baugh05}Baugh~C.~M., Lacey~C.~G.,
Frenk~C.~S., Granato~G.~L., Silva~L., Bressan~A., Benson~A.~J.,
Cole~S., MNRAS, 356, 1191 

\bibitem[Balogh et al. <2004>]{balogh04} 
Balogh M., Eke V., Miller C., Lewis I., Bower R.G., Couch W., Nichol
R., Bland-Hawthorn J., Baldry I.K., Baugh C., Bridges T., Cannon R.,
Cole S., Colless M., Collins C., Cross N., and Dalton G., de Propris
R., Driver S.P., Efstathiou G., Ellis R.S., Frenk C.S., Glazebrook K.,
Gomez P., Gray A., Hawkins E., Jackson C., Lahav O., Lumsden S.,
Maddox S., Madgwick D., Norberg P., Peacock J.A., Percival W.,
Peterson B.A., Sutherland W. \& Taylor K., 2004, MNRAS, 348,
1355-1372. 

\bibitem[Balogh et al. <2002>]{balogh02a} 
Balogh, M. L., Bower, R. G., Smail, I., Ziegler, B. L., Davies, R. L.,
Gaztelu, A., \& Fritz A., 2002, MNRAS, 337, 256B.  

\bibitem[Balogh et al. <2004>]{balogh04b} 
Balogh M.L., Baldry I.K., Nichol R., Miller C., Bower R. \& Glazebrook
K., 2004, ApJ, 615, 101 

\bibitem[Beijersbergen et al. <1999>]{bei99} 
Beijersbergen M., de Blok W.J.G. \& van der Hulst J.M., 1999, 351, 903-919.

\bibitem[Bell et al. <2003>]{bell}Bell~E.~F., McIntosh~D.~H., Katz~N., Weinberg~M.~D., 2003, ApJS, 149, 289

\bibitem[Benson et al. <2002>]{benson02}Benson~A.~J., Frenk~C.~S.,
Sharples~R.~M., 2002, ApJ, 574, 104B. 

\bibitem[Benson et al. <2004>]{bensonthickdisk}Benson~A.~J.,
Lacey~C.~G., Frenk~C.~S., Baugh~C.~M., Cole~S., 2004, MNRAS, 351, 1215

\bibitem[Berstein \& Jarvis <2002>]{bern02}Bernstein~G.~M., Jarvis~M., 2002, AJ, 123, 583

\bibitem[Blanton et al. <2003>]{blanton}Blanton~M.~R. et al., 2003, ApJ, 592, 819

\bibitem[Bower et al. <2006>]{bower06}Bower~R.~G., Benson~A.~J.,
Malbon~R., Helly~J.~C., Frenk~C.~S., Baugh~C.~M., Cole~S.,
Lacey~C.~G., 2006, MNRAS, 370, 645 

\bibitem[Bertin \& Arnouts <1996>]{ber96}Bertin E. \& Arnouts S., 1996, A\&A, 117, 393.

\bibitem[Bruzual \& Charlot <1993>]{bc93}Bruzual A. G., \& Charlot S., 1993, ApJ, 405, 538B.

\bibitem[de Jong <1996>]{deJ96}
de Jong R. S., 1996, A\&AS, 118, 557-573.

\bibitem[de Vaucouleurs <1961>]{dev61}
de Vaucouleurs G., 1961, ApJS, 5, 233.

\bibitem[Dressler <1980a>]{d80cat}
Dressler A., 1980, ApJ, 236, 351-365.

\bibitem[Efstathiou <1988>]{efs93}
Efstathiou G., 1993, {\it Les Houches Lectures: Observations of
Large-Scale Structure in the Universe}, Elsevier Science Publishers.  

\bibitem[Efstathiou, Ellis \& Peterson <1988>]{swml}
Efstathiou G., Ellis R. S. \& Peterson B. A., 1988, MNRAS, 232, 431.

\bibitem[Eggen, Lyden-Bell \& Sandage <1962>]{egg62}
Eggen O.J., Lynden-Bell D. \& Sandage A.R., 1962, ApJ, 136, 748.  

\bibitem[Fall <1979>]{fall79}Fall~S.~M., 1979, Nature, 281, 200

\bibitem[Fall \& Efstathiou <1980>]{fallef}Fall~S.~M., Efstathiou~G., 1980, MNRAS, 193, 189

\bibitem[Felten <1977>]{fel77}
Felten J.E., 1977, AJ, 82, 861-878. 

\bibitem[Fukugita et al. <1995>]{fuku95}
Fukugita M., Shimasaku K. \& Ichikawa T., 1995, PASP, 107, 945.

\bibitem[Gadotti \& Kauffmann <2007>]{gadotti07}Gadotti~D., Kauffmann~G., ``Multi-Band Bar/Bulge/Disk Image Decomposition of a Thousand Galaxies'', in ``Stellar Populations as Building Blocks of Galaxies'', proceedings of the IAU Symp. 241, La Palma, Spain, December 2006, A. Vazdekis, R. Peletier (eds.)

\bibitem[Gardner et al. <1996>]{gard96}
Gardner J.P., Sharples R.M., Carrasco B.E. \& Frenk C.S., 1996, MNRAS, 282.

\bibitem[Goto et al. <2002>]{goto02}
Goto T., Okamura S., McKay T.A., Bahcall N.A., Annis J., Bernardi M.,
Brinkmann J., G{\' o}mez P.L., Hansen S., Kim R.S.J., Sekiguchi M. \&
Sheth R.K., 2002, PASJ, 54, 515-525. 

\bibitem[Griffiths et al. <1994>]{grif94}
Griffiths R.E., Casertano S., Ratnatunga K.U., Neuschaefer L.W., Ellis
R.S., Gilmore G.F., Glazebrook K., Santiago B., Huchra J.P., Windhorst
R.A., Pascarelle S.M., Green R.F., Illingworth G.D., Koo D.C. \& Tyson
A.J., 1994, ApJL, 435, L19-L22.  

\bibitem[H\"aring \& Rix <2004>]{harrix}H\"aring~N., Rix~H.-W., 2004,
ApJ, 604, 89 

\bibitem[Hatton et al. <2003>]{hatton03}Hatton~S., Devriendt~J.~E.~G.,
Ninin~S., Bouchet~F.~R., Guiderdoni~B., Vibert~D., 2003, MNRAS, 343,
75 

\bibitem[Kauffmann, White \& Guiderdoni <1993>]{kwg93}Kauffmann~G.,
White~S.~D.~M., Guiderdoni, B. 1993, MNRAS, 264, 201 

\bibitem[Kauffmann, Charlot \& White <1996>]{kcw96}Kauffmann~G.,
Charlot~S., White~S.~D.~M. 1996, MNRAS, 283, 117 

\bibitem[Kauffmann \& Charlot <1998>]{kc98}Kauffmann~G., Charlot~S.,
1998, MNRAS, 297, 23 

\bibitem[Kauffmann \& Haehnelt <2000>]{kh}Kauffmann~G., Haehnelt~M., 2000, MNRAS, 358, 2121

\bibitem[Kauffmann et al. <2003>]{Kauffmann03}Kauffmann~G. et al., 2003 ,MNRAS, 341, 54

\bibitem[Kent <1985>]{kent85}
Kent, S. M., 1985, AJSS, 59, 115-159.

\bibitem[Kormendy \& Richstone <1995>]{kr95}Kormendy J., Richstone D., 1995, ARA\&A, 33, 581 

\bibitem[Krist <1995>]{krist95}
Krist J., 1995, {\it ASP Conf. Ser. 77: Astronomical Data Analysis
Software and Systems IV}, 349.

\bibitem[Lambas, Maddox \& Loveday <1992>]{lml92}Lambas~D.~G., Maddox~S., Loveday~J., 1992, MNRAS, 258, 404

\bibitem[Lupton et al. <2001>]{lup01}
Lupton R.H., Gunn J.E., Ivezi{\' c} Z., Knapp G.R., Kent S. \& Yasuda
N., 2001, ADASS, 10, 269L 

\bibitem[Magorrian et al. <1998>]{mag98}Magorrian~J., et al., 1998, AJ, 115, 2285

\bibitem[Marconi \& Hunt <2003>]{mh03}Marconi~A., Hunt~L.~K., 2003, ApJ, 589, L21

\bibitem[Marconi et al. <2004>]{marconi04}
Marconi A., Risalti G., Gilli R., Hunt L.~K., Maiolino R., Salvati M.,
2004, MNRAS, 351, 169

\bibitem[Malbon et al. <2006>]{malbon06}Malbon~R., Baugh~C.~M.,
Frenk~C.~S., Lacey~C.~G., 2006, MNRAS submitted (astro-ph/0607424) 

\bibitem[McLure \& Dunlop <2004>]{md04}
McLure R.~J., Dunlop J.~S., 2004, MNRAS, 353, 1390

\bibitem[Merritt \& Ferrarese <2001>]{mf01}Merritt~D., Ferrarese~L.,
2001, MNRAS, 320, L30 

\bibitem[Metropolis et al. <1953>]{metro53}
Metropolis N., Rosenbluth A., Rosenbluth M., Teller A. \& Teller E.,
1953, Journal of Chemical Physics, 21, 1087 

\bibitem[Mo, Mao \& White <1998>]{mmw}
Mo~H.~J., Mao~S., White~S.~D.~M., 1998, MNRAS, 295, 319

\bibitem[Moffat <1969>]{moff69}
Moffat A.F.J., 1969, AAP, 3, 455

\bibitem[Nakamura et al. <2003>]{naka03}
Nakamura O., Fukugita M., Yasuda N., Loveday J., Brinkmann J.,
Schneider D.P., Shimasaku K. \& SubbaRao M., 2003, AJ, 125,
1682-1688. 

\bibitem[Nelson et al. <2002>]{nelson02}
Nelson A. E., Simard L., Zaritsky D., Dalcanton J. J. \& Gonzalez
A. H., 2002, AJ, 567, 144. 

\bibitem[Press et al. <1992>]{numrec}
Press W.H., Teukolsky S.A., Vetterling W.T. \& Flannery B.P., 1992,
{\it Numerical Recipes in FORTRAN}, Cambridge University Press,
Chapter 10. 

\bibitem[Ratnatunga et al. <1999>]{rat99}
Ratnatunga K.U., Griffiths R.E. \& Ostrander E.J., 1999, AJ, 118, 86-107.

\bibitem[Salpeter <1955>]{sal55}
Salpeter E.E., 1955, ApJ, 121, 161

\bibitem[Sandage, Tammann \& Yahil <1979>]{sty}
Sandage A., Tammann G.A. \& Yahil A., 1979, ApJ, 232, 352-364.

\bibitem[Schechter <1976>]{sch}
Schechter P., 1976, ApJ, 203, 297-306.

\bibitem[Schechter \& Dressler <1987>]{sd87}
Schechter P.L. \& Dressler A., 1987, AJ, 94, 563S.

\bibitem[S\'{e}rsic <1968>]{sersic}
S\'{e}rsic J.-L., 1968, {\it Atlas de Galaxias Australes}, Cordoba:
Obs. Astronomico, Generalised $R^{1/4}$ law 

\bibitem[Shankar et al. <2004>]{shankar04}
Shankar F., Salucci P., Granato G.~L., De Zotti G., Danese L., 2004, MNRAS, 354, 1020

\bibitem[Shimasaku et al. <2001>]{shima01}
Shimasaku K., Fukugita M., Doi M., Hamabe M., Ichikawa T., Okamura S.,
Sekiguchi M., Yasuda N., Brinkmann J., Csabai I., Ichikawa S.,
Ivezi{\' c} Z., Kunszt P.Z., Schneider D.P., Szokoly G.P., Watanabe
M. \& York D.G., 2001, AJ, 122, 1238-1250 

\bibitem[Simard et al. <2002>]{sim02}
Simard L., Willmer C.N.A., Vogt N.P., Sarajedini V.L., Phillips A.C.,
Weiner B.J., Koo D.C., Im M., Illingworth G.K. \& Faber S.M., 2002,
ApJS, 142, 1 

\bibitem[Somerville \& Primack <1999>]{sp99}Somerville~R.~S.,
Primack~J.~R., 1999, MNRAS, 310, 1087 

\bibitem[Stoughton et al. <2002>]{sdssedr}
Stoughton C., et al., 2002, AJ, 123, 485S

\bibitem[Strateva et al. <2001>]{stra01}
Strateva I., et al., 2001, AJ, 122, 1861

\bibitem[Tasca \& White <2005>]{tw2}Tasca~L., White~S.~D.~M., 2005, in
``Multiwavelength mapping of galaxy formation and evolution'',
Proceedings of the ESO Workshop, eds. A.~Renzini and R.~Bender.,
Springer, Berlin, p.465 

\bibitem[Tasca \& White <2006>]{tw}Tasca~L., White~S.~D.~M., 2006, astro-ph/0507249

\bibitem[Trujillo et al. <2001>]{tru01}
Trujillo I., Aguerri J.A.L., Cepa J. \& Guti{\' e}rrez C.M., 2001, MNRAS, 328, 977-985 

\bibitem[Wadadekar et al. <1999>]{wad99}
Wadadekar Y., Robbason B. \& Kembhavi A., 1999, AJ, 117, 1219-1228

\bibitem[Yu \& Tremaine <2002>]{yt02}
Yu Q., Tremaine S., 2002, MNRAS, 335, 965

\end{thebibliography}
\end{document}